\documentclass[10pt]{article}
\pdfoutput=1
\usepackage{geometry}
\usepackage{graphics}
\usepackage{rotating}
\usepackage{subfigure}
\usepackage{textcomp}
 
\usepackage[latin1]{inputenc}
\usepackage{color}
\usepackage{array}
\usepackage{longtable}
\usepackage{lscape}
\usepackage{calc}
\usepackage{multirow}
\usepackage{nomencl}
\usepackage{hhline}
\usepackage{ifthen}
\usepackage{url}
\usepackage{lscape}

\begin{document}

\newlength{\gnumericTableWidth}
\newlength{\gnumericTableWidthComplete}
\newcommand{\is}{$s^{-1}$}

\newcommand{\RS}{\mbox{$\Lambda_S$}}
\newcommand{\RD}{\mbox{$\Lambda_d$}}
\newcommand{\RQ}{\mbox{$\Lambda_q$}}
\newcommand{\rz}{\mbox{$\lambda_\mu^+$}}
\newcommand{\Rq}{\mbox{$\Lambda_q$}}
\newcommand{\Rd}{\mbox{$\Lambda_d$}}
\newcommand{\RHE}{\mbox{$\Lambda_{He}$}}
\newcommand{\MHE}{\mbox{$\mu ^3$He}}
\newcommand{\qr}{\mbox{$\lambda_q$}}
\newcommand{\dr}{\mbox{$\lambda_d$}}
\newcommand{\qdr}{\mbox{$\lambda_{qd}$}}
\newcommand{\dqr}{\mbox{$\lambda_{dq}$}}
\newcommand{\precision}{1.5\,\%}
\newcommand{\mustopfrac}{85\,\%}
\newcommand{\degc}{$^\circ$C}
\newcommand{\mlan}{\mbox{$\mu$Lan}}
\newcommand{\gf}{$G_\mathrm{F}$}
\newcommand{\Xo}{$X_{0}$}
\newcommand{\dedx}{$dE/dx$}
\newcommand{\epl}{e$^{+}$}
\newcommand{\mupl}{\mbox{$\mu^{+}$}}
\newcommand{\mus}{\mbox{$\mu s$}}
\newcommand{\mup}{\mbox{$\mu^{+}$}}
\newcommand{\muc}{\mbox{$\mu p$}}
\newcommand{\mud}{\mbox{$\mu d$}}
\newcommand{\m}{\mbox{$\mu$}}
\newcommand{\pii}{\mbox{$\pi$}}
\newcommand{\mum}{$\mu^{-}$}
\newcommand{\pipl}{$\pi^{+}$}
\newcommand{\taumu}{$\tau_{\mu^+}$}
\newcommand{\om}{\mbox{$\omega$}}
\newcommand{\be}{\begin{enumerate}}
\newcommand{\ee}{\end{enumerate}}
\newcommand{\beq}{\begin{equation}}
\newcommand{\eeq}{\end{equation}}
\newcommand{\bi}{\begin{itemize}}
\newcommand{\ei}{\end{itemize}}
\newcommand{\ins}{\mbox{s$^{-1}$}}
\newcommand{\D}[2]{\frac{\partial #2}{\partial #1}}
\newcommand{\CIC}{cryo-TPC}
\newcommand{\capCIC}{Cryo-TPC}
\newcommand{\gA}{$g^{}_A$}
\newcommand{\gP}{$g^{}_P$}
\newcommand{\pionless}{\rotatebox{20}{\small $\backslash$}$\hspace{-0.6em}\pi$}
\newcommand{\dhr}{\mbox{$\hat{d}^R$}}
\newcommand{\pnpi}{\mbox{$^a$}}
\newcommand{\uiuc}{\mbox{$^b$}}
\newcommand{\PSI}{\mbox{$^c$}}
\newcommand{\uk}{\mbox{$^d$}}
\newcommand{\bu}{\mbox{$^e$}}
\newcommand{\ucl}{\mbox{$^f$}}
\newcommand{\ru}{\mbox{$^g$}}
\newcommand{\usc}{\mbox{$^h$}}

\vspace{.0cm}
{\raggedleft February 8, 2008 \\
\par}
\vspace{.5cm}
\begin{center}  

{\bf\LARGE Muon Capture on the Deuteron \\ 
\vspace{0.2cm}
\Large\em{The MuSun Experiment\\}}
\vspace{.5cm}

\Large
MuSun Collaboration
\vspace{.5cm}

\begin{figure}[h]
  \vspace{-15mm}
  \begin{center}
  \includegraphics[scale=0.6]{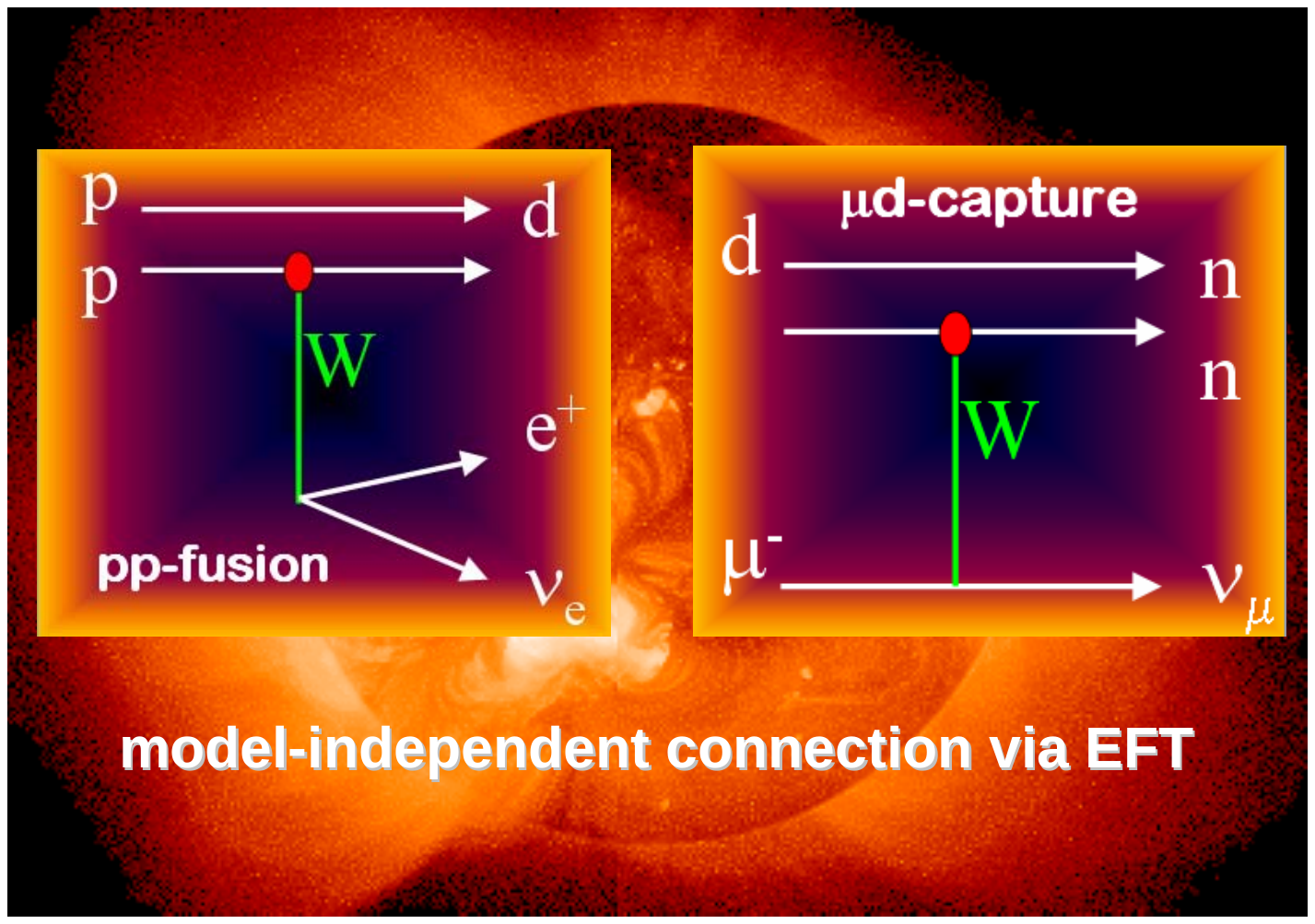}
  \vspace{-50mm}
  \end{center}
\end{figure}

\vspace{0mm}

http://www.npl.uiuc.edu/exp/musun

\end{center}

\begin{center}
\vspace{10mm}

V.A.~Andreev\pnpi, 
R.M. Carey\bu, 
V.A.~Ganzha\pnpi, 
A. Gardestig\usc,
T.~Gorringe\uk, 
F.E.~Gray\ru, 
D.W.~Hertzog\uiuc, 
M.~Hildebrandt\PSI, 
\underline{P.~Kammel}\uiuc, 
B.~Kiburg\uiuc, 
S.~Knaack\uiuc, 
P.A.~Kravtsov\pnpi, 
A.G.~Krivshich\pnpi,
K. Kubodera\usc, 
B.~Lauss\PSI,
M.~Levchenko\pnpi,
K.R.~Lynch\bu, 
E.M.~Maev\pnpi, 
O.E.~Maev\pnpi,
F.~Mulhauser\uiuc,
F. Myhrer\usc, 
\underline{C.~Petitjean}\PSI, 
G.E.~Petrov\pnpi, 
R.~Prieels\ucl, 
G.N.~Schapkin\pnpi,
G.G.~Semenchuk\pnpi, 
M.A.~Soroka\pnpi, 
V.~Tishchenko\uk, 
\underline{A.A.~Vasilyev}\pnpi, 
A.A.~Vorobyov\pnpi, 
M.E.~Vznuzdaev\pnpi,
P.~Winter\uiuc

\vspace{10mm}
 
\pnpi Petersburg Nuclear Physics Institute, Gatchina 188350, Russia

\uiuc University of Illinois at Urbana-Champaign, Urbana, IL 61801, USA

\PSI Paul Scherrer Institute, CH-5232 Villigen PSI, Switzerland

\uk University of Kentucky, Lexington, KY 40506, USA


\bu Boston University, Boston, MA 02215, USA 

\ucl Universit{\'e} Catholique de Louvain, B-1348 Louvain-la-Neuve, Belgium

\ru Regis University, Denver, CO 80221, USA

\usc University of South Carolina, Columbia, SC 29208, USA

\vspace{15mm}
Co-spokespersons underlined.

\end{center}

\newpage

\vspace{2cm}
{\bf Abstract:}
We propose to measure the rate \Rd\ for muon capture on the deuteron to better than 1.5\% precision. 
This process is the simplest weak interaction process on a 
nucleus that can both be calculated and measured to a high degree of precision.
The measurement will provide a benchmark result, far more precise than any current experimental 
information on weak interaction processes in the two-nucleon system. 
Moreover, it can impact our understanding of fundamental reactions of astrophysical 
interest, like solar pp fusion and the $\nu+d$ reactions observed by the Sudbury 
Neutrino Observatory. Recent effective field theory calculations have demonstrated, 
that all these reactions are related by one axial two-body current term, parameterized 
by a single low-energy constant. Muon capture on the deuteron is a clean and accurate way 
to determine this constant. Once it is known, the above mentioned astrophysical, as well
as other important two-nucleon reactions, will be determined in a model independent way at the 
same precision as the measured muon capture reaction.

At the moment the experimental situation on $\mu+d$ capture is inconclusive. An experiment
with 10\% errors agrees with theory, the most precise measurement with 6.2\% uncertainty
disagrees by three standard deviations from the best recent calculation, which has 1\% uncertainty.
If true, such a discrepancy would have major ramnifications on the above mentioned  
astrophysical processes.  The required significant improvement in precision expected with the 
MuSun experiment became feasible by the advanced techniques developed for the MuCap experiment. 
As in the case of that experiment, utmost care is required to eliminate
uncertainties due to muon atomic physics effects. Thus, while the general experimental strategy
is based on MuCap, a new cryogenic TPC operating at gas densities of 5\% of LH$_2$ at 30 K
will be developed to achieve optimal conditions for an unambiguous interpretation of the experiment.
The TPC will be filled with ultrapure deuterium and operated as a high resolution ionization chamber. 
The different physics requirements of the new MuSun experiment demand several upgrades to the MuCap detector,
including full analog readout of the cryo-TPC, the monitoring the muon chemistry by charged particle, neutron and 
$\gamma$ detection and an advanced D$_2$ gas purification system. 

\newpage

\tableofcontents

\newpage

\section{Beam Requirements and Beam Request}

Experimental Area:\\
$\pi$E3 equipped with a $\mu/e$ separator and the muon on request beam line setup using the MuLan kicker.\\[0.3cm]
Required beam properties:


\begin {itemize}

\item Particle:  $\mu^+$, $\mu^-$

\item Momentum:  30 - 50 MeV/c

\item Momentum width:  3$\%$ FWHM

\item Beam spot:  5 cm diameter max

\item Intensity:  $\approx 10^5 s^{-1}$ (requiring normal high intensity ring operation) 

\item Beam purity:   $\pi/\mu < 10^{-5}$,$e/\mu < 20 \%$

\end{itemize}
\vspace{0.3cm}
Duration of experiment:\\
Based on the experience with the MuCap experiment we would ask PSI after discussion with other area users, specifically the $\mu$SR facility, to provide space for the MuSun apparatus to be permanently positioned inside the $\pi$E3 area. If this is not possible, we emphasise the absolute necessity of 6~weeks preparation time before a run in a suitable mounting space in the experimental hall ($\sim$ 20 $m^2$) due to the complexity of the setup before an experimental run.


\begin{itemize}

\item  Preparation time in WEHA before first run 6 weeks;

\item  Test run with beam (5 weeks) in 2008;

\item  Engineering run and data taking (5 + 8 weeks) in 2009;

\item  Final production beam time is estimated to be 22 weeks. 

\end{itemize}
Further requests will depend on the results of the test and engineering runs.






\section{Questions of Safety}

\begin{enumerate}

\item There is no dangerous radioactivity involved (only some calibration sources).

\item The TPC detector is an active target filled with deuterium gas (volume $\sim$ 20 {\it l}, pressure $\le$ 10 bar at 30K). The usual hydrogen safety precautions shall be taken. The beryllium beam window 
is the weakest part of the hydrogen pressure vessel. It will be extensively tested during long term running.

\item The experiment will be operated in an air-conditioned climate tent with monitoring equipment for hydrogen, oxygen and flammable gases, similar to the equipment successfully operated over years within the MuCap experiment's climate tent.

\item Standard precautions for working with a cryogenic apparatus will be taken.
\end{enumerate}

\newpage

\newpage
\section{Physics Motivation}

The MuSun experiment will measure the rate \RD\ for the semileptonic 
weak process
\begin{equation}
\mu^- + d \to \nu_{\mu} + n + n
\label{eq:mudcap}
\end{equation}
to a precision of better than 1.5~\%. 
\RD\ denotes the capture rate from the doublet hyperfine state of 
the muonic deuterium atom in its 1S ground state. 
The measurement, based on novel techniques,  would exceed the precision of 
previous experiments by nearly an order of magnitude.
Here, we summarize the primary physics motivation, while more
details are provided in the next section.

\begin{itemize}

\item
Muon capture on the deuteron is the simplest weak interaction process on a 
nucleus which can both be calculated and measured to a high degree of precision.
The MuCap experiment, which published initial physics 
results~\cite{Andreev:2007wg} and successfully finished data taking in 2007, will determine
the singlet capture rate \RS\ of the basic process on a free nucleon 
$ \mu^- + p \to \nu_{\mu} + n$ to better than 1\%;
a prerequisite for precise calculations of muon capture. 
At the same time, modern effective field theories (EFTs) have been highly successful 
in calculating low-energy phenomena from first principles~\cite{Bernard:1995dp}.
Reaction~(\ref{eq:mudcap}) could serve as a benchmark of our
understanding of weak processes in the two-nucleon system. 
However, the best existing experiments~\cite{Bardin:1986,Cargnelli:1989} 
are not precise enough and the most precise result differs 
from modern theory~\cite{Ando:2001es,Chen:2005ak} by 2.9 standard deviations.

\item
Reaction~(\ref{eq:mudcap}) is closely related to fundamental reactions of 
astrophysical interest. These include the $p+p \rightarrow d +e^+ + \nu_e$ reaction, 
which is the primary energy source in the sun and the main sequence stars, and the 
$\nu + d$ reaction, which provided convincing evidence for solar neutrino
oscillation, as both its charged current and neutral
current modes are observed simultaneously at the Sudbury Neutrino 
Observatory~\cite{Aharmim:2007nv}. 
While the vector current interaction on the deuteron is scrupulously tested by a 
comprehensive set of experiments on electromagnetic observables, direct experiments
on the axial-vector interaction with the two-nucleon system are scarce and have not come even 
close to the required precision. 
The above mentioned astrophysical processes responsible for the slow burning of the stars 
are simply too feeble to be observed in the laboratory. 
Here again, the development of EFTs during the last years has led to an important 
model-independent connection. 
It was proved that, up to the required precision in the 
systematic chiral expansion, these weak reactions are related by a two-nucleon current term,
whose strength is parameterized by a single low-energy 
constant~\cite{Park:1998wq,Nakamura:2000vp,Ando:2001es,Ando:2002pv,Butler:2002cw}. 
The constant integrates all the short-distance physics, which is not well constrained and
considered the main theoretical uncertainty in these processes.
The proposed MuSun experiment can determine this constant precisely from  muon capture
on the deuteron and thus comes closest to calibrating these basic astrophysical
reactions under terrestrial conditions.

\item
The MuSun measurement is even more broadly related to different
physics via EFTs. The low-energy constant representing the coupling
of the axial current to the two-nucleon system resembles \gA\ in the one-nucleon sector. 
Analogously to the Goldberger-Treiman relation it relates the two-nucleon axial vector 
interaction to the coupling of a $p$-wave pion to two nucleons.
For example, Ref.~\cite{Gardestig:2006hj} points out 
that a precise measurement of $\mu+ d$ capture will significantly reduce the uncertainty in
the $nn$ scattering length a$_{nn}$= 18.59 $\pm$ 0.27(exp) $\pm$ 0.30(theory) fm 
extracted from radiative pion capture $\pi^-+ d \to \gamma + n + n $, 
by nearly eliminating the theory uncertainty. 
The difference between the $pp$ and $nn$ scattering lengths represents important data to quantify 
isospin symmetry breaking caused by the up and down quark mass difference in QCD.\\
On a more methodological aspect,  reaction~(\ref{eq:mudcap}) will allow a detailed 
comparison with existing calculations~\cite{Park:2002yp}, where the critical 2$N$ axial current is
determined from the more complex three-nucleon system (tritium beta decay).
Such a calculation can currently only be performed in a hybrid approach (using phenomenological 
wave functions), whereas the 2$N$ system can be successfully treated within a fully consistent
chiral perturbation theory  framework.
There is considerable practical interest in the verification of the reliability of the 
hybrid approach, which does not follow the strict chiral order counting, but is applicable 
in a wider range of few-body systems, like, e.g., the solar 
hep reaction~\cite{Kubodera:2004zm}, which is notoriously difficult to calculate.

\end{itemize}





\section{Muon Capture on the Deuteron}

\subsection{Theoretical Framework}

During the last decade, effective field theories (EFTs), especially 
chiral perturbation theory (ChPT), have been used
as a natural theoretical framework for calculating weak processes
on a single nucleon, the deuteron and even the three- and four-nucleon systems. 
ChPT inherits the relevant symmetries of QCD and its  
parameters are linked to matrix elements of QCD operators. 
At sufficiently low energy-momentum 
transfer, there has been tremendous theoretical progress, with strong 
experimental confirmation,
in applying EFTs to a variety of observables 
in the one-nucleon sector .

As developed in the pioneering work of 
Weinberg~\cite{Weinberg:1990rz,Weinberg:1991um,Weinberg:1992yk}, 
one can construct an EFT applicable to multi-nucleon systems. 
All EFTs rely on an expansion scheme in a small parameter $Q/\Lambda \ll 1$, where $Q$ is the 
four-momentum of the relevant process and $\Lambda$ indicates the relevant mass scale.
The expansion is in powers of $Q/\Lambda$, where the 
higher order terms give smaller corrections to the dominant lowest order terms. 
The {\it effective} nature of EFT is reflected in the presence of unknown coefficients, 
called the low-energy-constants (LECs), which parameterize the high-energy physics 
that, in generating the low-energy EFT, has been integrated out. 
In principle these LECs can be evaluated from QCD, but in practice they are determined 
from experimental data. Once the LECs are determined, the theory 
will make unambiguous predictions of observables for many different processes. 
For example, the $\mu^- + p$ capture rate has been evaluated including one-loop corrections, 
a level of precision  that includes several LECs which are 
determined empirically from other reactions~\cite{Ando:2000zw}.
The pseudo-scalar constant \gP, linked to the pion pole contribution in 
$\mu^- + p$ capture, is evaluated to high precision in ChPT which gives an
expression for \gP\ in terms of the fundamental LECs, $f_\pi$ and \gA\ and the nucleon
axial radius, which are  known.  

In the pion-less EFT (where the pion is considered a high-energy degree 
of freedom and has been integrated out) the expansion parameter is $ Q/m_\pi\ll1$, 
where $Q$ is a typical energy or momentum of the reaction and $m_\pi$ is the pion mass.
In ChPT, on the other hand, the expansion parameter is $ Q / \Lambda_\chi\ll1$,
where $\Lambda_\chi \simeq 4\pi f_\pi \simeq m_N $ with $m_N$ the nucleon mass.
At very low energies ($Q\ll m_\pi$) the pion-less EFT (\pionless EFT) is very useful. 
However, for $\mu^- + d$ capture $Q \simeq m_\pi$. Therefore ChPT, which is valid 
over a much larger energy-momentum region, is the better theory in this case. 
This conclusion is also obvious from Fig.~\ref{dp}, which shows the entire phase 
space available for $\mu^-+d$ capture.
The \pionless EFT applies only in the bottom left part of the figure, as 
indicated\footnote{Experimentally we could determine the reduced capture rate
for this region of $p_\nu \geq$  90 MeV/c by measuring both the total capture rate
and the higher energy Dalitz plot region by detecting neutrons with energies 
above 10 MeV. We are evaluating the overall physics motivation for such an 
expansion of the experiment, but that would be a separate proposal.}.
In contrast, ChPT is valid for the entire phase space, with some modifications perhaps
needed in the region closer to the $p_\nu=0$ point,
where the momentum transfer becomes large.
However, as shown in Ref.~\cite{Ando:2001es}, the contribution from this region to the 
total capture rate is marginal. Also, due to the larger range of allowed energies, 
ChPT applies to more processes than \pionless EFT.

At the desired level of precision the two-nucleon (2$N$) system has one new unknown 
LEC characterizing the short-distance 2$N$ axial current. 
This LEC is called $L_{1A}$ in the pionless theory~\cite{Chen:2005ak} and 
$\hat{d}^R$ in ChPT~\cite{Ando:2001es}. In the latter case the pion-exchange 
currents between the 2$N$s are explicitly evaluated, while they are embedded in 
$L_{1A}$ in the \pionless EFT. The coupling constant $\hat{d}^R$ enters in a range of 
important weak and pionic reactions, e.g., $\mu^- + d$, $p+p$ fusion~\cite{Park:2002yp}, $\nu+ d$ 
reactions~\cite{Nakamura:2002he}, and pion radiative capture on the deuteron~\cite{Gardestig:2006hj}. 
At present, $\hat{d}^R$ is determined from triton beta decay using so-called hybrid ChPT
(employing phenomenological wave functions).
Theoretically it is highly desirable to determine its value within the 2$N$ system and in a 
consistent ChPT framework. 
Once $\hat{d}^R$ is determined in this way, the solar $p+p$ fusion and 
$\nu d$ SNO reactions will be determined in a model independent way 
at the same precision as the measured $\mu^- + d$ capture reaction.
Recent developments have also made it possible to derive the deuteron and 
scattering state wave functions within chiral perturbation 
theory~\cite{Phillips:1999am,Gardestig:2005pp,Kaiser:1997mw,Epelbaum:1999dj,
Rentmeester:1999vw,Entem:2003ft}, without having to resort to the hybrid approach.
Thus all the relevant 2$N$ processes can be calculated ab-initio entirely within
the same consistent framework.

\subsection{Capture Rate Calculations, Status and Future}

\begin{figure}[hbt] 
\vspace{-1.cm}  
\begin{center}
\resizebox*{1.\textwidth}{0.5\textheight}{\includegraphics{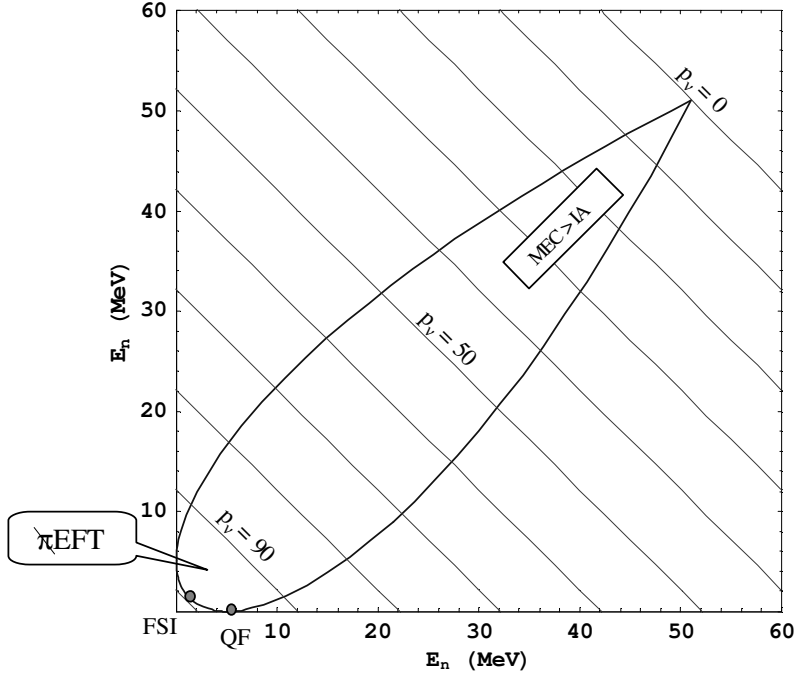}} 
\vspace{-1.6cm}
\caption{\label{dp} $\mu^- + d$ capture Dalitz plot as function of the neutron kinetic energy. Diagonal lines indicate constant neutrino momentum $p_\nu$ (MeV/c). Some interesting kinematic regimes are marked in the graph: final state interaction (FSI); quasifree (QF); $p_\nu \ge 90$~MeV/c, where pion-less EFT applies; small  $p_\nu$, where MECs dominate over impulse approximation.}
\end{center}
\vspace{-.4cm}
\end{figure}

Compared to the elementary $\mu^-+p$ capture, several additional features are 
important in the $\mu^-+d$ capture process. The properties of the 2$N$ system 
enter, in particular the deuteron wave function in the initial-, and the neutron-neutron
scattering length $a_{nn}$ in the final state. 
The above mentioned two-body axial currents contribute, making process~(\ref{eq:mudcap}) 
uniquely suited to study them in the 2$N$ system. 
The final three-body state allows a broad range of momentum transfer to the 
2$N$ system (see~Fig.~\ref{dp}). 
The total energy $Q= 102.1$~MeV in the final state, adequately described in 
non-relativistic kinematics, is the sum of the neutron CMS energy, two neutron relative energy 
$E_{nn}$, and the neutrino momentum $p_\nu$.
\begin{equation}
Q= \frac{p_\nu^2}{4 M_n} + E_{nn} + p_\nu.
\end{equation}
Thus the kinematics can be parameterized either by $E_{nn}$ or 
equivalently by   $p_\nu$.

Traditionally, muon capture has been calculated in the standard nuclear physics 
approach, essentially an impulse approximation calculation based on
realistic nucleon-nucleon potentials supplemented by explicit models of the two-body
meson exchange currents. The most sophisticated of these calculations were 
performed around 1990~\cite{Tatara:1990eb,Adam:1990kf, Doi:1989kv}. 
Thereafter the concept of effective field theories in the form of ChPT was expanded to the 2$N$ system. 
The latest calculations were performed in hybrid EFT~\cite{Ando:2001es} and most 
recently in \pionless EFT~\cite{Chen:2005ak}. 
Figure~\ref{f:deut_exp} summarizes the theoretical and experimental results.

Present calculations agree that the one-body operators are very well defined. 
The challenge lies in the short range part of the axial two-body current. 
In the meson exchange picture, it is dominated by the $\pi N N^*$ isobar 
current, which is not constrained by general symmetry principles. 
The systematic expansion of ChPT demonstrates that exactly the same 
combinations of low energy constants appear in the two body 
reactions $p+p$ fusion, $\nu+d$ scattering and $\mu^-+d$ capture. 
The hybrid EFT used tritium beta decay to constrain the unknown LEC. 
The \pionless EFT treats this LEC, called $L_{1A}$, as an unknown parameter 
and parameterizes the capture rate as
\begin{equation}
 \Lambda_d = a + b~ L_{1A}
\label{lamab.eq}
\end{equation}
where a and b weakly depend on the $E_{nn}$ Dalitz plot cut, provided $E_{nn}$ is limited to the 
kinematic region where $p_\nu<90$~MeV/c. For $E_{nn}=5$~MeV, a=239.2\ins and b=3.3\ins fm$^{-3}$.
The currently best estimate of $L_{1A}$ depends on hybrid ChPT calculations~\cite{Chen:2005ak}, see 
also Table~\ref{l1a.tab}. 
The hybrid ChPT calculation~\cite{Ando:2001es} obtains \Rd = 386 \ins\ and estimates an uncertainty of 1\%, 
resulting from a small dependence on the cutoff parameter, uncertainties in the tritium beta decay 
rate~\cite{Park:2002yp}, higher MEC contributions and
a phenomenological estimate of the contribution of the $L\ge1$ partial waves~\cite{Tatara:1990eb}. 
Reference~\cite{Ando:2001es} also notes that radiative
corrections still need to be calculated. 
The \pionless EFT paper concludes that its uncertainty of 2-3\% is dominated by N$^3$LO 
contribution, not yet calculated~\cite{Chen:2005ak}.
However, it would still rely on hybrid ChPT for a precise value for $L_{1A}$.

As mentioned above, the theoretical program supporting the MuSun experiment plans to calculate
$\mu^-+d$ ab-initio in a fully consistent approach, where both the 2$N$ wave functions
and the operators are derived within the recently developed ChPT framework. 
An advantage over the present \pionless EFT calculation is that the relevant kinematic range 
of the reaction is fully within the convergence radius of the theory and that the higher 
order effects are estimated to contribute less than 0.5\%. 

At the level of precision aimed at in the proposed $\mu^- + d$ 
capture experiment, an evaluation of the radiative corrections will be necessary. 
Traditionally, the evaluation of radiative corrections for a nucleon is based on 
either the quark picture  or the pre-EFT hadronic picture. 
In the former, the calculation is well defined at the quark-lepton 
level~\cite{Marciano:1985pd,Czarnecki:2007th}.
Meanwhile, calculations based on the pre-EFT hadron picture involve intrinsic model dependence.
ChPT provides a reliable systematic framework which respects all required symmetry properties 
and wherein one can enumerate and evaluate all the relevant Feynman diagrams up to a 
specified chiral order (see, e.g., \cite{Bernard:1995dp}).
The ChPT evaluation of radiative corrections of neutron $\beta$-decay was developed in 
Ref.~\cite{Ando:2004rk},  and this formalism can readily be applied to $\mu^- + p$ capture. 
This treatment involves unknown LECs associated with photon loops.
These LECs are constrained using the high-precision data on neutron $\beta$-decay and can then be 
used in $\mu^- + p$ capture.
Furthermore, ChPT can be applied in the calculations of $\mu^- + d$ radiative corrections.

\subsection{Experiment}
\begin{figure}[hbt]
\vspace{-0.5cm}  
\begin{center}
\includegraphics[angle=90,width=0.8\textwidth]{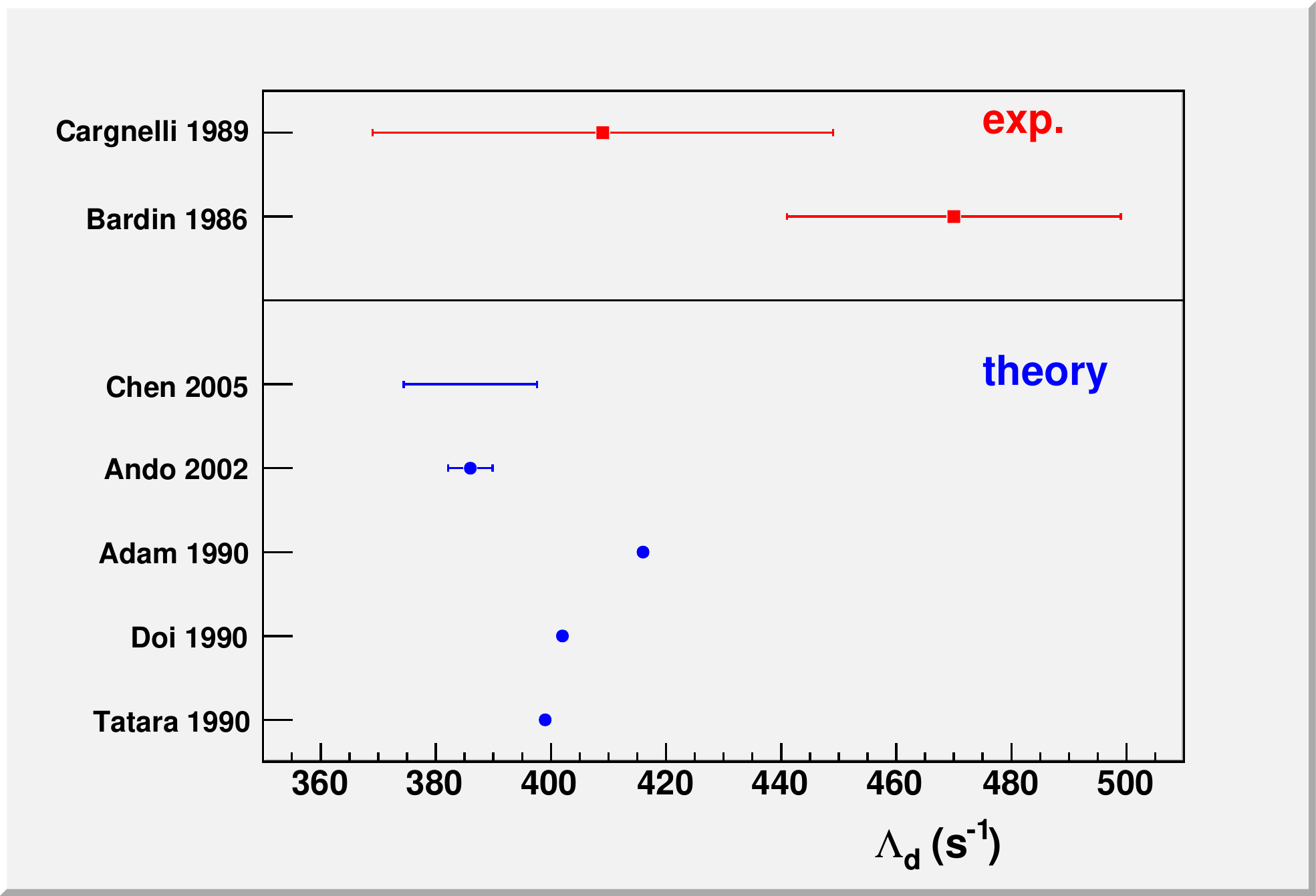} 
\vspace{-.0cm}
\caption{\label{f:deut_exp} Recent theoretical and experimental results on \RD. The theory improvements due to the EFT method are evident. For the first
time systematic theoretical error estimates are provided. 
The calculation of Ando 2002~\cite{Ando:2001es} claims an uncertainty of 1\%, albeit
based on a hybrid calculation. The error range for the pion-less EFT 
calculation of Chen 2005~\cite{Chen:2005ak}, applicable to the low momentum transfer region, 
includes the estimated uncertainty of uncalculated chiral orders. For the value shown in the figure,
$L_{1A}$ was adjusted to reproduce the rate of Ando et al. In the near future, \Rd\ will be calculated
with half the uncertainty of Ando 2002, i.e. 0.5\%,  based on a fully consistent ChPT formalism. This precision level is the challenge for the new MuSun experiment.}
\end{center}
\vspace{-0.cm} 
\end{figure}

The early counter experiments were based on the observation of capture neutrons. 
While a precise measurement of the 
absolute neutron emission rate is difficult even for 5.2 MeV neutrons resulting
from $\mu^- +p$ capture, it is even harder for the continuous spectrum of neutrons emitted in 
$\mu^- +d$ capture which is peaked around 1.5 MeV. Moreover, numerous 2.45 MeV neutrons are produced
by muon-catalyzed $dd$ fusion, representing a significant background. The interpretation
of the experiments requires an accurate knowledge of $d\mu^-$ hyperfine population at the
moment of capture, because the \mbox{V-A} structure of weak interactions suppresses capture from
the quartet relative to the doublet state. 
The first generation experiments~\cite{Wang:1965, Bertin:1974sr} tried to 
reduce the background from $dd$ fusion, by using hydrogen targets with small deuterium concentration.
Because of the fast $p\mu^-$ to $d\mu^-$ transfer rate, such a target acts as an effective
deuterium target. The price to pay for such an approach is that at high densities capture
occurs from a molecular state and the background from \MHE\ build-up is significant. For lower
densities, on the other hand, the $d\mu^-$ hyperfine population was essentially unknown at that
time and was conjectured as being pure doublet.

\begin{table}[htb]
\begin{center}
\begin{tabular}{ccccccc}
\hline 
 Ref. & $\phi(\%)$ 	& $c_D (\%)$  &T (K)& detection & statistics & \Rd (s$^{-1}$) \cr
\hline
 \cite{Wang:1965} 	& 100        &  0.32         &20&  neutron  &  615       &  365$\pm$96    \cr
 \cite{Bertin:1974sr} 	& 0.76        &  5       &300&  neutron  &  6295       &   ? $^{*)}$ \cr
 \cite{Bardin:1986} 	& 100         &  100        &20&  electron  &  5$\times$10$^8$    &  470$\pm$29   \cr
 \cite{Cargnelli:1989}	& 4          &  100         &45&  neutron  &  $\approx$ 9000   &  409$\pm$40  \cr
\hline
\end{tabular}
\caption{Experiments on $\mu^- +d $ capture. 
Density $\phi$ normalized to LH$_2$ density. Experiments performed in D+H mixtures, with $c_D$ given in column 3.  
$^{*)}$~Result uncertain, because experimental conditions
corresponds to mostly statistical, not doublet d$\mu^-$ hyperfine population, as originally assumed. }
\end{center}
\vspace{-.5cm}
\end{table}

The situation changed with the discovery~\cite{Kammel:1984vu} of a strong temperature dependent 
hyperfine effect in resonant $dd\mu^-$ formation, which made the $d\mu^-$ hyperfine population 
directly experimentally accessible. This and subsequent experimental and theoretical work showed that the 
conditions in experiment~\cite{Bertin:1974sr} were closer to statistical, than doublet. As a consequence
the reported experimental rate of \Rd = 445$\pm$60 \ins\ would have to be corrected upwards by a factor of up to
three, according to the weight of the $\mud$ doublet state in the statistical mixture. 
The resulting contradictory situation stimulated a new generation of experiments based on 
pure deuterium targets.

One innovation~\cite{Bardin:1986} was the detection of electrons with a liquid target at Saclay, 
which avoided neutron detection and background from $dd$ fusion. 
Lifetimes of both $\mu^{-}$ and $\mu^{+}$ were measured with an accuracy of a few times 10$^{-5}$.
The muon lifetime measurement started 1~$\mus$ after the beam burst, after muons stopped in wall 
materials were already captured. The final rate included a corrections of $\Delta \Rd$ = 12 \ins\ for protium
 ($c_P$=0.13-0.18\%) and of  $\Delta \Rd = (60\pm16)~s^{-1}$ for $\mu^- {}^{3}$He  capture. The total uncertainty
quoted is $\pm$29 \ins, corresponding to $\pm$25(stat) and $\pm$16(sys) \ins.

The Vienna PSI experiment~\cite{Cargnelli:1989} on the other hand used neutron detection, but 
reduced the gas density and temperature to suppress fusion neutrons and performed supplemental 
analyses with high neutron threshold above the 2.5 MeV fusion neutrons. 
In the low density target the stopping fraction of muons in
D$_2$ was $(75.7\pm1.7)\%$. The physics background, consisting of carbon stops, diffusion,
photo neutrons and fusion neutrons, exceeded the physics signal by a factor of 1.4.  The uncertainties 
in its subtraction dominated the averaged final error of $\pm$40 \ins, with statistics and uncertainty in
neutron detection efficiency contributing  $\pm$15 and  $\pm$20 \ins, respectively.   

The current  overall experimental situation on \RD\ (see Fig.~\ref{f:deut_exp}) is quite 
unsatisfactory. The best two measurements were performed almost two decades ago. 
They have uncertainties of 6.2-10\% and are only marginally consistent,
with the more accurate experiment deviating from theory by  2.9 $\sigma$.

\subsection{Connections to Neutrino and Astrophysics}

As mentioned in the introduction the reactions
\begin{eqnarray}
p+p &\rightarrow& d +e^+ + \nu_e  \label{pp.eq} \\
\nu_e+d &\rightarrow& e^- + p + p \label{cc.eq} \\
\nu+d &\rightarrow& \nu + p + n \label{nc.eq}
\end{eqnarray}
are of fundamental physics interest. Reaction~\ref{pp.eq} is the primary solar fusion 
process which is one of the key inputs that controls the solar model. 
Reactions~\ref{cc.eq},~\ref{nc.eq} are the charged and neutral current reactions (CC, NC) 
detected by SNO and their comparison provides direct evidence for neutrino oscillation and 
the NC process serves as the measurement of the total $^8$B neutrino flux from the sun~\cite{Aharmim:2005gt}.
As these processes have eluded quantitative measurements, there has been a tremendous theoretical 
effort to calculate them with ever increasing precision. Let us focus on the $\nu+d$ reaction. 
Up to 2001, the calculations~\cite{Nakamura:2000vp} were performed within the standard nuclear physics
approach, where two body-current effects were estimated from two experimental sources, namely the 
tritium beta decay rate $\Gamma^\beta$ and the $n+p \rightarrow d+\gamma$ cross section. 
The results of the two methods differed by 3\%, which was adopted as the theoretical uncertainty. 
In 2002 these results were updated~\cite{Nakamura:2002jg} 
and the uncertainty estimate reduced to 1\% by arguing that the $n+p \rightarrow d+\gamma$ 
constraint should be discarded, because it refers to a vector transition only. 
In 2003 these calculations were corroborated by the above mentioned EFT inspired hybrid 
approach~\cite{Ando:2002pv}, which uses phenomenological wave function together with EFT 
derived operators and again employs $\Gamma^\beta$ to control the 2$N$ part. 
The hybrid approach is, in principle, subject to criticism concerning consistency
in chiral power counting, off-shell ambiguities etc., although these effects are estimated
to be small~\cite{Park:2002yp}. 
Possible approaches that are formally consistent are the \pionless 
EFT~\cite{Butler:2000zp} and ChPT.
Both these approaches, however, are not predictive, as the missing LECs
$L_{1A}$ and \dhr, respectively, are not determined at this point. 
Very recently, a model dependent analysis of low energy $\nu d$ cross sections has 
estimated the $\nu d$ cross sections to be accurate to 2--3\%~\cite{Mosconi:2007tz}.

The SNO experiment~\cite{Aharmim:2005gt} has adopted a 1.1\% uncertainty in the cross sections
used in their data analyses to determine the $^8$B neutrino fluxes. 
In view of the discussion above this
appears optimistic and basically rests on validity of the hybrid approach at this precision level.

As regards the $pp$ fusion process, the situation is similar. 
The standard solar model~\cite{Bahcall:2004fg} adopts a 0.4 \% uncertainty in the $pp$ $S$-factor, 
in accordance with a standard nuclear physics~\cite{Schiavilla:1998je}
and hybrid EFT calculations~\cite{Park:2002yp}, both relying on the $\Gamma^\beta$ constraint.
These considerations make assumptions regarding the dynamics of the three-nucleon system which 
can be avoided if we stay completely within the two-nucleon sector.

\begin{table}[hb]
\begin{center}
\begin{tabular}{lccc}
\hline
 			& method      		&  $L_{1A}$  (fm$^3$) 	&  comment\\
\hline
{\bf two-body} 	&			&			&     \\
 	&reactor $\bar{\nu} +d$	&  3.6 $\pm$ 5.5~\cite{Butler:2002cw}   	& i)  \\
	& ES, CC, NC in SNO     & 4.0  $\pm$   6.3~\cite{Chen:2002pv}   	& ii) \\
&{\bf MuSun proposal}   		        &{\bf $\pm$1.25} 	&    \\
\hline
\hline
{\bf three-body}  	&			&			&		\\	
			& tritium beta decay & 4.2 $\pm$ 3.7~\cite{Butler:2002cw}, 4.2 $\pm$ 0.1~\cite{Chen:2002pv} & iii)\\
\hline
{\bf other} 		& 			& 			& 		\\
\		 	& helioseismology  	&4.8  $\pm$ 6.7~\cite{Brown:2002ih} & iv)  \\
\hline
\end{tabular}
\caption{$L_{1A}$ determinations compiled in Refs.~\cite{Butler:2002cw,Chen:2002pv}. 
The only theoretical clean and fully self-consistent calculations can be performed in the 2-nucleon
system, where the MuSun experiment will have a major impact. The $L_{1A}$ determination
from tritium decay claims high accuracy, but the extraction can only be done in
the hybrid EFT approach with phenomenological wave functions in the more complex three-body system. 
i) The best experiment determines $L_{1A}$ to $\pm$ 8.1 fm$^3$ only. The averaging procedure is 
questionable as indicated by the small global $\chi^2$; ii) the error is expected to be reduced 
to $\sim 5fm^3$ with the final SNO data;  iii) Ref.~\cite{Butler:2002cw} includes the uncertainty in
higher-order theoretical systematics, while Ref.~\cite{Chen:2002pv} does not, which leads to the large
difference in the error estimate for $L_{1A}$ ;
iv)  subject to other solar model uncertainties. }
\label{l1a.tab}
\end{center}
\end{table}

In view of the importance of these cross sections and the ongoing discussion described above,
several attempts have been made to fix the 2$N$ contribution independent of  $\Gamma^\beta$. These 
efforts can be conveniently parametrized in terms of the LEC $L_{1A}$ and in the future with
\dhr, which is more appropriate if one also includes muon capture (c.f. table~\ref{l1a.tab}). 
MuSun will lead to a decisive improvement, using information
solely from the theoretically clean two-nucleon sector.

\begin{figure}
\begin{center}
\includegraphics[width=0.7\textwidth]{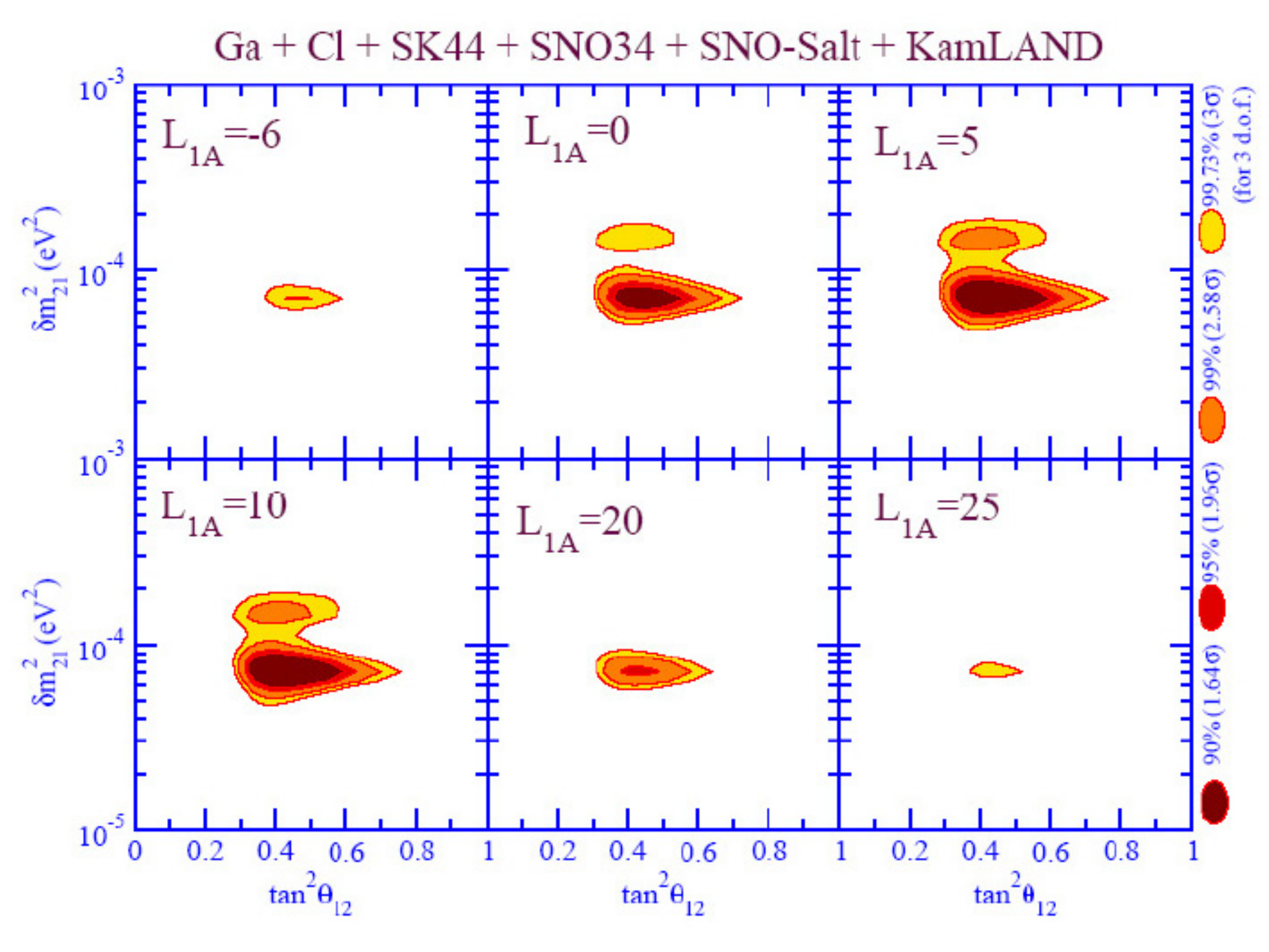}
\end{center}
\vspace{-5mm}
\caption{Dependence of solar neutrino mixing parameters on assumed value of
$L_{1A}$.  Reproduced from~\cite{Balantekin:2004zj}.}
\label{balantekin}
\end{figure}

Refs.~\cite{Balantekin:2003ep,Balantekin:2004zj} study how the solar neutrino mixing parameters 
vary with the assumed value of $L_{1A}$.  Figure~\ref{balantekin} shows that there 
are subtle but nevertheless visible changes in the $(\theta_{12},\delta m^2_{21})$ exclusion regions 
as $L_{1A}$ is varied over its plausible range from 0 to 10~fm$^{-3}$.  
These changes primarily correspond to a uniform reduction in the fit
probability, as the solar neutrino data becomes less self-consistent for 
extreme values of $L_{1A}$.  The authors' conclusion is that the uncertainty 
in the solar neutrino mixing parameters from $L_{1A}$ is somewhat smaller than 
that arising from $\theta_{13}$. However, as we enter a phase of precision neutrino
physics, several current uncertainties will be tightened by future experiment, enhancing
the importance of $L_{1A}$. 
As examples the uncertainty in $\theta_{13}$ will be reduced to the sub-percent level and the 
final SNO-III phase~\cite{Poon:2005qu} is expected to reduce the total experimental
uncertainty in the charged-current flux from 6.3 \% to 4.0 \%, which is comparable to the theoretical
uncertainties discussed above. 
An updated analysis along the lines of Refs.~\cite{Balantekin:2003ep,Balantekin:2004zj}
would be very interesting, which includes constraints from future neutrino experiments
and also the effect of $L_{1A}$ on the $^8$B  (and $pp$) flux within the standard solar model.    
As regards the solar $pp$ neutrino flux, the solar luminosity constraint strongly reduces its 
sensitivity to the $pp$ cross section. 
For constraining alternate fusion sources in the sun or testing, whether the sun is in a steady 
state, or calculating the $^8$B flux, improved knowledge of the $pp$ reaction is more
than a symbolic achievement. Finally, studying few-nucleon systems also helps us to improve
our understanding of electroweak phenomena in more complex nuclei, which feature in a wide 
variety of astrophysical phenomena including neutrino-nucleosynthesis.

\subsection{Other Physics Connections}

\subsubsection*{Nuclear Physics}
As already mentioned, the LEC $\hat{d}^R$ also connects $\mu^- + d$ capture to $\pi^- d \to\gamma nn$. 
This radiative pion capture process is  considered to be the most 
reliable (experimentally and theoretically) method of extracting the $nn$ 
scattering length ($a_{nn}$)~\cite{Machleidt:2001rw}, while the alternative 
method of using the three-nucleon process $nd\to nnp$ is plagued with 
inconsistent results~\cite{GonzalezTrotter:2006wz,Huhn:2001yk}.
The difference of the $nn$ and $pp$ scattering lengths is used to constrain the
charge-symmetry-breaking pieces of the modern high-precision phenomenological
nucleon-nucleon potentials, which in turn are needed for detailed understanding
of the lighter nuclei ($A<20$)~\cite{Wiringa:1994wb,Pieper:2001mp}.
The theoretical error in $a_{nn}$ extracted from $\pi^- d\to\gamma nn$ can 
with ChPT methods be reduced to $\pm0.05$~fm, i.e., to the 
0.3\% level~\cite{Gardestig:2005pp,Gardestig:2006hj,Gardestig:2006jd}.
However, this precision can be reached only if the short-range physics is 
constrained by $\hat{d}^R$.
Thus a precision measurement of $\mu + d$ would also help in establishing a 
precise value of $a_{nn}$ completely within the two-body sector and ChPT.

\subsubsection*{Hydrogen TPCs}

With the MuCap TPC and the MuSun cryo-TPC we will have developed a range of 
high-density, thin walled time projection chambers covering an equivalent pressure range 
of 5-100 bar at room temperatures. The chambers operate with ultra-pure hydrogen at a
purity level of ppb. The MuCap TPC operates at lower pressure with gas amplification. 
The MuSun TPC will be a cryo ionisation chamber with full analog readout
and excellent energy resolution. As hydrogen and deuterium are basic target elements, 
these new instruments might find interesting applications in nuclear/particle physics.
In the realm of muon physics, we are considering future experiments on the hep
process and on rare fusion reactions of astrophysical interest.

\subsection{Extraction of  (L$_{1A}$, \dhr) from $\mu+d$ Capture}

We have already made the more general statement that the solar $p+p$ fusion and $\nu d$ SNO 
reactions will be determined in a model independent way at the same precision as the measured 
$\mu^- + d$ capture reaction.

Here we use Eq.~\ref{lamab.eq} for estimating more specifically how the \Rd\ measurement 
can determine $L_{1A}$. Equivalently, we use $L_{1A}$ as a convenient device to estimate how 
different uncertainties affect the $\mu^- d$ capture rate. The discussion could equally be 
framed in terms of $\hat{d}^R$, as the \Rd\ dependence on $\hat{d}^R$ is well approximated by a 
linear relation, equivalent to Eq.~\ref{lamab.eq}. 
Simple error propagation leads to
\beq
\delta L_{1A} \approx \frac{a}{b} \frac{\delta \Rd}{\Rd}
\eeq

The fractional uncertainty $\frac{\delta \Rd}{\Rd}$ consists of the measurement error (1.2\%) and 
a theoretical uncertainty. The latter consists of an estimated uncertainty of 0.5\% 
in the ChPT calculation and  an uncertainty introduced using \gP\ from the \RS\ measurement (0.7\%). 
The dependency on the neutron scattering length $a_{nn}$ leads to 
$\frac{\delta\Lambda}{\Lambda}= 0.9\% $, if the currently accepted value of 
$a_{nn}= -18.6 \pm 0.4$~fm is being used. 
Together these add up to
\beq  
\frac{\delta \Rd}{\Rd} = \pm 1.7 \% = \pm 1.2\% (exp) \pm 1.25\% (theory) 
\eeq
The resulting precision in $L_{1A}$= 1.25 fm$^3$. 
If only an overall $\frac{\delta \Rd}{\Rd}$ of 2\% is obtained, the uncertainty in $L_{1A}$ would 
increase to 1.5 fm$^3$, still far better than any other 2-body information in table~\ref{l1a.tab}.

\subsection{Muon Capture, the Big Picture}

\begin{figure}[hbt] 
\vspace{-1.cm}  
\begin{center}
\resizebox*{1.\textwidth}{0.35\textheight}{\includegraphics{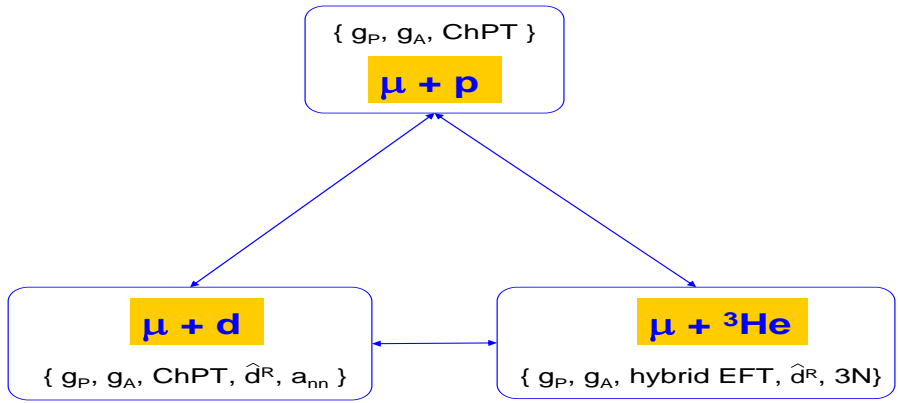}} 
\vspace{-2cm}
\caption{\label{muoncapture.fig} Relations between basic muon capture reactions on the nucleon
and A = 2, 3 nuclei. This is further discussed in the text.}
\end{center}
\vspace{-.4cm}
\end{figure} 

The MuSun experiment is part of our systematic program to achieve a new level of precision
in confronting the theories of weak interactions, QCD and few body physics with precision muon capture
experiments on $\mu^-+p \rightarrow \nu_\mu+p$, $\mu^-+d \rightarrow \nu_\mu + n + n$ and 
$\mu^- + ^3$He$\rightarrow \nu_\mu + t$. We limit the discussion to these three cases, as the 
nuclear physics of nuclei with A$>$3 becomes increasingly difficult to control at the 
required precision level, notwithstanding the significant progress achieved recently in this field. 
Moreover, it is clear that the understanding of the
basic reactions above is required for predicting muon capture observables in heavier nuclei.

The different parameters and theories relating the above three muon-capture reactions are illustrated
in Fig.~\ref{muoncapture.fig}. As is typical for EFTs, many LECs are determined already in the 
single-nucleon system and carry over unchanged when more nucleons are added. 
However, in the two-nucleon system, there appear two new parameters, i.e., $\hat{d}^R$/$L_{1A}$ and
$a_{nn}$, both of which are well defined and can be determined precisely.
The one- and two-nucleon systems can be treated entirely within the framework of ChPT, but when going
to three nucleons, the currently only available option is to use phenomenological wave functions,
i.e., hybrid ChPT. The three-nucleon system also introduces additional LECs and the complications 
of three-body dynamics. 

Quantities like  \gA, \gP\ and \dhr/$L_{1A}$ encode basic properties of the nucleon 
or two-nucleon system, when interacting with the axial-current. They are equally important
as indispensable input to precision calculations. For example, \gA\ is needed for all calculations 
of weak Gamow-Teller transitions, \gP\ quantitatively tests our understanding of basic QCD symmetries
and \dhr/$L_{1A}$ allows the calculation of immeasurable and fundamental neutrino processes. 

From the modern nuclear physics perspective, the triad of muon capture experiments is an important 
component of a main thrust to derive nuclear physics from QCD. This worldwide effort utilizes 
different probes to precisely determine the low-energy constants and test the internal consistency of 
the theory. As just one example, it is expected that the new generation of neutron decay experiments 
will resolve the current controversy  on the neutron lifetime and \gA. 
Before our program started at PSI, muon capture was hardly a contender  in this worldwide effort, 
as the experimental precision was not competitive. 
This has changed dramatically during the last decade. 

Our first experiment~\cite{Ackerbauer:1997rs} on $\mu^- + ^3$He$\rightarrow \nu + t$ 
achieved a precision of 0.3\%, unique in this field of physics and equal to the precision 
of the tritium decay rate~\cite{Simpson:1987zz}. 
Efforts to achieve similar precision theoretically are still ongoing and, recently, a calculation
of radiative corrections~\cite{Czarnecki:2007th} revised the elementary particle model calculation, 
which was previously considered to be very accurate, by 3\%. 
The analyses of a first data set~\cite{Andreev:2007wg} on $\mu^-+p \rightarrow \nu+n$ achieved 
a first precise result on \gP, nearly independent from the uncertainties from muonic molecule 
formation, which have plagued earlier experiments for many years. 
The final result on the capture rate \RS\ is expected to have $<$1 \% uncertainties and will 
establish a critical test of a basic ChPT prediction,  where \gP\ is an accurately defined 
derived quantity.
At this level the extraction of \gP\ would be affected by the new neutron lifetime 
experiment~\cite{Serebrov:2007ve}, which, if taken as the new standard, would imply a 
$\approx$ 0.8\% shift in the theoretical prediction for \RS.
The new MuSun experiment will be the cleanest way to determine $\hat{d}^R$/$L_{1A}$ required
for precision calculations of basic astrophysical reaction and will shed light on the 
precision of ChPT versus hybrid EFT, relevant for many reactions. 

Once this experimental program is completed, we plan a combined, consistent analysis of our results
on all three reactions using the best theoretical input available at that 
time~\footnote{Depending on the evolving overall picture, we will then assess whether 
additional dedicated, but very difficult experiments of polarization observables in muon 
capture are justified, where the $\mu^-+d$ is a promising candidate.}.

\section{Experimental Strategy}
 

\subsection{Overview}

In order to achieve the goal of this experiment two main conditions have to be met.
\begin{itemize}
\item
The measurement must be performed at conditions, such that the experimental result leads
to an unambiguous extraction of \Rd, independent from muonic atomic physics uncertainties.

\item
The measurement must achieve an overall precision of \precision\ or better of \Rd\ (6 \ins), which is 
nearly an order of magnitude 
higher than achieved in previous experimental work.

\end{itemize}

\subsubsection*{Muon Kinetics and Optimal Target Conditions}
The muon induced atomic and molecular processes ({\em muon kinetics}) are
quite different for negative muons in deuterium compared to the muon kinetics
in pure protium (relevant for the MuCap experiment). 
The hyperfine transition rate \qdr\ of the upper $d\mu(\uparrow \uparrow)$ quartet 
to the $d\mu(\uparrow \downarrow)$ doublet state is slow. The V-A structure
of weak interactions, however, disfavors capture from the quartet state (\Rq=12 \ins) compared
to capture from the doublet state (\Rd=386 \ins), so that the experimentally observed 
capture yield is largely proportional to the population of the doublet state. 
The $d\mu$ system has been intensively studied as the prototype for resonant muon-catalyzed 
fusion~\cite{Breunlich:1989vg}.  For a clear interpretation and for the accumulation of
 sufficient capture 
statistics,  the target conditions should be chosen such that the 
$d\mu$ doublet state dominates and the  population of states can be verified {\em in-situ} 
by the observation of muon-catalyzed fusion reactions.
Our optimization indicates excellent conditions at $\phi=5\%$ of liquid hydrogen density and T=30 K, 
which we define as the baseline of the experimental proposal.  
On the positive side, complications from muon capture in the $dd\mu$ molecule are nearly absent,
as it is short lived. Moreover, the pronounced $\mu  d + p$ diffusion problem
does not exist like in the case of the MuCap protium measurement, as the elastic cross section for $\mu d+d$ scattering is
large. However, $\mu^3$He formation, isotopic purity, and chemical impurities still need careful
attention. Different from the isotropic decay from the singlet $\mu p$ state in MuCap, the $\mu$
doublet and quartet state can remain polarized, which might lead to a time dependent decay asymmetry
(c.f. appendix~\ref{msr.tex}). 
 
\begin{figure}[tbh]
  \begin{center}
  \includegraphics[scale=0.6]{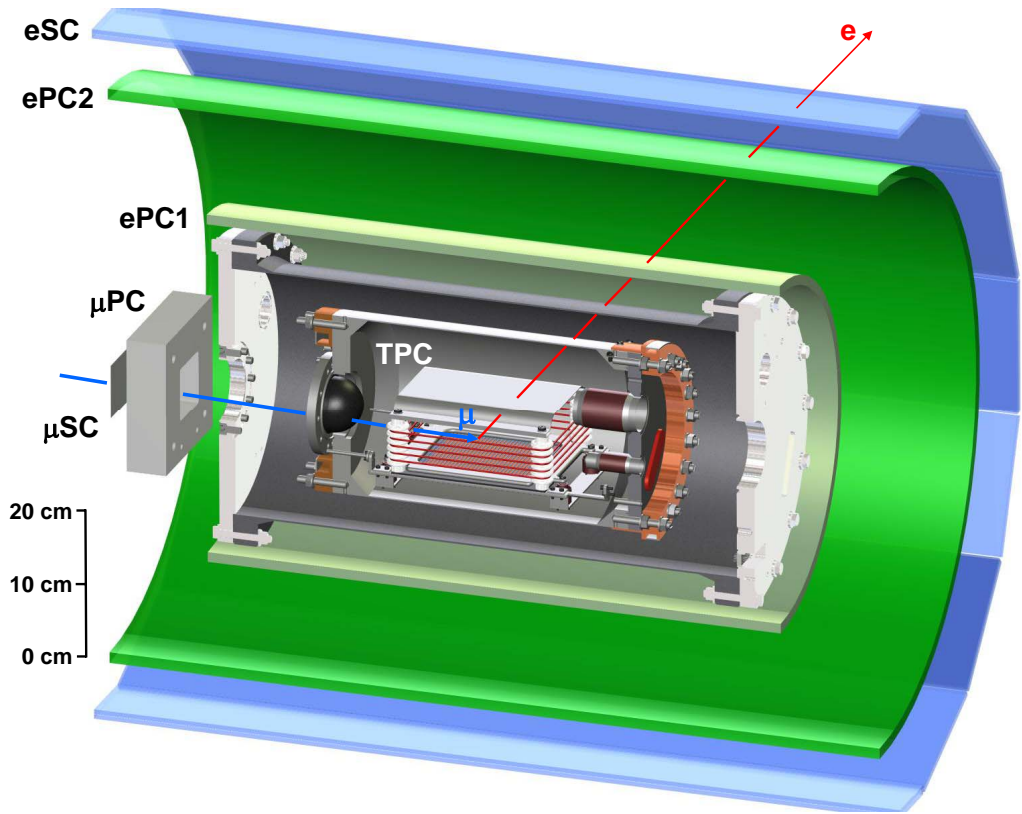}
  \caption{Simplified cross-sectional diagram of the MuSun detector.
   The detector components are described in the text.}
  \label{setup.fig}
  \end{center}
\vspace{-4mm}
\end{figure}

\subsubsection*{Experimental Technique for $<\precision$ Measurement}

The gain in experimental precision relies on the fundamental techniques developed for MuCap.
Muons will be stopped in an active gas target consisting of a cryogenic ionization chamber operated as time projection chamber (TPC or \CIC) with chemically and isotopically ultra-pure deuterium. The reconstruction of the muon stopping point in 3 dimensions eliminates the otherwise overwhelming background from muon
stops in wall materials. The capture rate is determined using the lifetime 
technique; that is, from the difference between the measured disappearance rate 
$\lambda_\mu^- \approx \lambda_\mu^+ + \RD$ of negative muons in hydrogen and 
the $\mu^+$ decay rate~$\lambda_\mu^+$, where it is assumed that free $\mu^-$ and $\mu^+$ 
decay with identical rates according to the CPT theorem.

The experimental setup is illustrated in Fig.~\ref{setup.fig}. 
Incident muons first traverse a plastic scintillator~($\mu$SC) and a multiwire 
proportional chamber~($\mu$PC), and then pass through a 150 $\mu$m kapton 
window into the insulation vacuum and second, a 0.4-mm-thick hemispherical 
beryllium window to enter an aluminum pressure vessel filled with 
ultra-pure, deuterium gas at a pressure of 0.5~MPa and 30 K temperature.  
In the center of the vessel is the \CIC\ (sensitive volume $10\times10\times10$~cm$^3$),
which tracks incoming muon trajectories and thus enables the selection of muons that 
stop in the  gas at least 5~mm away from chamber materials.  
Monte Carlo simulations indicate that approximately \mustopfrac\ of the muons passing through the 
$\mu$SC stop within this fiducial volume.  The ionization electrons produced by incoming muons 
drift downwards at velocity 4~mm/$\mu$s in an applied field of 10~kV/cm, 
towards a multi-pad plane of the \CIC. Signals 
from the \CIC\ are recorded deadtime free with custom build FADCs.
The chamber is surrounded by two cylindrical wire chambers~(ePC1, ePC2), each 
containing anodes and inner/outer cathode strips, and by a hodoscope 
barrel~(eSC) consisting of 16 segments with two layers of 5-mm-thick 
plastic scintillator. This tracking system  detects outgoing decay electrons 
with $3\pi$ solid angle acceptance.  All data are recorded in a trigger-less, 
quasi-continuous mode to avoid deadtime distortions to the lifetime spectra. 

The MuSun technique heavily builds on the R\&D, equipment investments, techniques and analysis refinements
developed for the MuCap and \mlan\ experiments. The electron tracking system, the beam counters, 
the sophisticated vacuum and purification system and a large part of the electronics and data acquisition 
can be taken over from the MuCap experiment. The fast electric kicker, crucial for achieving pile-up free 
high event rates, and custom build FADCs are provided from the \mlan\ experiment. The main distinctive 
features of the MuSun experiment are demanded by physics requirements and include:
\begin{itemize}
\item
High density cryogenic ionization chamber operating as a TPC filled with ultra-pure 
deuterium to define
the muon stop, identify impurities and to observe muon induced processes. 

\item
Excellent energy resolution of the \CIC\ and full analog readout with FADCs to
monitor the charged particles induced by fusion and impurity capture processes.

\item
Advanced purity monitoring system with new particle detectors and chromatographic
methods.

\item
Neutron detectors to monitor the muon kinetics via capture neutrons and fusion products 
and separate impurity from fusion signals in the TPC.

\end{itemize}

\subsection{Kinetics}
\begin{figure}[t]
  \begin{center}
  \includegraphics[scale=0.5]{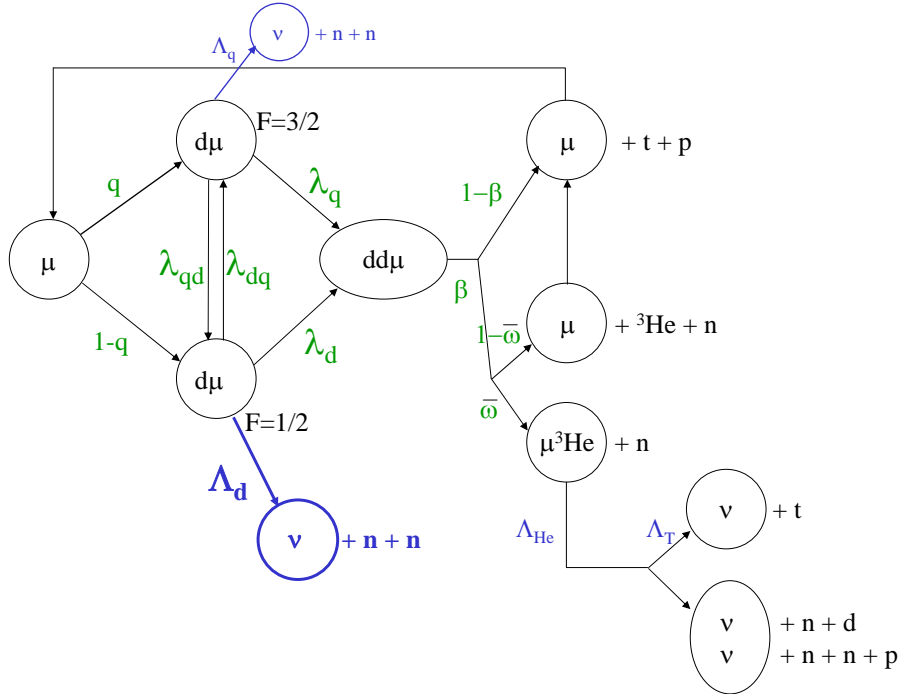}
  \vspace{-5mm}
  \caption{Simplified muon kinetics in pure D$_2$.  }
  \label{kinetics.fig}
  \end{center}
\end{figure}

\begin{figure}[tbh]
  \begin{center}
  \includegraphics[scale=0.35]{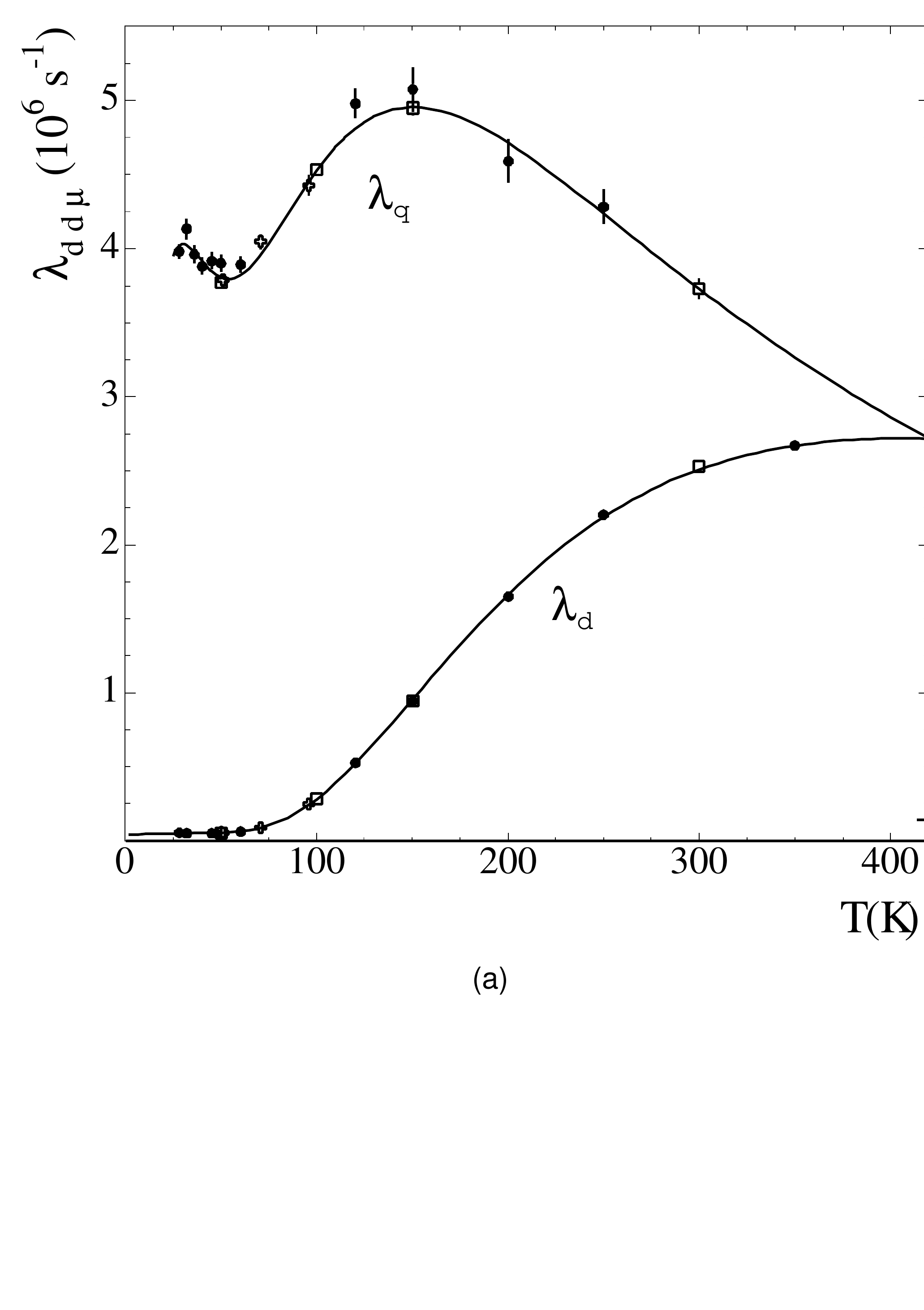}\hfill
  \includegraphics[scale=0.35]{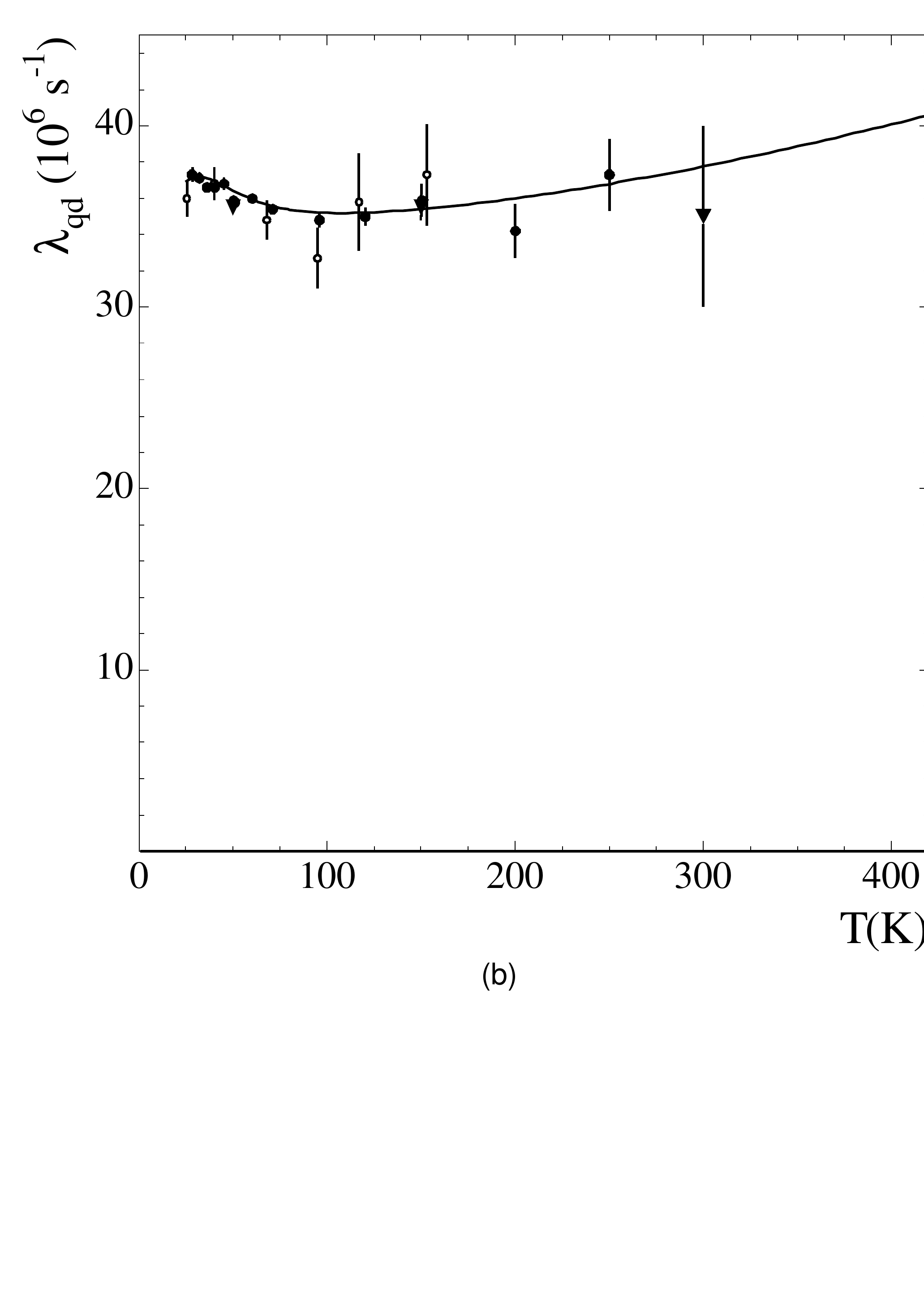}\\
  \vspace{-26mm}
  \caption{(a) Molecular $dd\mu$ formation rates \qr\ and \dr\ and (b) the hyperfine transition rate \qdr. The experimental data comes from \cite{Bal07, Kammel:1982tj, Kammel:1984vu, Zmeskal:1990}. For both panels, the plots are taken from \cite{Bal07} with some slight modifications.}
  \label{rates.fig}
  \end{center}
\end{figure}

Fig.~\ref{kinetics.fig} shows a simplified scheme of the muon induced kinetics in 
pure deuterium\footnote{The simplifications include: 
The effective dd\( \mu  \) fusion rate has been omitted, since it is 
nearly instantaneous ($\le$ 1 ns) at the time scales considered. The hyperfine state dependence of
the branching ratio $\beta$ has been ignored. Small corrections
to the kinetics are induced by the finite thermalization time of \( \mu  \)d
atoms.}. 
Because of its unique importance for understanding muon-catalyzed fusion, resonant 
molecule formation, and weak interactions, this reaction chain has been scrupulously 
studied both experimentally and theoretically. 
The latest experimental results are presented in Ref.~\cite{Bal07} which also
includes many experimental details relevant for the present proposal. 
The current knowledge of the relevant parameters is compiled
in Fig.~\ref{rates.fig} and Table~\ref{rates.tab}. As is conventional, all density dependent kinetic
rates have been normalized to LH$_2$ density N=4.25 $\times$ 10$^{22}$ atoms/cm$^3$, and the density $\phi$ is
expressed relative to this value. 

\clearpage

\begin{table}[tbh]
\begin{center}
\begin{tabular}{|c|c|c|c|}
\hline
description			& quantity			& \multicolumn{2}{|c|}{value }  \\
    				&       			&  300K		& 30K 		\\
\hline 
initial quartet fraction& q 					& \multicolumn{2}{|c|}{2/3 } 	\\
\hline  
hf transition $q \rightarrow d$ & $\lambda _{qd}(\mu s^{-1})$	& 35(5)   	& 37.0(4) 	\\
\hline 
hf transition $q \rightarrow d$ & $ \lambda _{dq}(\mu s^{-1})$  &  $^{1)}$      &         	\\
\hline 
dd$ \mu  $ form. rate from q  &$ \lambda_{q}(\mu s^{-1}) $ 	& {$ \sim 3.75 $} & 3.98(5) 	\\
\hline 
dd$ \mu  $ form. rate from d  &$ \lambda _{d}(\mu s^{-1}) $ 	& 2.549(23) 	& 0.053(3)  	\\
\hline 
effective fusion fraction     &$ \beta  $        		& 0.590(6)      & 0.517(15) 	\\
\hline 
sticking probablity           & $ \overline{\omega}~^{2)} $     &   \multicolumn{2}{|c|}{0.1206(6)}   	\\
\hline 
$ ^{3} $He total capture rate & $ \RHE(s^{-1}) $  	& \multicolumn{2}{|c|}{2216(70)}	\\
\hline 
$ ^{3} $He partial capture rate & $ \Lambda _{T}(s^{-1}) $	& \multicolumn{2}{|c|}{1496.0(40)}    	\\
\hline 
$ \mu  $d quartet capture rate & $ \RQ (s^{-1})  $	& \multicolumn{2}{|c|}{$\sim 10 $  }	\\
\hline 
$ \mu  $d doublet capture rate & $ \RD (s^{-1})  $	& \multicolumn{2}{|c|}{$ \sim 400 $}\\
\hline 
\end{tabular}
\caption{Kinetic parameters. All values given with error bars are directly determined experimentally, others
theoretical. $^{1)} \lambda _{dq}\sim \frac{q}{1-q} e^{-\frac{\Delta}{kT}}$,
with the \( \mu  \)d hyperfine splitting energy \( \Delta  \)=0.0485 eV and
k=8.6174 10\( ^{-5} \) eV/K (small deviations possible due to back-decay); $^{2)}$ it is convenient
to define the effective sticking fraction $\om = \beta \overline{\om} $. }
\label{rates.tab}
\vspace{-.5cm}
\end{center}
\end{table}

The vector $N(t)$ for the populations of the \mud\ quartet, \mud\ doublet, and $\mu ^3$He states
\beq
   N(t)= \left ( \matrix{ N_q(t) \cr N_d(t) \cr N_{He}(t) \cr  } \right ) 
\eeq
with initial conditions
\beq
   N(t=0)= \left ( \matrix{ q \cr 1-q \cr 0 \cr  } \right ) 
\eeq
obeys the following kinetic equation
\beq
  \frac{dN(t)}{dt}=K N(t),
\label{eq:sol}
\eeq 
where
\beq
 K = \left (\matrix{
  - \rz - \Rq - \phi \qdr - \phi \qr(1-q(1-\om))  &  \phi \dqr +\phi \dr q (1-\om)  &    0         \cr 
  \phi \qdr + \phi \qr (1-q)(1-\om)          & - \rz - \Rd  - \phi \dqr - \phi \dr(1-(1-q)(1-\om))   &   0  \cr  
          \phi \qr \om                       & \phi \dr \om                                       & - \rz - \RHE  
    } \right ) 
\label{eq:K}
\eeq

The observable time distributions include el$(t)$ for the electrons, fus$(t)$ for the $^3$He fusion products, $cap_n(t)$ for the neutron from $\mu+d$ capture and $cap_T(t)$ for the tritons from $\mu+^3$He capture.
\begin{eqnarray}
el(t)\equiv & \frac{dN_e}{dt}= & \rz \sum_i N_i(t) \label{el.eq}\\
fus(t)\equiv & \frac{dN_{He}}{dt}= & \beta (\phi \qr N_q(t) + \phi \dr N_d(t)) \label{fus.eq}\\
cap_n(t)\equiv & \frac{dN_{n}}{dt}= & 2 ( \Rq N_q(t) + \Rd N_d(t)) \label{capN.eq}\\
cap_T(t)\equiv & \frac{dN_{T}}{dt}= & \Lambda_{T} N_3(t) \label{capT.eq}\\
\end{eqnarray}

\subsection{Optimization of the Target Conditions}

\label{targetoptimization.sub}
\begin{figure}[htb]
\begin{center}
\begin{tabular}{cc}
\resizebox*{0.45\textwidth}{!}{\includegraphics{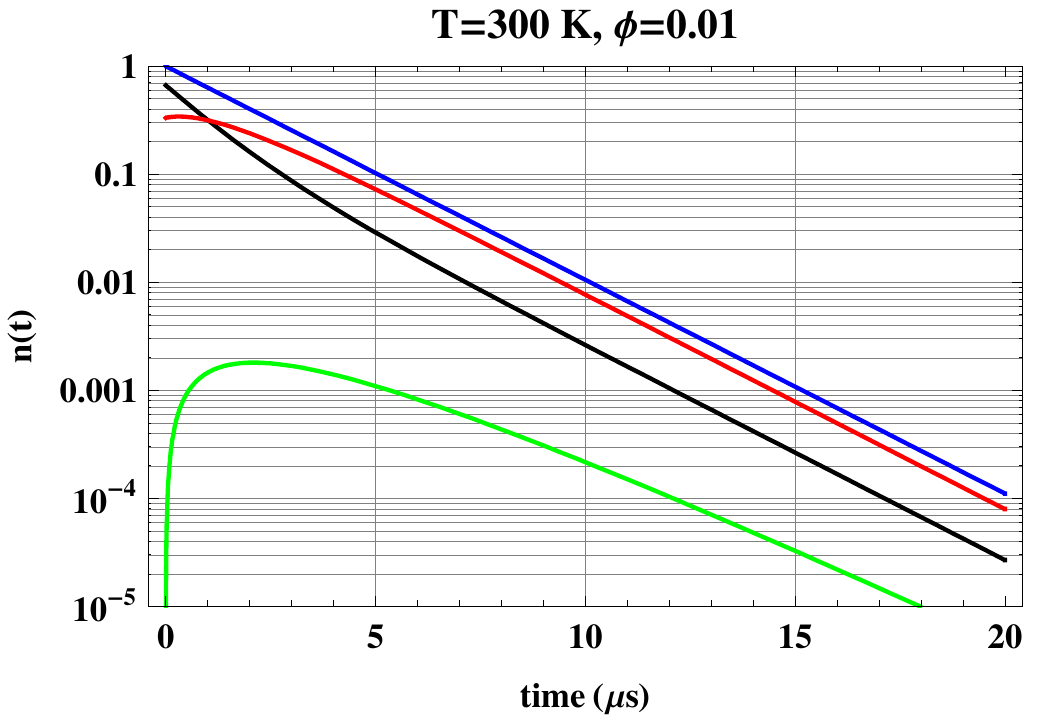}} & 
\resizebox*{0.45\textwidth}{!}{\includegraphics{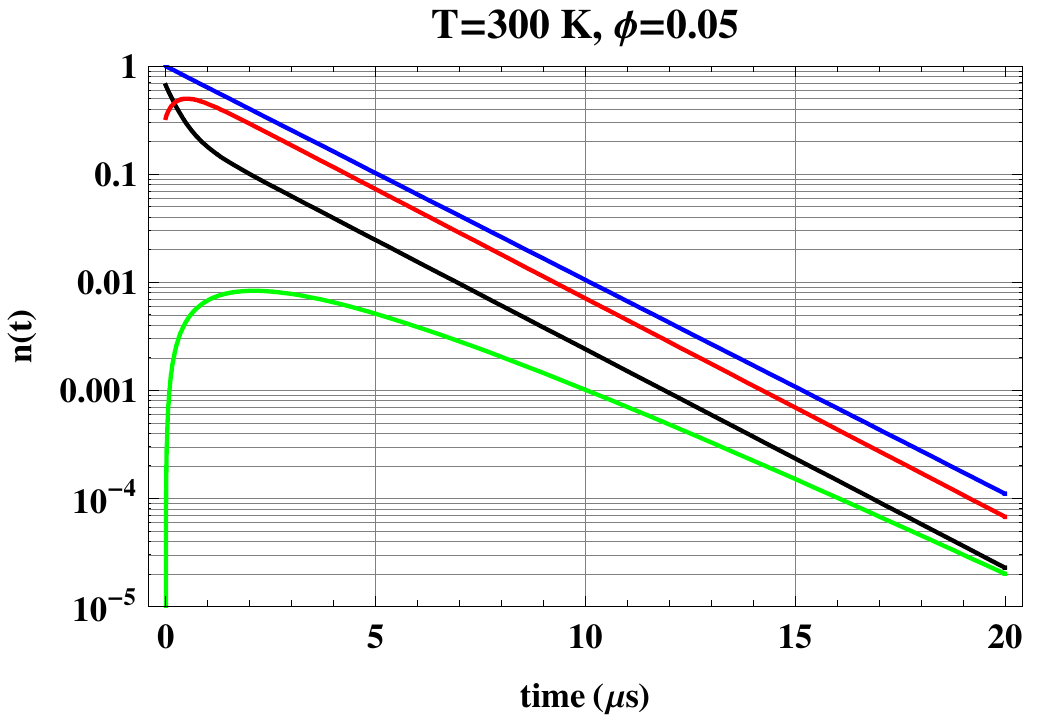}} \\
\vspace{.5cm}
\resizebox*{0.45\textwidth}{!}{\includegraphics{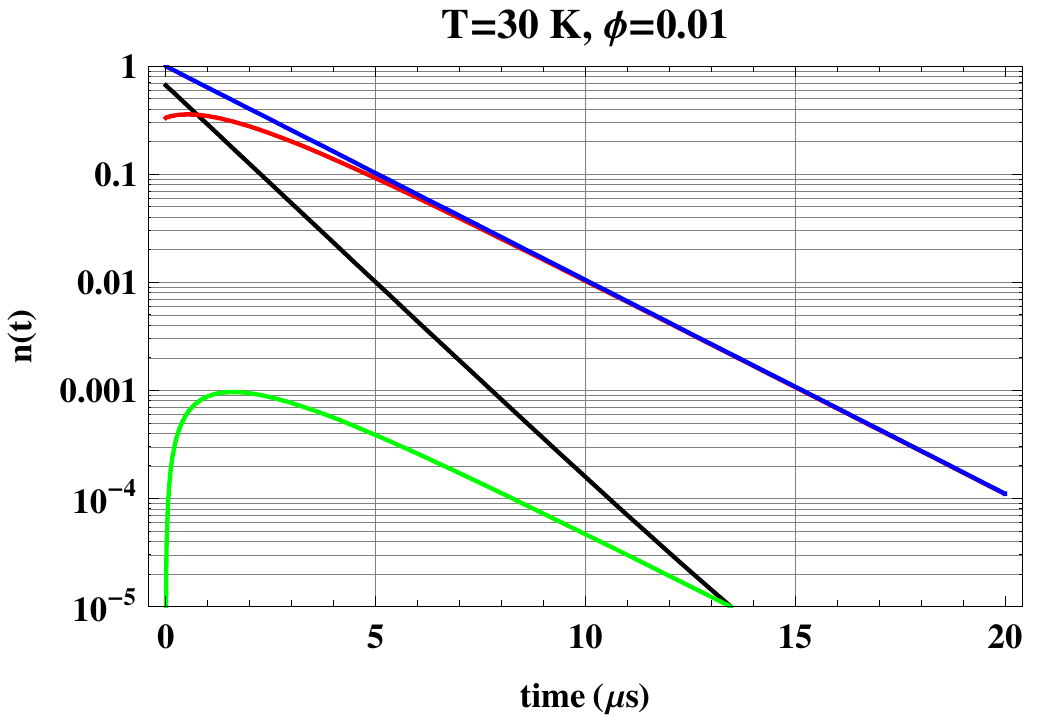}}  & 
\resizebox*{0.45\textwidth}{!}{\includegraphics{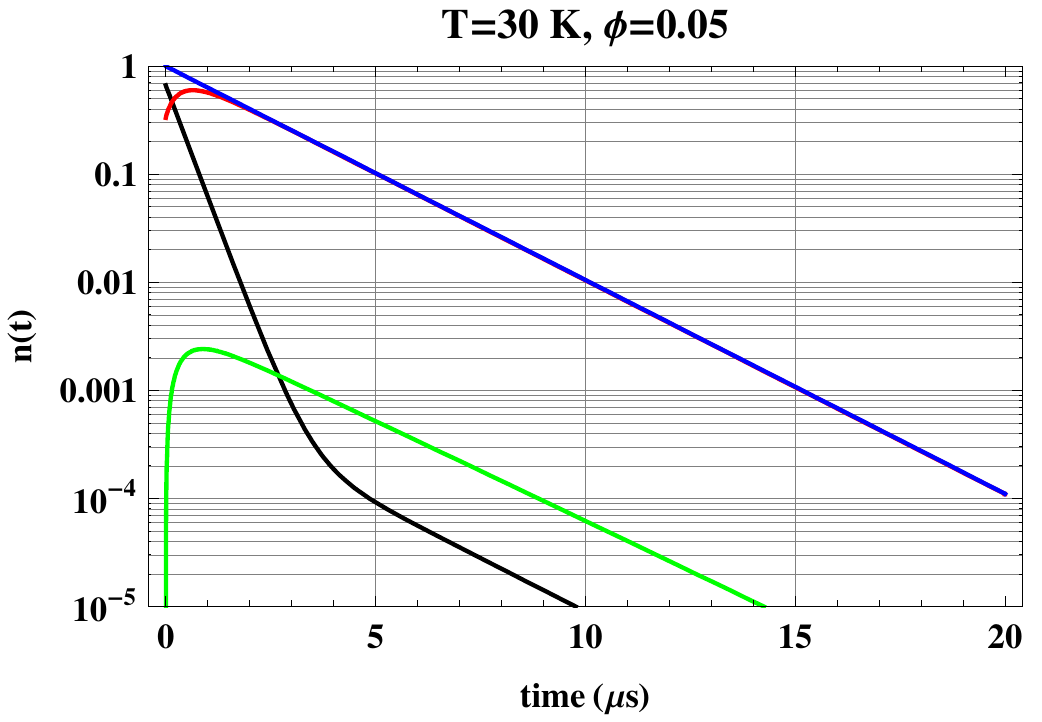}}\\
\end{tabular}
\vspace{-.5cm}
\caption{Time distributions of relevant states (blue=$\mu d$, red=$\mu d(\uparrow \downarrow)$, black=$\mu d(\uparrow \uparrow)$, green=$\mu^3$He) for different deuterium densities
$\phi$ and temperatures T. The bottom right panel illustrates the optimized running conditions for the MuSun experiment. }
\label{time.fig}
\end{center}
\end{figure}

Fig. \ref{time.fig} shows time distributions calculated by numerically solving the full system of
the linear differential equations (\ref{eq:sol}). The goal of the experiment is the determination
of the \mud\ doublet capture rate, thus (i) the doublet population should be maximized 
(or at least well defined) and (ii) background from \MHE\ minimized.
To optimize (i), the density should be increased compared to the density $\phi$=0.01 of the MuCap
experiment. This accelerates the hyperfine transition according to the rate $\phi \qdr$.  
Lower temperatures provide a significant advantage because the smaller rate \dr\ leads to less 
quartet population (via recycling) and to less \MHE\ production (ii). Moreover, at low temperatures
the $dd\mu$ formation rates \qr\ and \dr\ are dramatically different, making it easy to monitor
the hyperfine populations via the fusion time distribution $fus(t)$. This fact is clearly
demonstrated in Fig. \ref{rates.fig}, where the hyperfine transition rate is determined with
high precision at low temperatures whereas the experimental uncertainty increases to 15\%
at T=300\,K. Finally, T=30\,K allows for 5 times higher density, while 
keeping the operating pressure comparable to the MuCap conditions, so that
walls and entrance window thickness need not be increased. In summary, based on the physics 
requirements and practicability of the target design, the conditions $\phi$=0.05 and T=30\,K 
indicated in the right, lower panel of Fig. \ref{time.fig} were chosen as the baseline design of this experiment.

For a systematic study of the error contribution from all relevant quantities we used two methods, namely the first moment method and direct fits to Monte Carlo generated data sets. While we will describe the two studies in the following two subsections, it shall be noted that in both cases the set of investigated parameters was $\alpha = \{\qr, \dr, \qdr, \RHE, \omega\}$. An additional set $\alpha' = \{\lambda_\mu^+, \Rq\}$ was studied for the fitting method, only. These parameters were then varied individually over a reasonable range in order to study their effects on the experimental observable of interest, i.e. the muon disappearance rate $\lambda_\mu^-$ (or equivalently the difference in appearance rates $\lambda_\mu^- - \lambda_\mu^+)$. For this, we chose the experimental values from literature and allowed a $\pm2\sigma$ variation with $\sigma$ being the experimental uncertainty. This was compared to the variation of \Rd\ by $1\%$ as this corresponds to the best achievable precision for MuSun. The bottom line of both studies is that they are in line with each other and that the overall effect of the kinetic uncertainties in the various parameters is at a negligible level for the final determination of \Rd\ to a precision of $<\precision$.

\begin{figure}[htb]
\begin{center}
\resizebox*{0.48\textwidth}{!}{\includegraphics{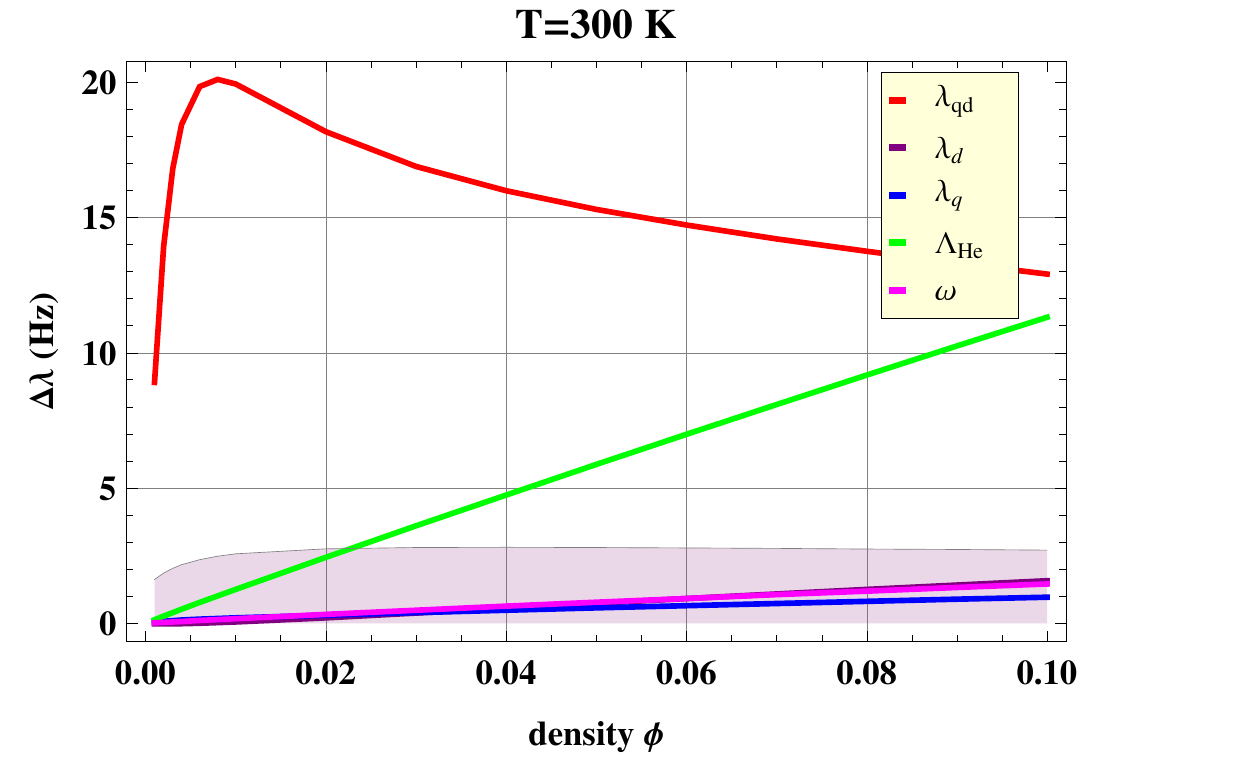}} \resizebox*{0.48\textwidth}{!}{\includegraphics{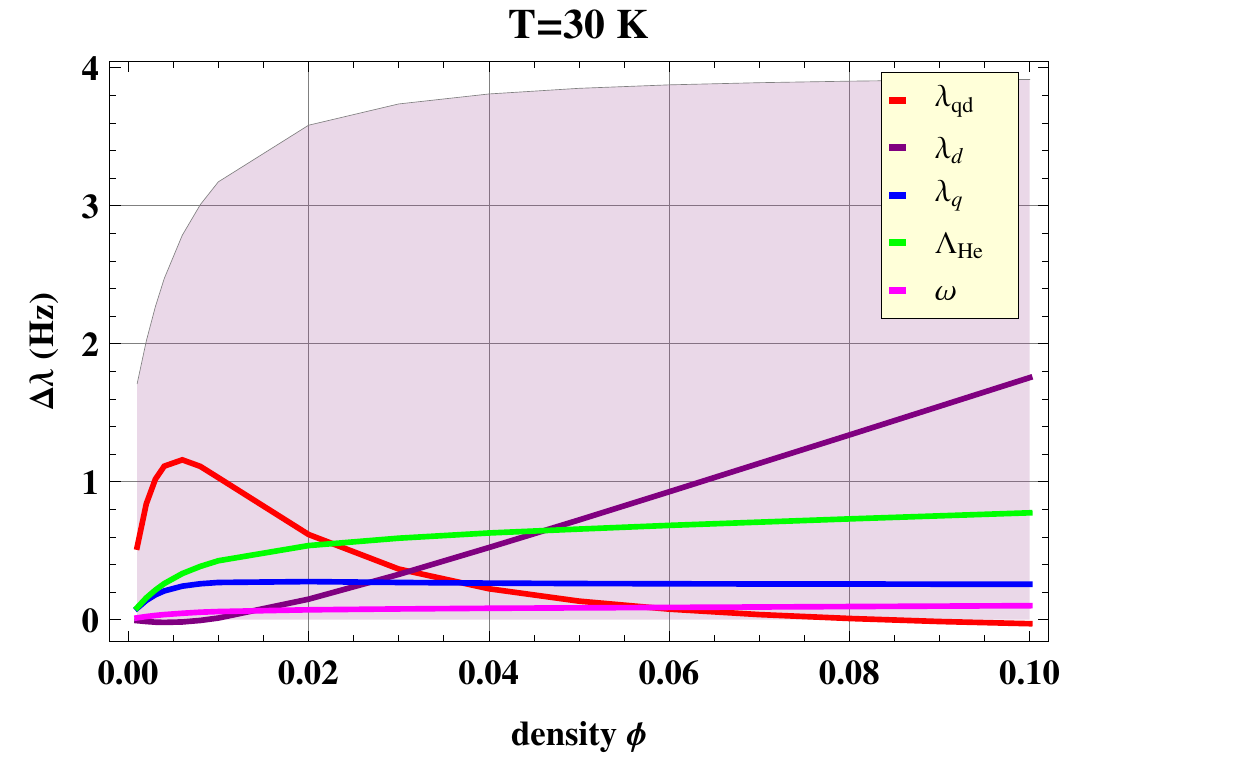}} 
\caption{Effect of a $2\sigma$ variation of the kinetics parameters on the difference $\delta \lambda$ in the observed disappearance rates. The shaded area indicates the variation of $\delta \lambda$ for a variation of \Rd\ by 1\%.}
\label{alpha.fig}
\end{center}
\end{figure}

\subsubsection{First Moment Method} 
For this method we define
\begin{equation}
\delta \lambda \equiv \frac{\int_0^\infty N(t) dt}{\int_0^\infty t\, N(t) dt} - \rz
\end{equation}
where $\delta \lambda$ is an approximation for the difference in the disappearance rates  $\lambda_\mu^- - \lambda_\mu^+$.  In Fig.~\ref{alpha.fig}, the effect on the observable $\delta \lambda$ for changes of all parameters in the set $\alpha$ is shown as function of the density. In addition, the shaded area represents the effect of a fractional variation of \Rd\ by 1\% (i.e. the measurement goal). The left plot shows this study for T=300K and the right for T=30 K.
 
The density dependence is easy to understand. At the limit of $\phi\ll 0.01$ the $\mu d$ hfs states remain in a nearly statistical mixture, therefore the dependence on $\qdr$ is reduced. However, the observed capture rate approaches the statistical rate 
$\approx \frac{\Rd}{3}$ and therefore, the sensitivity to \Rd\ decreases as well. At $\phi \gg 0.01$ the hyperfine transition becomes fast, nearly all $\mu$ atoms
are in the doublet state and, as a consequence,  the sensitivity to  $\qdr$ is small again.
Clearly, sufficient precision
is difficult at 300 K, while the uncertainties are below 0.25\% at 30 K and $\phi=0.05$, thus at an almost negligible level.

\begin{figure}[bth]
\vspace{-0mm}
\begin{center}
\subfigure[]{\includegraphics[width=0.42\textwidth]{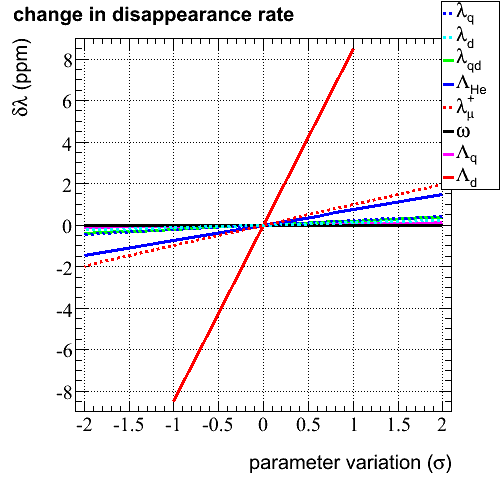}}
\subfigure[]{\includegraphics[width=0.42\textwidth]{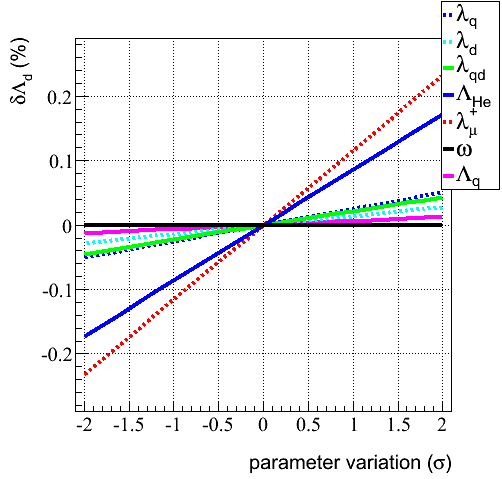}}
\vspace{-5mm}
\caption{(a) Sensitivity of the muon disappearance rate to the rate $\Lambda_\mathrm{d}$ and the set of kinetics parameters $\alpha$ and $\alpha'$. (b) Fractional change in the ``best fit'' rate $\Lambda_\mathrm{d}$ due to variation of the parameters. The $\pm2\sigma$ experimental uncertainties were: $\delta\lambda_\mathrm{q} = \pm 0.1\times10^6$ s$^{-1}$ (blue dashed line), $\delta\lambda_\mathrm{d} = \pm 0.006\times10^6$ s$^{-1}$ (cyan dashed line), $\delta\lambda_\mathrm{qd} = \pm 0.8\times10^6$ s$^{-1}$ (green solid line), $\delta\Lambda_\mathrm{He} = \pm 140$ s$^{-1}$ (blue solid line), $\delta\lambda_\mathrm{\mu^+} = \pm  2$ ppm (red dashed line) and $\delta\omega = \pm 0.0037$ (black solid line). The rate \Rd\ was varied by $\pm 1\%$ and the unmeasured rate \Rq\ was varied by $\pm 20\%$ (magenta solid line).}
\label{fullfit.fig}
\end{center}
\end{figure}
 
\subsubsection{Full Kinetic Fits}
The impact of the experimental uncertainties in the kinetics parameters
on the extraction of the rate \Rd\ from the measurement of the electron time spectrum was also studied by a Monte-Carlo simulation for the MuSun condition of T=30\,K and $\phi=0.05$.
First, the analytical solutions obtained from the
coupled differential equations describing the 
$\mu d$ chemistry were used to generate electron time spectra.
Next, the doublet rate \Rd\ was extracted 
via a fit of the electron time spectrum to the aforementioned analytical solutions 
with all kinetics parameters fixed at the measured values.
Last, by varying the values of the kinetics parameters
when generating the electron time spectra, 
but fixing the values of the kinetics parameters
when fitting the electron time spectra,
the sensitivity of the rate \Rd\ to the set of kinetics parameters $\alpha$ and $\alpha'$ could be determined.

As can be seen from the results shown in the Fig.~\ref{fullfit.fig}, the sensitivity of the extracted rate \Rd\ to the experimental uncertainties in the kinetics parameters (including correlations between kinetics parameters) is well below our best expected precision of $\pm$1\% in \Rd. In addition, no significant sensitivity for variable fit start times in the range of 0 to 1\,$\mu$s was found.

\subsection{Observables}

\begin{figure}[htb]
\begin{center}
\resizebox*{0.45\textwidth}{!}{\includegraphics{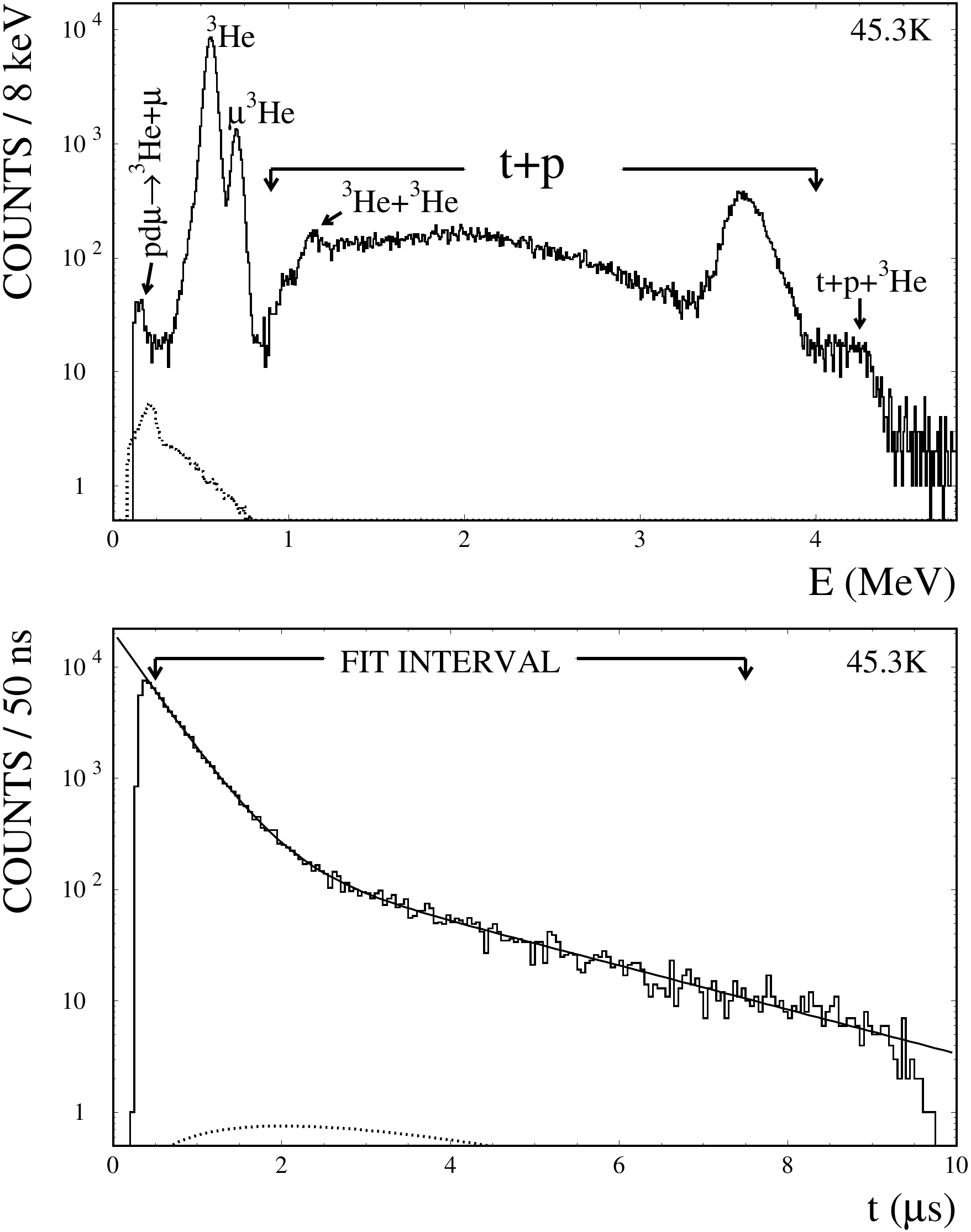}} 
\caption{Charged particle spectra after muon stop from Ref.~\cite{Bal07}. The conditions were T=45.3 K (which is very close to the operating point of MuSun), $\phi$=0.0524 and nitrogen impurity level c$_N \approx$ 41 ppb. Impurity capture background indicated by dotted line.}
\label{recoil.fig}
\end{center}
\end{figure}

The primary observables for the precision measurement of the muon decay rate are the muon track in the TPC 
and the decay electron vector reconstructed by the electron tracker. In addition, critical information 
for the unambiguous interpretation of the experiment is derived from the detection of charged particles
from fusion and impurity capture within the TPC and with external neutron and gamma detectors. 
Table~\ref{recoil.tab} and Fig.~\ref{recoil.fig} show the properties of these charged particles in the TPC 
as observed under similar conditions in a previous experiment~\cite{Bal07}. The events below 3.6 MeV are 
due to proton tracks escaping the sensitive volume. Two small peaks at 1.2 MeV and 4.2 MeV indicate the pile up 
of two subsequent fusion signals within 200 ns for $^3$He$ + ^3$He and $(t + p) + ^3$He, respectively. Events below
0.45 MeV are due to $pd$ fusion and gas impurities. 
Complementary information on fusion and capture neutrons is provided by neutron detectors, which
have small efficiency, but provide excellent time resolution.

\begin{table}
\begin{center}
\begin{tabular}{rrrr}
\hline
		&E (MeV)	&E$_{obs}$ (MeV)	&R(mm)    \\
\hline
$^3$He		&0.82		&0.6			&0.18     \\
$\mu$$^3$He	&0.80		&0.75			&0.6      \\
t		&1.01		&			&1\\
p		&3.02		&			&16 \\
\hline 
\end{tabular}
\caption{Recoil energies and range of fusion products at $\phi$=0.05. The observed energies are lower
due to charge recombination. }
\label{recoil.tab}
\vspace{-.5cm}
\end{center}
\end{table}


\section{Experimental Setup}

\subsection{Cryogenic Time Projection Chamber }
\label{CIC.sec}

\subsubsection{Main Design Considerations}

\begin{figure}[htb]
\begin{center}
\vspace{-30mm}
\resizebox*{1.3\textwidth}{!}{\includegraphics{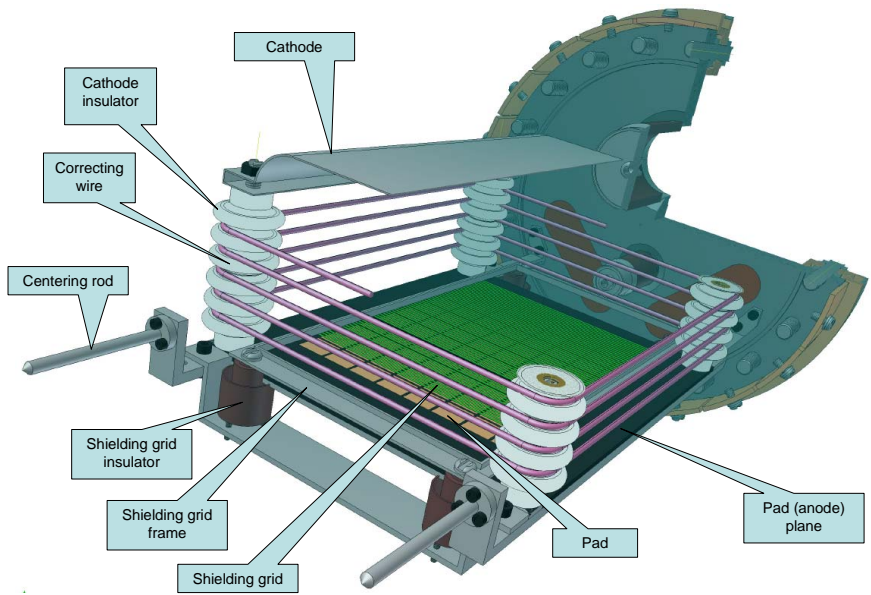}} 
\vspace{-50mm}
\caption{\capCIC\ layout with the main components described by the labels.}
\label{CIC.fig}
\end{center}
\end{figure}

The baseline design of the cryogenic TPC is presented in Fig.~\ref{CIC.fig}. We define the chamber coordinate system with the $x$-axis horizontal and transverse to the beam direction, the $y$-axis pointing vertical up and the $z$-axis along the beam direction. At the gas density $\phi$=0.05 it will operate without gas amplification as a cryogenic time-projection chamber (\CIC).
The main design criteria are as follows.

\begin{figure}[htb]
\begin{center}
\resizebox*{0.45\textwidth}{!}{\includegraphics{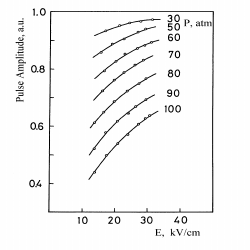}}\resizebox*{0.45\textwidth}{!}{\includegraphics{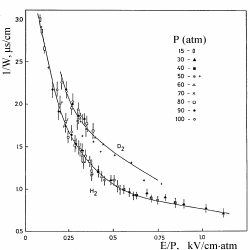}} \\
\caption{Recombination and drift velocity from Ref.~\cite{Bal07}. The MuSun conditions correspond to P=50 atm, E=10 kV/cm and E/P= 0.2 kV/cm-atm.}
\label{properties.fig}
\end{center}
\end{figure}

The size of the sensitive pad area is determined by the properties of the $\pi$E3 muon beam and the 
scattering in the beryllium entrance window. GEANT Monte Carlo simulations, tuned to the 
observed MuCap stopping distributions, predict that \mustopfrac\ of incoming muons will stop
over a horizontal $100\times 100$ mm$^2$ pad area (along the $z$ (beam) and $x$ (transverse to beam) axis). 
The chamber height of 80-100 mm (vertical axis $y$) was chosen to contain most of the beam.
This choice is critical, as it determines the required drift high voltage $U_d$.
With $U_d$=100 kV, the electric field to pressure ratio is E/P=0.2~kV/cm-atm (same as for MuCap).
According to Fig.~\ref{properties.fig} this results in an acceptable recombination R$_\alpha$ = 0.82
and a drift velocity $v_d \approx$ 0.4 cm/$\mu$s. Gas gain is difficult to achieve as the field on the
anode wires' surfaces would need to be increased by about 
5 times due to the higher gas density compared to the MuCap conditions. However, intrinsically an ionization chamber is more robust and capable of higher resolution than a proportional chamber. Therefore, gas amplification is not necessary if an overall resolution
of 30-50 keV can be achieved in the pad readout.

\begin{figure}[htb]
\begin{center}
\begin{tabular}{ccc}
\resizebox*{0.3\textwidth}{!}{\includegraphics{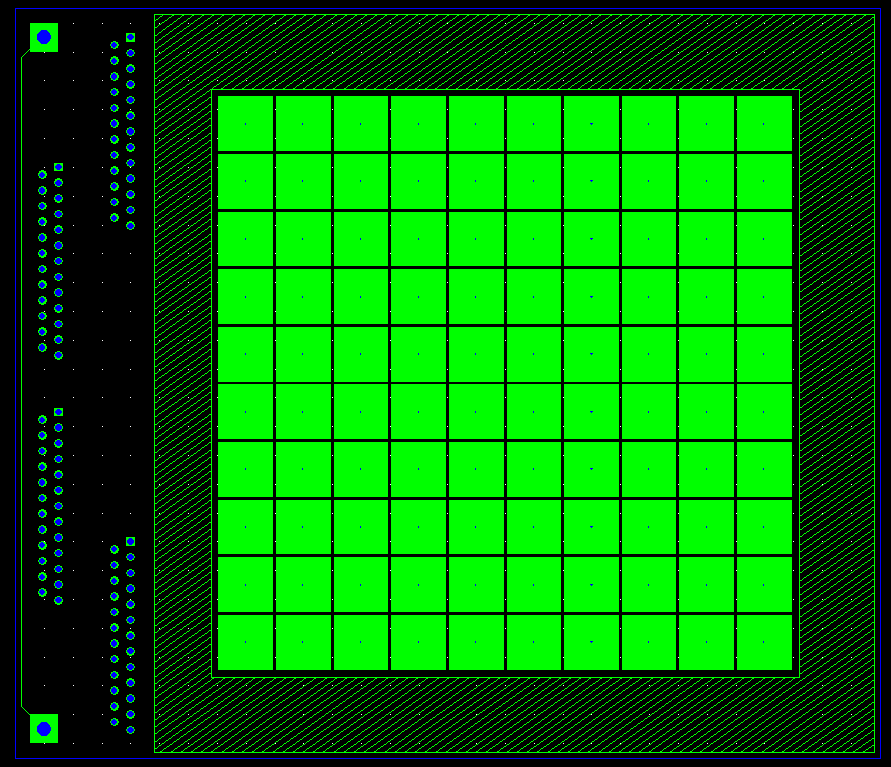}} &
\resizebox*{0.3\textwidth}{!}{\includegraphics{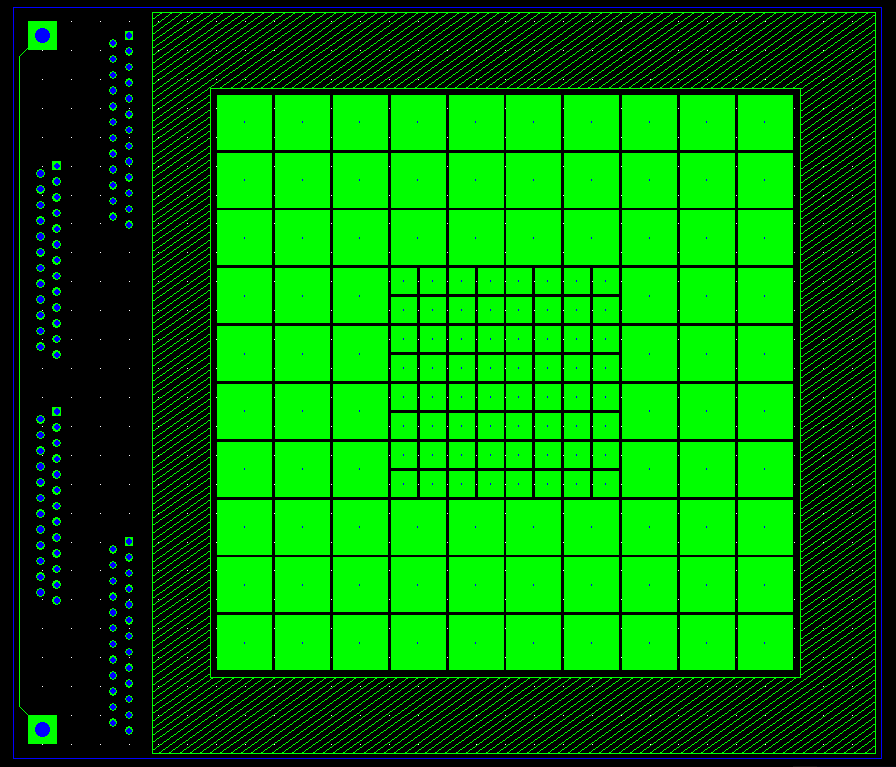}} &
\resizebox*{0.3\textwidth}{!}{\includegraphics{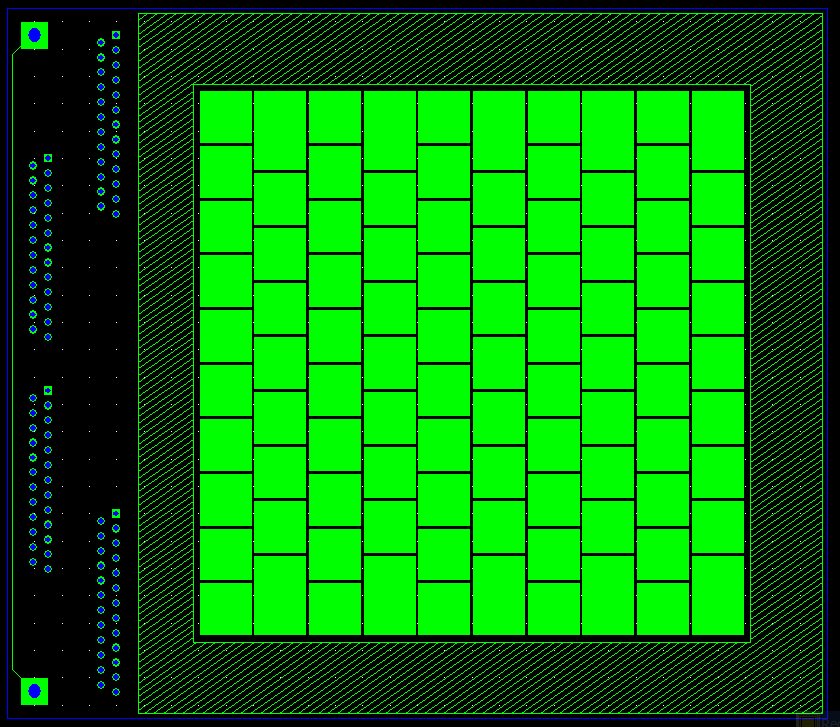}} \\
\end{tabular}
\caption{Pad layout variants under consideration. Left: Baseline $10\times 10$ mm$^2$, Center: $5\times 5$ mm$^2$ in center, Right: ``Brick wall'' for improved $x$ tracking}
\label{choices.fig}
\end{center}
\end{figure}

\begin{figure}[htb]
\begin{center}
\resizebox*{0.6\textwidth}{!}{\includegraphics{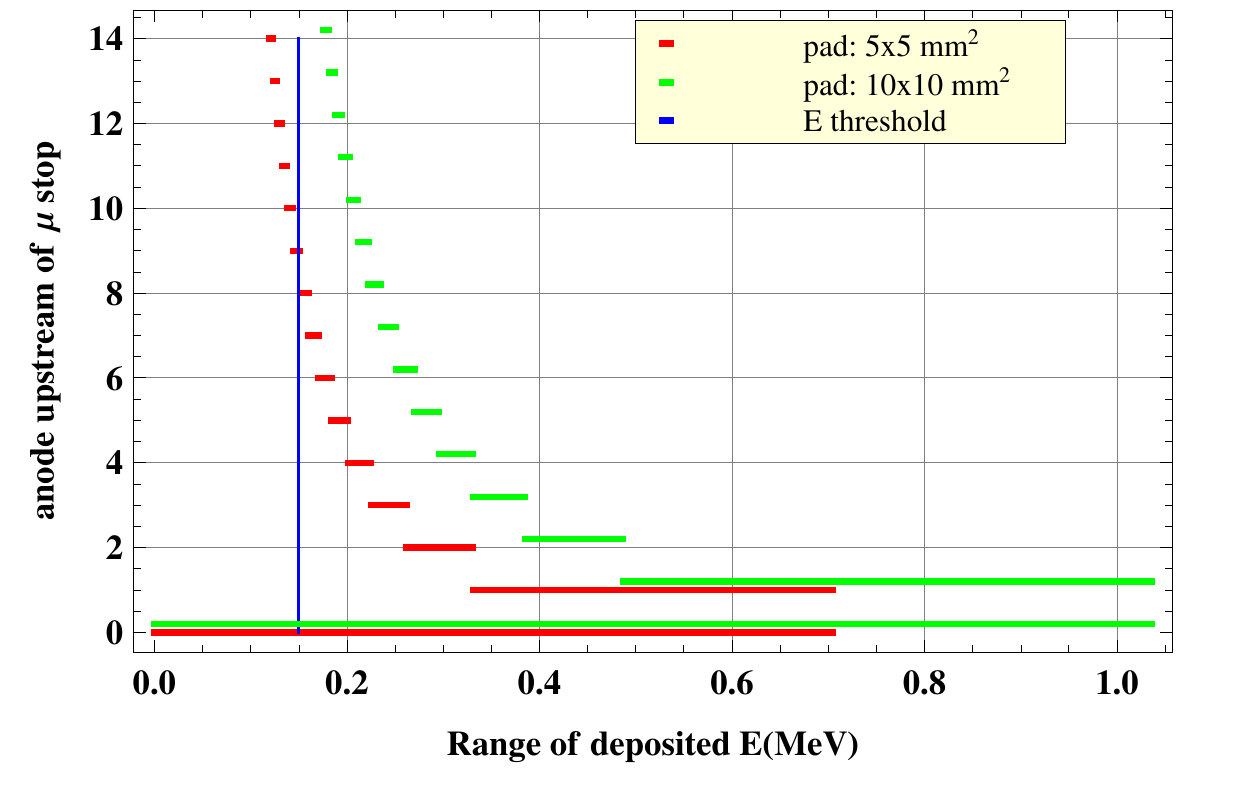}} 
\caption{Muon energy deposition on anode pads. The anode number 0 corrsponds to the muon stopping pad, whereas the higher anode number are the pads upstream of this muon stop position along the incoming muon path.}
\label{edep.fig}
\end{center}
\end{figure}
\clearpage

For the baseline configuration we divide the readout plane into 100 pads of each $10\times 10$ mm$^2$ in size. An R\&D program
including Monte Carlo studies as well as experiments with prototypes is foreseen to define the
final optimized layout. Some choices currently considered are indicated in Fig.~\ref{choices.fig}.
A total number of 100-200 pads would be well matched to our existing custom made FADCs and the DAQ for full
analog readout. During this R\&D studies, the following requirements need to be optimized:

\begin{itemize}

\item
The muon energy deposition relative to the stopping pad is shown in Fig.~\ref{edep.fig}. With $10\times 10$ mm$^2$ pads, all muons entering the chamber deposit enough energy per pad so that they are detected above threshold along their entire path. In comparison, a $5\times 5$ mm$^2$ pad size would give higher spatial resolution, but the 
muons are only detected up to $\approx$ 35 mm upstream of the pad where the muon stops. Both pad sizes are acceptable for 
imposing muon - decay electron vertex cuts to suppress background, as this impact cut will not be tighter than 
10 mm to avoid time-dependent effects from $\mu d$ diffusion.
 
\item
The energy resolution will deteriorate with the square root of the number of pads accumulating the signal, as
it is dominated by the readout cable capacitance and preamplifier noise.  For $^3$He fusion recoils with a range of 0.18 mm, about twice as many fusion recoils will escape a pad for 5 mm 
compared to 10 mm pad length. In both cases, the effect is at the percent level only.  

\item
Impurity detection might be the strongest argument for higher granularity in at least part of the pad plane. 
As discussed below the detection of proton emission after muon capture on trace levels of nitrogen contaminants
might provide a definitive signature to distinguish impurity capture from the frequent $^3$He$ + n$ fusion channel.

\item
The systematic effects due to muon fusion overlaps would be reduced by higher granularity. As this effect leads to systematic distortions of the time spectrum, it should be minimized by an optimal detector geometry.

\end{itemize}

\subsubsection{Technical Design}

\begin{figure}[htb]
\begin{center}
\vspace{-0mm}
\resizebox*{0.7\textwidth}{!}{\includegraphics{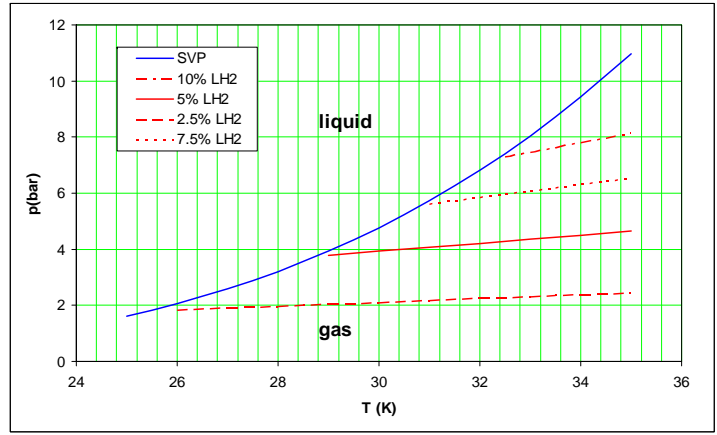}}
\vspace{-10mm} 
\caption{Saturated vapor pressure curve and isochores for cryogenic deuterium.}
\label{eos.fig}
\end{center}
\end{figure}

The \CIC\ is filled with ultrapure deuterium and works at low temperatures T$=25-35$\,K  and densities of $\phi = 0.05-0.08$ relative to liquid hydrogen density. According to the thermodynamic properties of gaseous deuterium in Fig.~\ref{eos.fig} this corresponds to a pressure range of P$=4-6.5$\,bar.
Fig.~\ref{CIC.fig} shows the main parts of the chamber and it specifications are given in table~\ref{CIC.tab}.

\begin{figure}[htb]
\begin{center}
\vspace{-20mm}
\resizebox*{1.3\textwidth}{!}{\includegraphics{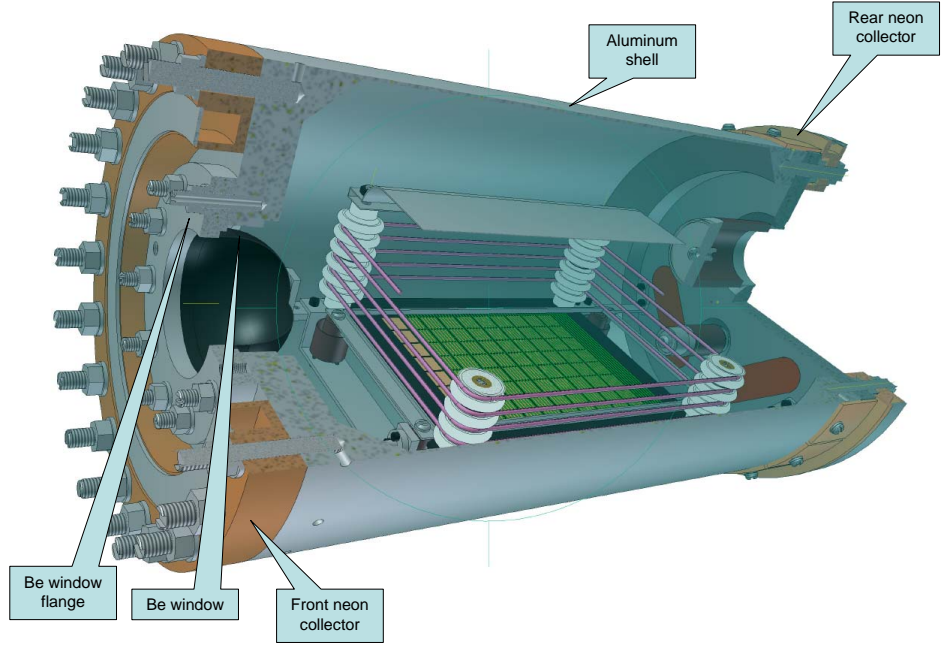}}
\vspace{-25mm} 
\caption{\capCIC\ with pressure vessel.}
\label{rear.fig}
\end{center}
\end{figure}

\begin{table}[htb]
\begin{center}
\begin{tabular}{llr}
\hline 
system 		& parameter 			& value \\
\hline
Drift cathode   & potential $U_d$               & -80 to -100 kV \\
                & cathode-grid distance 	& 80 mm		\\
Grid		& potential $U_g$		& -3 to -5 kV   \\
		& grid-anode distance           & 1 mm          \\
                & wire pitch                    & 400 $\mu$m    \\
		& wire diameter			& 55 $\mu$m     \\
		& wire material                 & stainless steel\\
Anode           & area                          & $100 \times 100$ mm$^2$ \\
		& pad size                      & $10\times 10$ mm$^2$   \\
		& total number pads                       & 100  \\
Performance     & drift velocity                & 0.4 cm/$\mu$s  \\
 		& drift time cathode to grid    & 25 $\mu$s      \\		 
    		& drift time grid to anode      & $\approx$ 150 ns \\
                & energy resolution             & 30-50 keV      \\
                & energy threshold              & 150-200 keV    \\
\hline
\end{tabular}
\caption{\capCIC\ specifications}
\label{CIC.tab}
\end{center}
\vspace{-.5cm}
\end{table}

\clearpage

At the top of the chamber is the high voltage cathode. Its tent-like shape helps to establish a constant electrical 
field between the cathode and the first field shaping wire. The final spacing and geometry of the field shaping wires will be determined using standard electric field calculation software in order to obtain a homogeneous field distribution in the drift region. The drift volume is separated by a (Frisch) grid  from 
the detection pad plane.
The grid is mounted on 4 ceramic insulators with possibility to vary distance between the anode pad plane and 
the grid from 0.5 mm to 1.5 mm. For the moment, we foresee a distance of 1 mm still being subject to further optimization.
The anode pad plane is fixed on a supporting frame with centering pins on the front end. These will be 
connected to the front flange  of the aluminum shell to protect the \CIC\ against vibrations.

The grid consists of the frame and the wires which are soldered on the long side of
the frame (wires along $x$ direction). It works as an electrical screen for the positive charge in the \CIC\ and is
transparent for the electrons drifting from the cathode to the anode plane. The grid wires will have a diameter
 of 55 $\mu$m, a pitch of 400 $\mu$m and will be strung with a tension of 60 g. For a total number of  325 wires
this adds to a total distributed load of 19.5 kg. The internal stress distribution and deformation of the
 frame was calculated with the finite element code ANSYS. The calculated values of deformations guarantee 
that at all temperatures the wires sagging will be less then 80 $\mu$m, corresponding to 8\% of the 1 mm 
distance between pad plane and grid, which is below the homogeneity requirement for the electrical field.
\begin{figure}[htb]
\vspace{-0mm}
\centering
\includegraphics[width=0.40\textwidth]{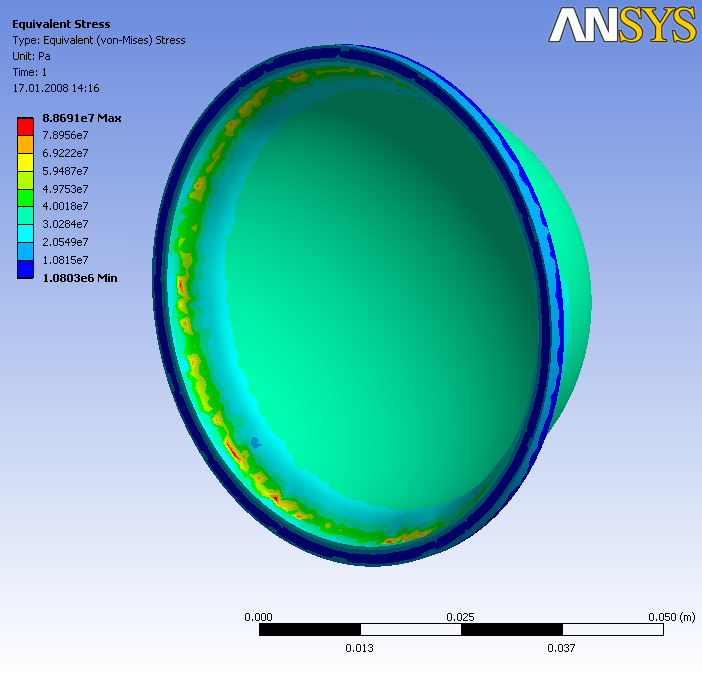}
\hspace{0.06\textwidth}
\includegraphics[width=0.40\textwidth]{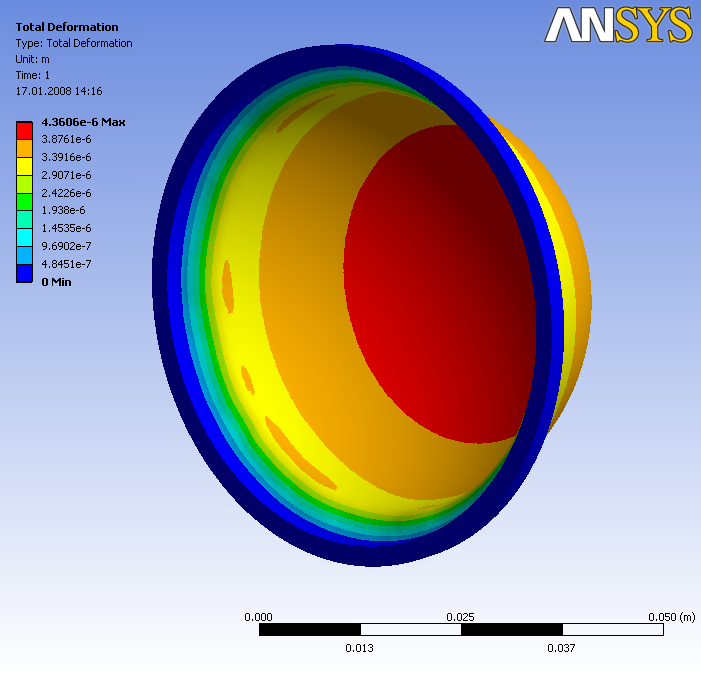}
\includegraphics[width=0.45\textwidth]{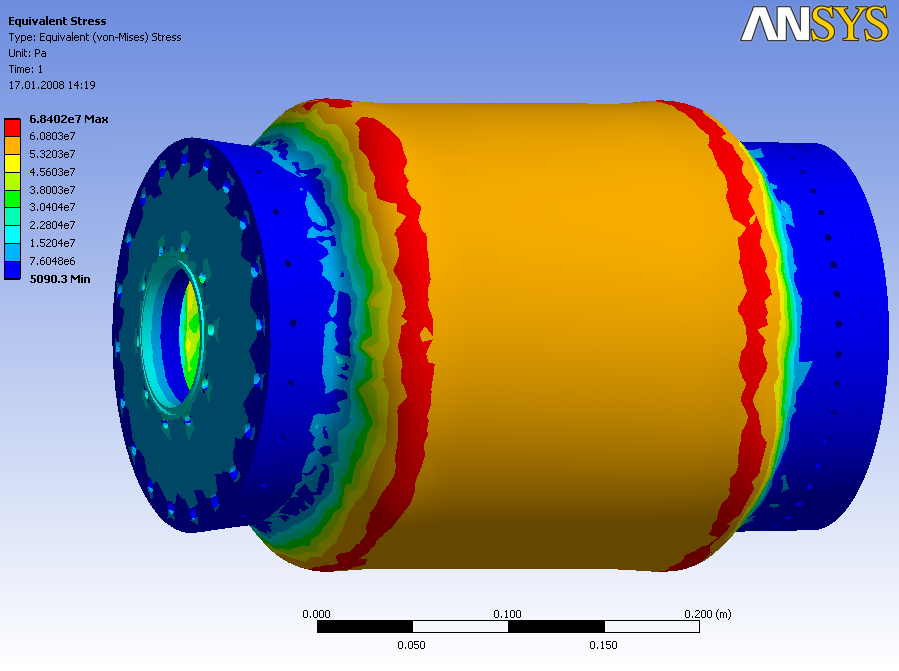}
\includegraphics[width=0.45\textwidth]{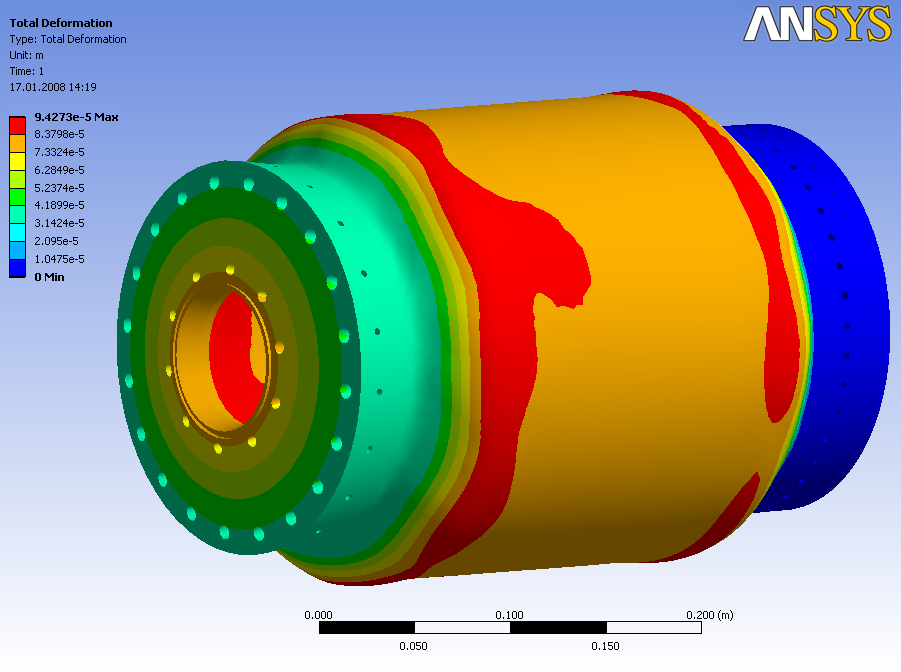}
\caption{ANSYS calculations of the stress (left plots) and the deformation (right plots) for the window and the entire pressure vessel. The maximum scale (red areas) for the stress and deformation are: 88.7 MPa and 0.005\,mm for the window and 68.4\,MPa and 0.1\,mm for the entire vessel.}
\label{stress.fig}
\end{figure}

Onto the rear flange (Fig.~\ref{rear.fig}) of the \CIC\ vessel several ceramic feedthroughs are welded: The 100 kV feed-through for the cathode, 
the 10 kV feed-through for the grid, and two 50-pin signal connectors for the signals from the cathode pad plane.
The \CIC\ active target is placed  inside an aluminum  shell which has to be thin to transmit the decay
electrons, but thick enough to provide the necessary strength against the pressure in the chamber. 
ANSYS calculations (Fig.~\ref{stress.fig}) indicate that this can be achieved with a side wall thickness of 2.5 mm leading to a 
maximum stress of 68 MPa. Different aluminum alloys give different 
numbers from 150-300 MPa. We have about 2-3 times reserve of the aluminum shell strength.
 The spherical beryllium window for the input muons and two neon collectors for the cooling of the target will be installed on the front flange. 
The window is a beryllium semi-sphere welded into the stainless steel flange and it has a diameter of 61 mm and a thickness 0.4 mm. Calculations of the stress distribution and 
deformation distribution are shown in Fig.~\ref{stress.fig}.  All calculations were done at a pressure of 10 bar 
and room temperature. The maximum stress is 88.7 MPa. The hot-pressed beryllium maximum stress value 
is 200-300 MPa, so that there is a reserve of about 2-3 times in the stability of the window. 
We are planning to carry out a number of tests to evaluate the window performance at cryogenic temperatures. 
The mechanical properties of the system of the stainless steel flange and the beryllium window at low temperatures are improved, but we have to evaluate the thermal stress. For the connection of the beryllium window to the chamber and the rear flange 
we are using cog-groove type compaction with an indium wire in between. This type of connection is well known 
for cryogenic applications and suitable for our pressures.  The choice of the Al alloy 
and the overall design will be finalized in accordance to the PSI/Swiss safety requirements and procedures. 


\subsection{Cryogenics and Gas System}
\begin{figure}[htb]
\centering
\vspace{-10mm}
\resizebox*{1.3\textwidth}{!}{\includegraphics{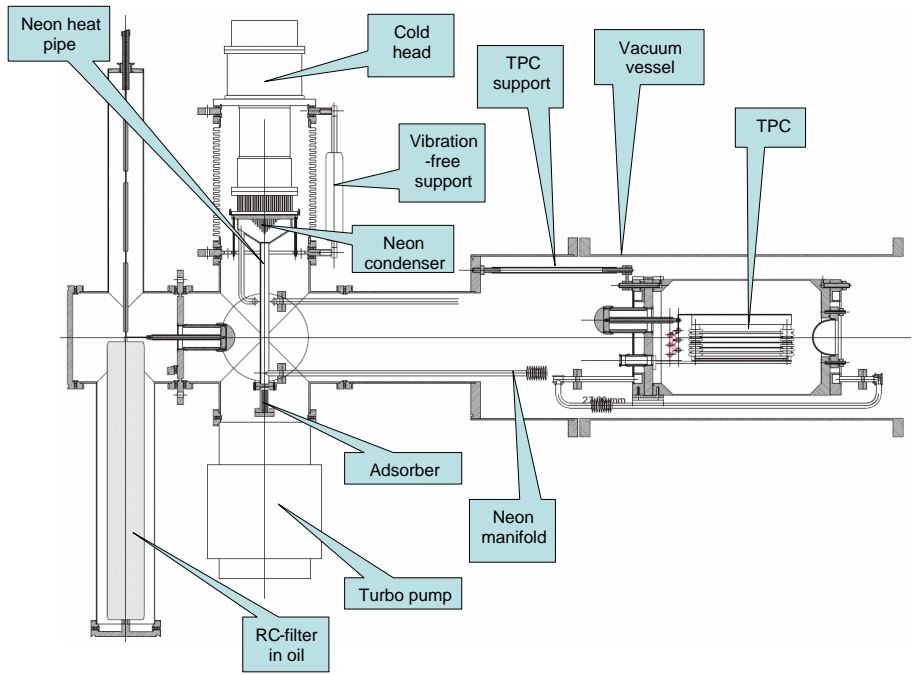}}
\vspace{-55mm} 
\caption{Overall cryogenic system as described in the text.}
\label{cryo.fig}
\end{figure}

For the cooling system of the \CIC\ (see Fig. \ref{cryo.fig}), we are planning to use the cold head {\tt COOLPOWER} 140T by Leybold. At 25-35 K, the cold head can produce about 30 W cooling power, which is sufficient. We propose to use a heat pipe system with neon. From the cold head's lower flange the cooling power is transported through flexible copper elements (to suppress the transmission of vibrations) to the neon condenser where the neon gas is liquified. Running down a vertical tube, the liquid neon is collected in two heat exchangers that are mounted on both ends of the chamber with good thermal contact to the body of the chamber. Each heat exchanger is equipped with a heater (for precise temperature stabilization) and a temperature sensor (Pt500 or Pt1000). Additional heater and temperature sensors on the neon condenser enable to control the necessary level of liquid neon in the vertical tube. The liquid neon vaporizes in the heat exchangers and returns to the condenser. To exclude any transmission of vibration from the cold head via vacuum insulation, the vacuum  flange of the cold head is connected to the vacuum chamber via a bellow with a vibration-free support.

\subsubsection{Chemical Purity}

It is impossible to provide the required ultra-high chemical purity of deuterium only by means of initial purification. Various contaminants (especially air components and moisture) will migrate to the gas from chamber elements and sealings even under cryogenic conditions. The existing circulating cryogenic system for continuous ultra-high hydrogen purification~\cite{Ganzha:2007uk} from MuCap will be used for the removal of contaminants in the deuterium gas. This system was successfully used to provide ultra-high purity of deuterium-depleted hydrogen in the MuCap experiment and an excellent purification power during several MuCap runs between 2004-2007 was achieved. The quality of the purification was monitored by gas chromatography and by means of an online humidity sensor. The achieved results~\cite{Ganzha:2007uk} corresponded to MuCap demands (7$\pm$1 ppb of N$_2$; $<$ 5 ppb of O$_2$, $\approx$10 ppb of H$_2$O). The system consists of three base units: compressor, purifier and automated control system. The compressor has to provide a constant flow of hydrogen through the purifier with a rate high enough to support the specified purity of the gas in the TPC. Since the compressor works as a cryopump it has the following advantages: big reliability, and a large range of gas flow rate. A cryopump is based on the ability of a special substance (adsorbent) to absorb considerable amounts of gas while cooled and relieve them upon subsequent heating. In our case, activated carbon is used as the adsorbent and its cooling and heating are provided by liquid nitrogen and electrical heaters, respectively.
 
The principle of operation of the purifier is based on prevalent (in comparison with the main component, hydrogen) adsorption of contaminations (nitrogen, oxygen, water etc.) in an adsorption filter. Synthetic zeolite is used as the adsorbent for this goal. To increase the rate of purification, the filter is strongly cooled with liquid nitrogen (the same as for the cryopump). The adsorption method of purification guarantees a high level of purification at the very wide range of species.
The automated control system consists of a microprocessor control block, a set of sensors and devices (remotely controlled valves, mass-flow controllers and heaters), and a control PC. This control unit organizes the operation of the system in fully automatic mode. It provides all sensor interfaces and low-level control procedures. The advantage of the system is the ability to provide relative pressure stability at the level of better than 0.1\%. The system is designed following all hydrogen safety precautions and it protects the detector from overpressure damage in case of vacuum problems. This general-purpose cryogenic adsorption method allows using the system for cleaning of deuterium as well as protium without major changes. 

The cryopump can support circulation of deuterium at the same level as protium and the same contaminants (O$_2$, N$_2$, water) are removed.
To provide the final extra-cleaning, an additional adsorption filter will be introduced in the new cryogenic scheme. This filter is mainly identical to the filters of the purifier but it is cooled by liquid neon. Its temperature corresponds (or even slightly lower) to the temperature of the detector ($\approx$30 K) which is about 37 degrees lower than the temperature of the filters of the purifier. This temperature difference provides sufficiently better cleaning conditions on the final stage of purification.

\subsubsection{Isotopic Purity}

\begin{figure}[htb]
\begin{center}
\vspace{-10mm}
\resizebox*{1.2\textwidth}{!}{\includegraphics{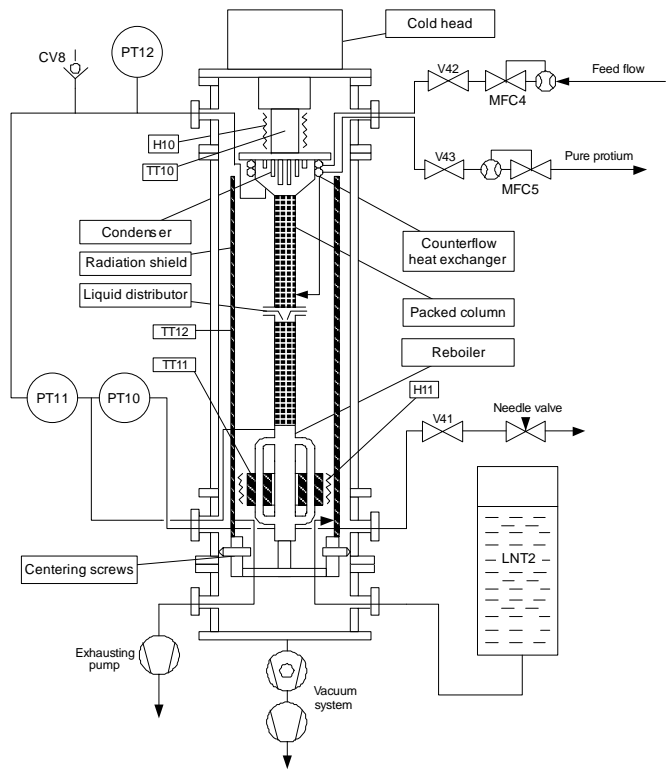}} 
\vspace{-25mm}
\caption{The isotope separation system. The main components are the cold head with the condensor at the top, the packed column where the actual separation takes place, and the reboiler at the bottom.}
\label{dru.fig}
\end{center}
\end{figure}

The cryogenic separation facility (Deuterium Separation Unit, Fig.~\ref{dru.fig}) will be used to provide a good initial isotopic purity. This setup was designed and built in 2006 for MuCap. Based on the achieved results with H/D mixtures, we expect to be able to produce pure deuterium with less than 1 ppm protium contamination. We briefly review the system as used.

The facility uses the well-known rectification method to separate isotopes of hydrogen using the difference in saturation vapor pressure of separating species above the surface of a liquid mixture. It can be considered as a multi-step distillation with the use of a column filled with special material (either a set of perforated plates or a particular packing) to increase a phase contact surface. A condenser is placed on the top of the column to liquify the vapor mixture. This liquid, called reflux, then returns into the top of the column. The vapor can also be partially taken away from the top of the column as a pure product of the process. The reflux drains down along the column, moistening the packing. An amount of the liquid suspended on the packing is a column holdup. The lower end of the column is equipped with a reboiler. A separated mixture boils in the reboiler forming the vapor, which rises upward along the column and interacts with the counter flow of draining reflux. The liquid is saturated with the high-boiling component, and the gas with the low-boiling one. For the hydrogen isotopic separation the cryogenic modification of the rectification method is used. A separation column of 2.2 cm inner diameter and 155 cm overall packing height is cased into a vacuum heat insulation volume. Liquification of the distillated gas is provided by the 20\,W cold head connected to the condenser. The specially designed control system provides the necessary algorithms for the column operation. The schematic drawing of the separation unit is shown in Fig.~\ref{dru.fig}. As a result of the column operation using natural hydrogen as the source gas, H$_2$ with a concentration of HD molecules less than 6 ppb was obtained; that is the most isotopically clean protium in the world. The result was confirmed by direct measurements with a large compact accelerator at the Institute of Particle Physics, ETH-Zurich, Switzerland~\cite{Suter:2007}.

\subsection{Detectors}

Whereas the essential detector, the \CIC, was described in a separate section~\ref{CIC.sec}, we will now review the other relevant components, their role with respect to the measurement and the design criteria. Most of these components come from the existing MuCap experiment, while some additional detectors will be added for MuSun. A schematic view of parts of the full system was shown in Fig.~\ref{setup.fig}.

\subsubsection{Entrance Detectors}

Entrance muon detectors provide timing of incident beam muons and enable the selection of events where only one muon entered the target (``pileup protection''). These detectors are thin to avoid degrading the beam, which would adversely affect the spatial stopping distribution of muons
within the hydrogen gas.
The elements between the beam window and the entrance window of the pressure
vessel containing the D$_2$ are
described below in the order encountered
by a beam particle.

The first detector is a 500-$\mu$m-thick muon scintillator, the $\mu$SC, to provide a fast
timing signal.  
After the $\mu$SC is the $\mu$SCA, a scintillator with a 35-mm-diameter hole
in the middle to allow most beam particles to pass.  Its purpose is to
veto muons that are too far off the beam axis.  Immediately behind the
$\mu$SCA and aligned to it is a lead collimator also with a 35-mm-diameter hole.

A multiwire proportional chamber, the $\mu$PC, follows the lead
collimator.  The $\mu$PC has two anode planes, each with 24~wires,
and 25-$\mu$m-thick aluminized mylar cathode planes.
The anode planes are oriented such that one provides horizontal ($x$)
positions of beam particles, and the other provides vertical ($y$) positions.
The $\mu$PC improves the pileup-protection efficiency compared to the $\mu$SC alone.

We are considering to place a thin double sided silicon strip detector inside the insulation
vacuum to facilitate the beam targeting onto the Be window into the D$_2$ pressure vessel.

\subsubsection{Electron Detectors}

MuSun will use the MuCap electron tracker which consists of two cylindrical chambers (ePC1, ePC2) and a scintillator hodoscope (eSC).

The electron tracking detectors are two concentric,
cylindrical multiwire proportional chambers, each with readout of
anodes and cathode strips, to give the complete $(\phi, z)$ positions
(in cylindrical coordinates) of an electron track at two different radii.
The smaller chamber (ePC1) sits just outside the pressure vessel.
The larger chamber (ePC2), with about twice the diameter as the 
smaller one, sits somewhat inside the scintillator hodoscope barrel (eSC).
Anode wires run parallel to the cylinder axis, and cathode strips
wrap around the chamber making an angle of $\approx 45$ degrees with
the anodes.  The inner and outer cathode planes wind in opposite
directions, providing redundancy if the anode and both cathode planes
of a chamber are required.
Physical parameters of ePC1 and ePC2 are given in Table~\ref{tab:epc_phys_pars}.
\begin{table}
\centering
\begin{tabular}{|l|r|r|}
\hline
Parameter                         & ePC1          & ePC2      \\
\hline
Number of anode wires             &  512          & 1024      \\
Number of inner cathode strips    &  192          &  320      \\
Number of outer cathode strips    &  192          &  320      \\
Operating voltage                 &  $+2.6$~kV    & $+2.8$~kV \\
Half-gap                          &   4~mm        & 4~mm      \\
Diameter at anodes                &  384~mm       & 640~mm    \\
Active length                     &  580~mm       & 800~mm    \\
Anode spacing                     &  2.356~mm     & 1.963~mm  \\
Inner cathode screw angle         &  43.81~deg.  & 44.31~deg. \\
Outer cathode screw angle         &  46.19~deg.  & 45.74~deg. \\
\hline
\end{tabular}
\caption{Physical parameters of the electron proportional chambers (ePCs).}
\label{tab:epc_phys_pars}
\end{table}

The anode and cathode planes are fully instrumented with chamber-mounted,
charge-integrating-preamp-discriminator cards.  Although mounting the preamps
directly on the chambers reduced electronic noise into the sensitive
preamplifiers, it was necessary to shield the cards and the entire chamber
from external electromagnetic interference via thin copper meshes.
The outputs of the preamp/discriminator cards are connected to custom
data acquisition modules through 40-wire twisted pair
cables (32 wires are used for signals, the remaining
for threshold setting and preamp power),
and each cable is wrapped in braided wire shielding.
The custom data-acquisition modules, called compressors,
are based on FPGA circuitry.
The discriminated signals from the ePC electronics are transmitted
as low-voltage differential signals (LVDS) to the compressors,
which encode them into time--channel words that are saved in a buffer.

Fast timing of electrons is the purpose of the eSC, a scintillator hodoscope
comprising sixteen segments,
each with an active area of 90$\times$15~cm$^2$ placed with the long
axis parallel to the beam axis, together forming a barrel with
a diameter of 78~cm.
Each eSC segment has two 5-mm-thick scintillating plastic layers with photomultiplier tubes on both ends. The total of 64 photomultiplier signals are input via discriminators to
data acquisition modules (CAEN V767 time-to-digital converters) that
record the time of each leading edge with 1.25~ns precision. In addition, the full analog signals are read out by custom built 8-bit wave form digitizer (WFD) boards. The time difference between detection by the upstream and downstream photomultipliers provides some information about where the particle
hit along the length of the segment.
All four photomultipliers on a given segment are required to be
in coincidence in the data analysis. The 4-fold coincidence
reduces the level of random noise from the eSC as well as afterpulsing, generally leaving
only signals from real particles.

\subsubsection{Neutron Detectors}

\begin{figure}[htb]
\begin{center}
\begin{tabular}{cc}
\resizebox*{0.3\textwidth}{!}{\includegraphics{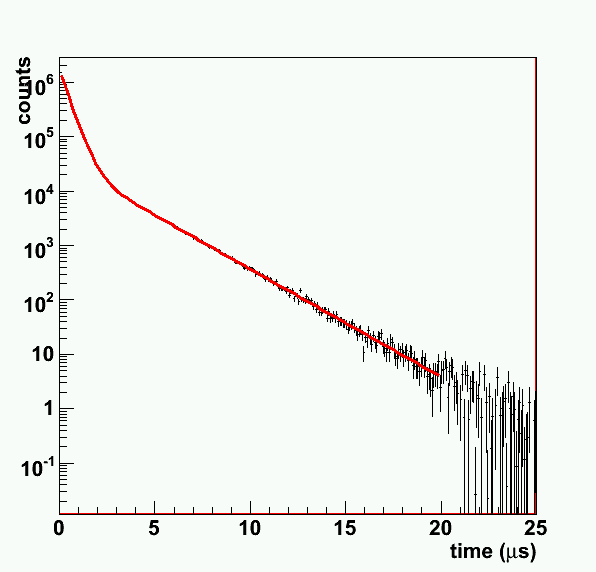}} &
\resizebox*{0.3\textwidth}{!}{\includegraphics{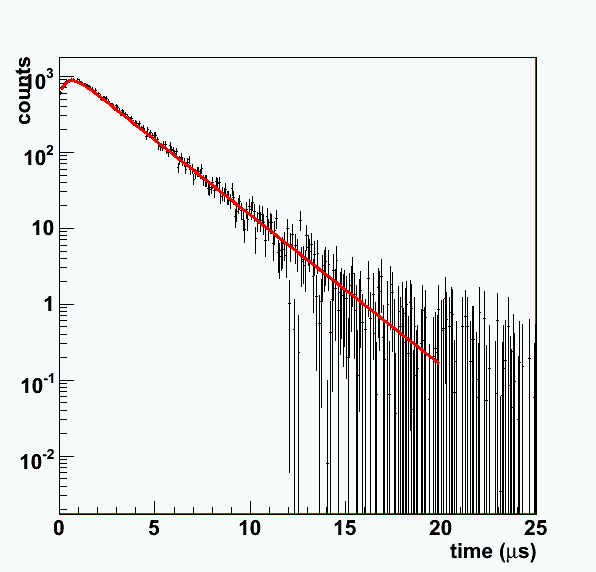}} \\
\end{tabular}
\caption{Simulated time distribution of fusion neeutrons (left) 
and capture neutrons (right) for $T = 30$~K and $\phi = 0.05$.
Also shown are the results of the fits to the neutrons time spectra
from the solution of the $\mu^-$d kinetics equations.}
\label{neutrons.fig}
\end{center}
\end{figure}

Muon stops in deuterium yield two distinct sources 
of neutrons: (i) fusion neutrons following 
$d \mu d$ molecule formation and the subsequent 
$\mu$-catalyzed $d + d \rightarrow ^3$He$ + n$
fusion reaction, and (ii) capture neutrons following 
the $\mu + d \rightarrow n + n +\nu$ capture reaction 
from the two $\mu d$ atomic spin states (F=$\frac{1}{2},\frac{3}{2}$).
The $d \mu d$ fusion neutrons are monoenergetic with a kinetic
energy of 2.45~MeV. The $\mu$d capture neutrons---although peaked at energies of 1-3~MeV---include an energetic component that reaches 53~MeV. 
The time dependence of the fusion neutrons and capture neutrons
are shown in Fig.\ \ref{neutrons.fig}.
Additionally, neutrons are emitted following muon capture on 
impurity atoms such as nitrogen, and coincident detection 
of capture recoil and capture neutrons
may assist in separating these rare `impurity' events from
the large fusion background. 

We consider two possible configurations for 
neutron detection: (i) a ``dedicated'' neutron counter setup with 
neutron counters situated immediately outside the
TPC vessel, and (ii) a ``parasitic'' neutron counter setup
with neutron counters situated outside the eSC array similar to the one already used in the MuCap run. The parasitic neutron setup would record neutrons from the full sample of several $10^{10}$ muon stops
in deuterium, whereas the dedicated neutron setup would record
neutrons from a subset of $10^{9}$ muon stops in deuterium.
As a benchmark design for this setup we assume the use of 8 DEMON neutron counters in an identical configuration as used for MuCap. These detectors comprise cells of NE213 organic 
scintillator coupled to 13~cm diameter XP4512B photomultipliers. Each cell 
is 16~cm in diameter, 20~cm in length and contains 4~liters 
of liquid scinitillator. Each cell has a 6.35~mm thick Al 
entrance window and a 21.5~mm thick cylindical walls. The detectors
are read out with custom built 12-bit FADCs, which allow offline 
reconstruction and optimization of their neutron - gamma separation.

\subsubsection{Gamma Detectors}

Auxilary gamma detector are considered for tagging the capture events on
impurities. We discuss them briefly in section~\ref{systematics.sec} on systematics.

\subsubsection{Electronics and Data Acquisition}


The data acquisition (DAQ) system will provide the read out, event building
and data storage for the entrance muon detectors, outgoing electron detectors,
cyrogenic TPC, neutron detectors and various slow control items
such as the HV systems, beamline system, etc. An online analysis layer
will enable both monitoring and diagnostics of the incoming data.

A substantial amount of acquisition infrastructure will be
inherited from the existing MuCap experiment.  Specifically, the readout 
apparatus for the electron detector will be carried forward, while we 
propose upgraded electronics for the new TPC.  The system provides for the 
untriggered readout of data blocks during the livetime of the DAQ, each typically $\sim$0.1~s, 
from many different types of electronics.  It has been demonstrated to
function with total data flow rates up to $\sim$15~MB/s with livetime 
fractions of 80\% or better.  The software is based on the MIDAS framework 
developed at PSI and TRIUMF, and it includes both online compression and 
real-time analysis. 

Details of the various components of the acquisition system are given below:

\begin{itemize}

\item 
{\underline{Electron readout system}}

Data from the electron detector scintillators are recorded in two parallel
data streams.  Leading edge discriminators are used to produce digital 
signals, and these hit times are recorded by a CAEN V767 time-to-digital 
converter (TDC).  This VME module has 128 channels; with an external clock 
speed of 25~MHz, it provides 1.25~ns time resolution.  In addition, a set 
of 500~MHz waveform digitizers 
(WFD), originally developed by the Boston University group for the MuLan/MuCap 
experiments, will be used to record all of the pulse shapes from the electron 
scintilators.  Only the CAEN TDC data were available for the first published 
result, but a comparison with the WFD is underway as part of the full analysis 
effort.

Electron proportional chamber hits are digitized with a system of custom 
multichannel time-to-digital converters.  Named ``COMET'', these devices 
have a time resolution of 20~ns and are able to compress clusters of 
simultaneous hits in nearby wires into a single data word.  While these boards 
are housed in VME crates, they do not transfer data via the standard VME 
protocol; rather, they send it into the acquisition system through a Struck 
SIS3600 latch module.  

\item 
{\underline{TPC readout system}}

A new acquisition sub-system will be implemented for the cryogenic 
TPC. It will provide for the digitization and the readout 
of all pulses on all pads of the TPC. The distribution 
of hits in space and time will enable the three dimensional 
tracking of both incoming muons and charged products 
(e.g. protons, deuterons, etc) from impurity capture. 
Additionally,  the pulse-shape digitization will enable 
particle identification based on energy loss, and be
important for the discrimination between muon stops, 
fusion events and capture events.

The baseline design for the cryogenic TPC is a 10$\times$10
array of anode pads. Each pad will be readout via a custom pulse 
splitter card to two 8-bit waveform digitizer channels (the BU waveform digitizers
that were developed for the MuLan experiment). The combination 
of low gain output and a high gain output 
from the splitter cards will provide the required
energy resolution and dynamic range for both low amplitude
and high amplitude pulses. The digitizers will operate at 
approximately 50~MHz to enable sufficient spatial resolution 
in the TPC drift direction.\footnote{Most likely the digitizers
will be operated at a higher clock frequency with either firmware 
ADC summation or software ADC summation yielding an effective
20~MHz rate.}

The setup will utilize a total of 200 digitizer channels, or 
50 digitizer modules. The digitizers will be distributed over
four VME crates and readout by four rack-mounted frontend processors.
The data will be transferred from the FIFO memories of the waveform 
digitizer channels to the random access memories of the frontend
processors via Struck SIS3100/1100 bridges.

For the estimation of the data rate from the cryo-TPC  
we have assumed an incoming muon rate of 30~kHz, an average 
of 10 pads per incoming muon, a total of 24 ADC samples per pad hit, 
and an average of 2.5 bytes per ADC sample. This yields 
a data rate of about 15~MB/sec in total and about 4~MB/sec 
per VME crate. Lossless compression of the incoming data-stream
is expected to reduce this data volume by one third or better.

\item 
{\underline{Other readout systems}}

Two additional CAEN TDCs are used to record signals from the muon entrance 
counters and other miscellaneous sources.  Finally, the DEMON neutron 
detectors are read out by custom 12-bit, 170~MHz WFDs.

\end{itemize}

A total of eleven front-end crates contribute data to be stored.  Four of these
contain single-board VME computers, while six (those holding the BU 
waveform digitizers that will instrument the TPC and eSC) 
have dedicated PCs connected through Struck SIS1100/3100 VME interfaces; the 
12-bit WFDs interface directly to Ethernet.  Their operations are coordinated
by MIDAS remote procedure calls passed over the gigabit Ethernet network
that interconnects them.  Corresponding blocks of data from all of
these sources are merged together by an event-builder process running on a 
dual-processor PC, which also applies lossless compression techniques.
It then records a copy on a local tape drive and also transfers a copy to 
PSI's central archive system. An online analysis layer -- receiving
a fraction of events -- will be used for integrity checking and online 
histogramming.

\subsection{Monte Carlo Simulation}

\begin{figure}[h]
\begin{center}
\vspace{-0mm}
\includegraphics[height=11cm]{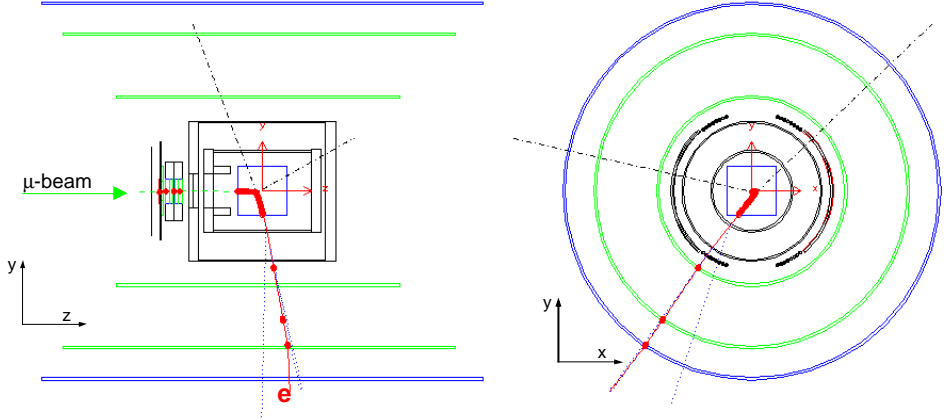}
\vspace{-20mm}
\caption{\label{gmu}
Simulated event showing the different detector components 
included in the present GEANT Monte Carlo. 
The red/green line shows the muon 
passing the muon entrance counters and stopping in the TPC.
An electron (solid red line) and two neutrinos (dashed black lines) are
created at the decay vertex. The electron 
is observed in the ePCs and the eSC, where the timing of the
decay is recorded.}
\end{center}
\end{figure}

We have developed a GEANT-based Monte Carlo program
in order to simulate the MuSun setup.  
A sketch of the setup as it is generated by the simulation program 
together with a typical $\mu$-e decay event is shown in Fig.~\ref{gmu}
for different views of the setup.

\begin{figure}
\begin{center}
\subfigure[\label{MC-stop-energy-left}]{\includegraphics[angle=0,width=0.45\textwidth]{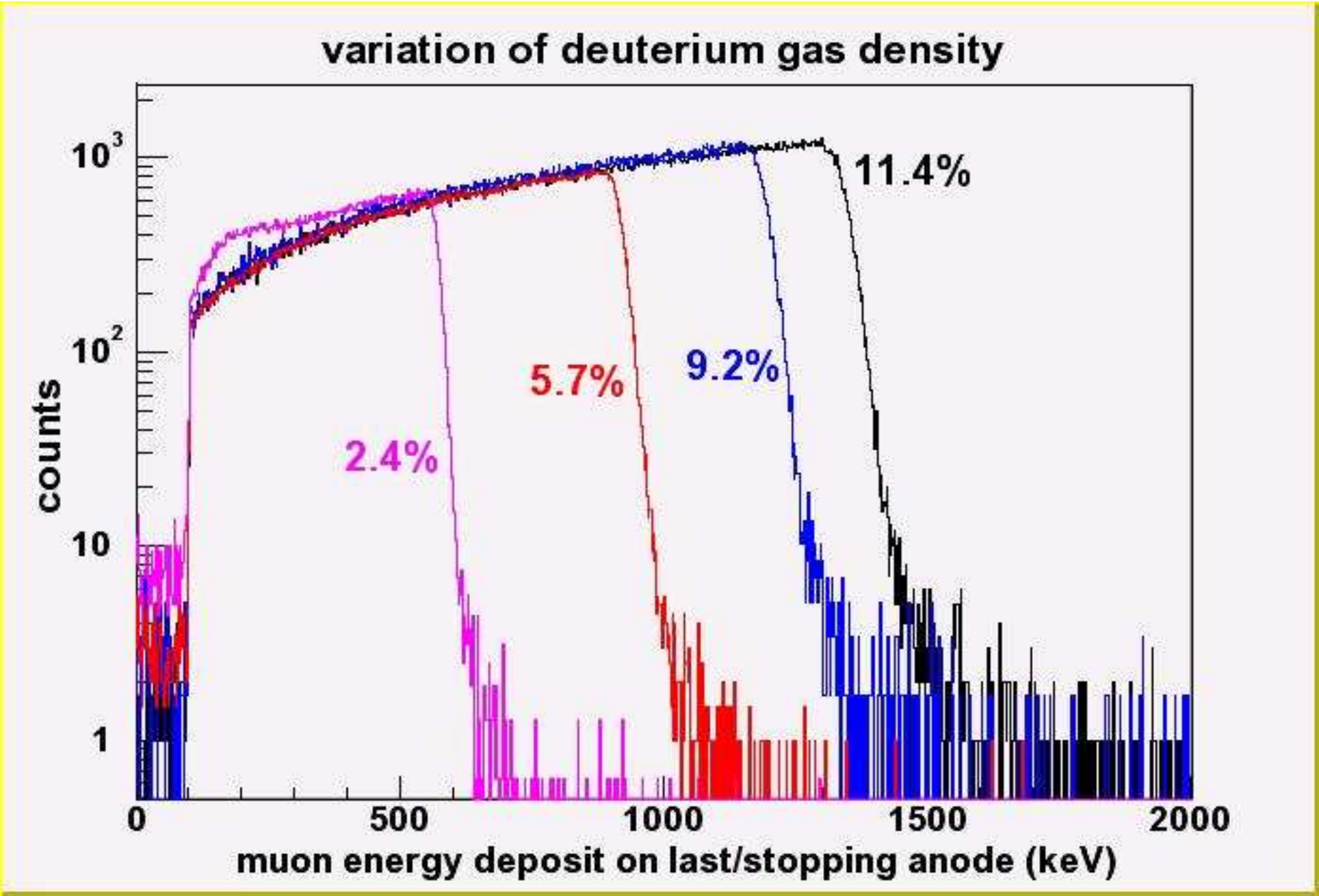}}
\subfigure[\label{MC-stop-energy-right}]{\includegraphics[angle=0,width=0.45\textwidth]{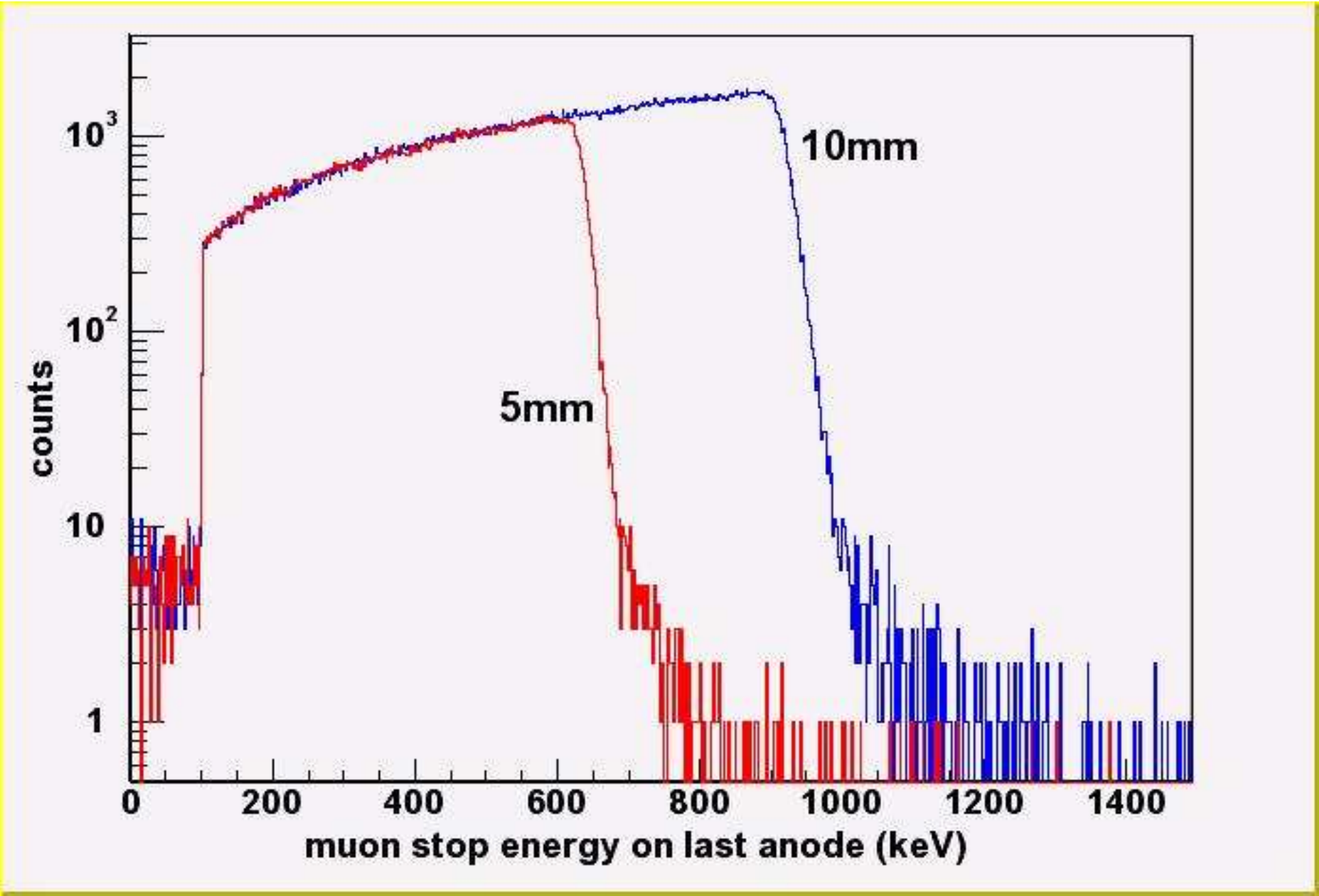}}
\end{center}
\caption{(a) The energy deposition of beam muons on the stopping anode depending on
the deuterium target density given in percent of liquid hydrogen density
as calculated for a TPC with $10\times10$~mm$^2$ padsize.
\newline
(b) The energy deposit of beam muons on the stopping anode 
with two differently sized TPC pad sizes, $5\times5$~mm$^2$ and $10\times10$~mm$^2$
at a target density of 5.7$\%$ of liquid hydrogen density.
}
\label{MC-stop-energy}
\end{figure}

The Monte Carlo is based on our development for MuCAP,
where it has been extensively used.
The program is currently used as a design tool 
to guide the construction of various experimental sub-units.
One study shows the large effect of the target density 
on the energy deposit of the beam muons on the stopping
anode as shown in Fig.~\ref{MC-stop-energy-left} for 
a 10~mm pad structure of the TPC.  
A muon typically deposits a large energy corresponding 
to the Bragg peak on the stopping anode.
This critical study will ultimately 
define the optimal pad size for a given deuterium density,
which was optimized for reasons given by muonic atoms 
and molecular kinetics.
A similar comparison can be done for different pad sizes 
as shown in  Fig.~\ref{MC-stop-energy-right}.

Beyond design studies, we intend to use the Monte Carlo 
to study specific systematic effects.
The overall setup and idea of the simulation is 
a three-stage program sequence 
which starts with GEANT ``data''. 
The next stage applies specific detector properties, 
and then stores the final Monte Carlo events in the data format
as will be delivered from electronic units. 
The resulting file can then be analyzed with standard analysis routines,
which are identically used for ``real'' data.

\section{Statistics and Systematics}

\subsection{Statistics and Rates}\label{statistics.sub}

The MuSun experiment will measure \Rd\ to $<1.5\%$. However, as it will become clearer from the error estimate in the later subsection~\ref{systematics.sec}, a precision of 1.2\% seems to be achievable and will be used as the basis here. Thus $\delta\lambda =\lambda_\mu^- - \lambda_\mu^+$ has
to be measured to a precision of 4.8 \ins. If we assume the final expected
1 ppm error in the $\mu^+$ decay rate from MuLan, $\delta  \lambda_\mu^+$= 0.455 \ins, MuSun then has to measure the $\mu^-$ decay rate to $\delta \lambda_\mu^-$= 4.8 \ins. This can be achieved with  
a statistics of 1.8$\times$10$^{10}$ fully reconstructed $\mu^-$. A similar, but somewhat smaller statistics
of $\mu^+$ events will be collected, which is a powerful systematics check, as most instrumental related systematics 
are identical for $\mu^-$ and $\mu^+$ measured with the same apparatus and hence cancel.

\begin{table}[ht]
\begin{center}
\begin{tabular}{lcc}
\hline
Run	       		&   	                 & weeks beam time\\
\hline
300 K, phase 1          &                        & 10$^*$\\ 
\hline
30 K, phase 2           &                        &    \\
commissioning and setup &                        & 8$^{**}$  \\   
primary data taking 	& N$_-$: $1.8 \times 10^{10}$ & 10\\
                        & N$_+$: $1.2 \times 10^{10}$ & 6 \\
systematics and calibration &                    & 6  \\
\hline
total               	&			 & 40   \\	
\end{tabular}  
\caption{Statistics and beam time estimates. N$_-$ (N$_+$) are the statistics
of fully reconstructed  $\mu^-$ ($\mu^+$) decay events $\mu \rightarrow e \nu \bar{\nu}$ after all selection cuts have been applied. $^*$We will split these weeks into two blocks of 5 weeks, one at the end of 2008 and the second at the beginning of 2009. $^{**}$These weeks include the commissioning of the full setup and individual setup time in case of a non permanent location of the MuSun detector in the $\pi$E3 area. See also the comment on this in the text.}
\label{tab:statistics}
\end{center}
\end{table}

In phase 1 of the run plan the new pad TPC will be commissioned and several key systematic and physics issues will be investigated using a room temperature setup. 
Phase 2 is the full experimental setup operating at cryogenic temperatures. 
The setup time for this complex experiment, including pumping and cooling times and slow HV ramping is significant, if the apparatus has to be craned in and out before each experiment. If we can establish a dedicated area for the experiment as is requested in section~\ref{requesttopsi.sec}, we will save at least 4 weeks of beam time and reduce considerably the risks related to repeated setup and dismantling stages each year. The measurement schedule will be detailed in the beam time request later in this document. Once the full setup is commissioned, the experiment will require two 12 weeks data taking runs.

\begin{table}[ht]
\begin{center}
\begin{tabular}{lrl}
\hline
Selection criteria					& Relative event fraction & Rate (kHz)  \\ 
                                                        & per selection step &    \\
\hline
$\mu SC$ entrance scintillator 				&   	            & 27  \\
 \& full pile-up protected $\mu SC \times \mu PC$       & $\epsilon_1$=0.81  & 22  \\   
 \& stop in TPC fiducial volume                 	& $\epsilon_2$=0.45 & 9.7 \\
 \& fully reconstructed electron 			& $\epsilon_3$=0.61 & 5.9 \\      
\hline
\end{tabular}  
\caption{Relative event fractions between each selection step and resulting event rates for the different cuts applied.}
\label{tab:rates}
\end{center}
\vspace{-5mm}
\end{table}

The translation of statistics into measuring time is based on the realized rates achieved in MuCap. Typically 2$\times$10$^9$ events were collected in one week. Table~\ref{tab:rates} compiles the factors contributing to the final rates. We are
optimizing the new setup to improve $\epsilon_2$, because of the higher gas density. Moreover, the experiment
will benefit from the increase in proton current at PSI anticipated during the next years.

\begin{figure}[ht]
  \begin{center}
  \includegraphics[scale=0.9]{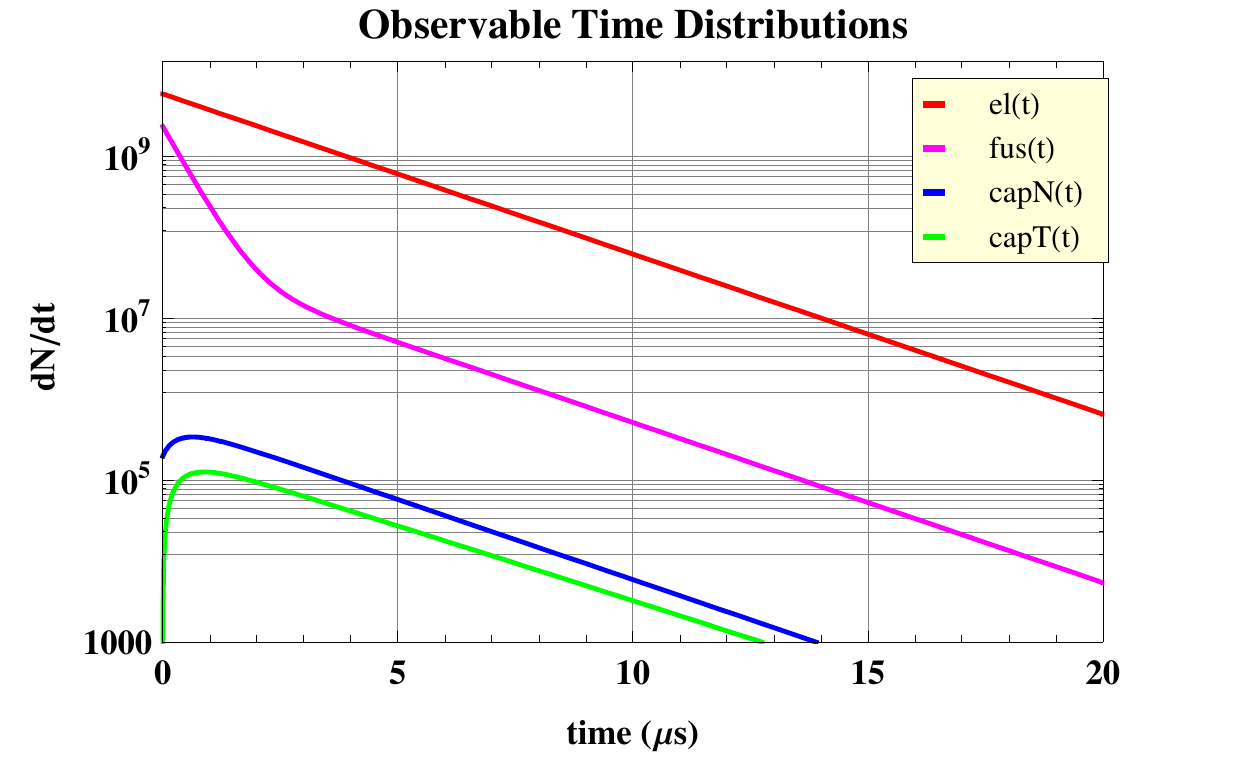}\\
  \vspace{0mm}
  \caption{Observable time distributions for total good muon statistics of N$=3\times 10^{10}$.}
  \label{obs.fig}
  \end{center}
\end{figure}

The MuSun experiment will also derive essential information from the time distributions of capture and fusion products. Their statistics are estimated based on a total number of stopped muons $N$ in the TPC calculated as 

\begin{equation}
N = \frac{N_-}{\epsilon_3}
\end{equation} 

which, for N$_-=1.8\times 10^{10}$, amounts to N = $3\times 10^{10}$. A rough estimate of the expected time distributions and statistics are given in Fig.~\ref{obs.fig} and table~\ref{obs.tab}.

\begin{table}[ht]
\begin{center}
\begin{tabular}{llccc}
\hline
Process 	   		 	& Distribution  & Yield/$\mu$   & Efficiency estimate & Total observed events\\
\hline
$\mu \rightarrow e \nu \bar{\nu}$	& $el(t)$		& 0.9992	&  0.61		&    1.8$\times$ 10$^{10}$ \\	
$dd\mu \rightarrow ^3$He$ + n + \mu$	& $fus(t)$	& 0.0305	&  1.00		&    9.1$\times$ 10$^{8}$ \\	
$\mu+d \rightarrow n + n + \nu$		& $cap_n(t)$	& 0.0015	&  0.01		&    4.5$\times$ 10$^{5}$ \\	
$\mu + ^3$He$ \rightarrow t + \nu$	& $cap_T(t)$	& 1.2$\times$ 10$^{-5}$	& 1.00  &    3.6$\times$ 10$^{5}$ \\
$\mu + N \rightarrow C^* + \nu$                  &              	&       &       &    3.0$\times$ 10$^{5}$ \\
\hline
\end{tabular}  
\caption{Total number of events for different processes based on  N$=3.5\times 10^{10}$ and estimated detection 
efficiencies (column 3). The impurity capture events are based on typical MuCap conditions of $10^{-5}$ observed  
captures/muon. The impurity level should be reduced at T=30~K compared to the MuCap roomtemperature conditions.}
\label{obs.tab}
\end{center}
\end{table}

At $T = 30$~K and $\phi = 0.05$ the time spectrum (Fig.\ \ref{obs.fig}) of 
fusions shows two components: (i) a relatively strong, 
short time constant component with a lifetime that is 
governed by the $\mu$d hyperfine transition rate, and 
(ii) a relative weak, long time constant component
with a lifetime that is governed by the muon dissappearance rate.
Encoded in this time dependence of the fusion products 
are the $d \mu d$ molecular formation rates 
from the two hyperfine states (\qr\ and \dr)
and the hyperfine transition rate between the two hyperfine 
states (\qdr). Consequently, the detection of fusion products 
should enable the determination of the kinetics parameters 
\qdr, \dr\ and \dr\ that are important in the extraction 
of the $\mu$d doublet capture rate 
\RD\ from the decay electron time spectrum.

An interesting feature of muonic deuterium is the
combination of a very large hyperfine dependence of the muon capture rate
with the similar magnitudes of the hyperfine transition 
rate and the muon disappearance rate.
Consequently, the yield of capture neutrons (Fig.\ \ref{obs.fig})
first rises with time due to the
hyperfine transition rate and then falls with time
due to the muon disappearance rate. In principle, the detection
of capture neutrons thus offers a method of 
determining both the hyperfine transition rate \qdr\ 
and the capture rate from the quartet state \RQ\
that are important in the extraction of the $\mu$d doublet capture
rate from the decay electron time spectrum.

Initial simulations based on the detection of fusion products in the \CIC\ reveal that sensitivities to
\qr, \dr\ and \qdr\ of several parts-per-thousand or better were achieved using the time information. 
For the capture neutrons 
the fitting procedure of simulated data results in a determination of
\qdr\ to $\pm$3\% and to \RQ\ to $\pm$8~s$^{-1}$.
The measurement of these parameters to such accuracies
is well beyond the needed precision for the extraction of 
the $\mu d$ doublet capture rate from the decay electron time spectrum. But given the 
long history of difficult and controversial interpretation of basic muon capture experiments
due to muon-induced uncertainties, over-constraining the muon-induced kinetics will
increase the confidence in the extraction of the weak capture rate \RD.

\subsection{Systematics}
\label{systematics.sec}

Table~\ref{tab:systematics} separates the systematic issues for the MuSun experiment into three categories.
For each category the achieved precision of the first MuCap result is given in column 2, the anticipated precision of 
the final MuCap results (based on data already taken and ongoing analyses) is estimated in column 3 and 
the projected uncertainty of the MuSun experiment is presented in the last column. 
 For systematics common to both experiments, only cases which differ between MuSun and MuCap are discussed
  below.

\begin{table}[ht]
\begin{center}
\begin{tabular}{cccc} \hline
\bf Topic 			&\bf MuCap 2007 	&\bf MuCap Final&\bf MuSun	\\	
\hline
Statistics			&	12.59	&	3.7	&	3.4		\\	
\hline
\multicolumn{4}{l}{\bf Similar for MuCap and MuSun}				\\	
\hline
chemical impurities		&	5.0	&	2	&	2$^*$   \\	
analysis methods		&	5	&	2	&	2	\\	
$\mu+p$ scattering		&	3	&	1	&	1$^*$  \\	
$\mu$ pilup veto inefficiency	&	3	&	1	&	1	\\ \hline	
\multicolumn{4}{l}{\bf MuCap only}					\\	\hline
$\mu$d diffusion			&	1.6	&	0.5	&	 	\\	
$\mu$p diffusion			&	0.5	&	0.5	&		\\	
muon-induced kinetics		&	5.8	&	2	&		\\ \hline 	
\multicolumn{4}{l}{\bf MuSun only}					\\	\hline
$\mu$d diff			&		&		&	0.5     \\	
$^1$H contamination		&		&		&	0$^*$  \\	
fusion processes		&		&		&	1$^*$  \\ 	
muon-induced kinetics		&		&		&	0.5$^*$ \\ \hline	
\bf total sys error		&	11.41	&	3.8	&	3.3	\\
\bf total error		&	13.32	&	5.3	&	4.7	\\ \hline	
\end{tabular}
\end{center}
\caption{Comparison of the systematic uncertainties (in \ins). Specific MuSun issues, different from the
MuCap experimental conditions, are marked by $^*$ and evaluated in this section.}
\label{tab:systematics}
\end{table}

\subsubsection{Clean Muon Stop}

\begin{figure}[ht]
  \begin{center}
  \fbox{\includegraphics[scale=0.24]{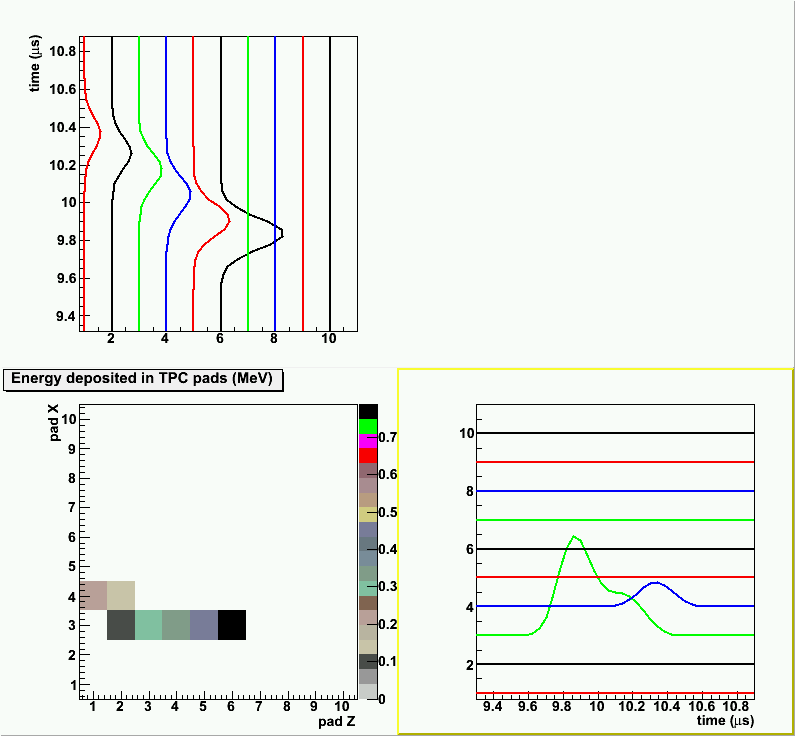}}
  \hspace{2mm}
  \fbox{\includegraphics[scale=0.24]{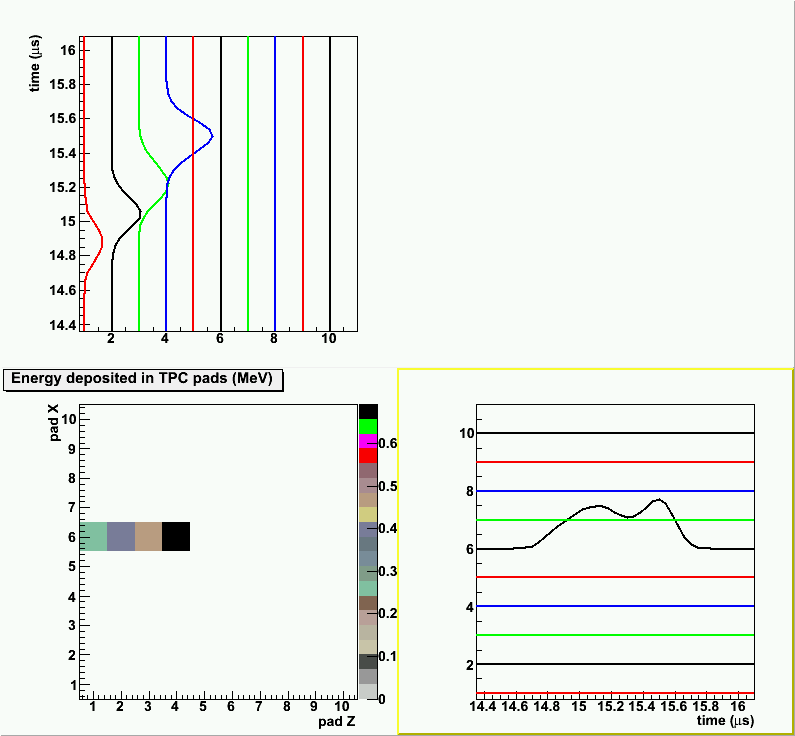}}\\
  \vspace{3mm}\hspace{0.2mm}
  \fbox{\includegraphics[scale=0.24]{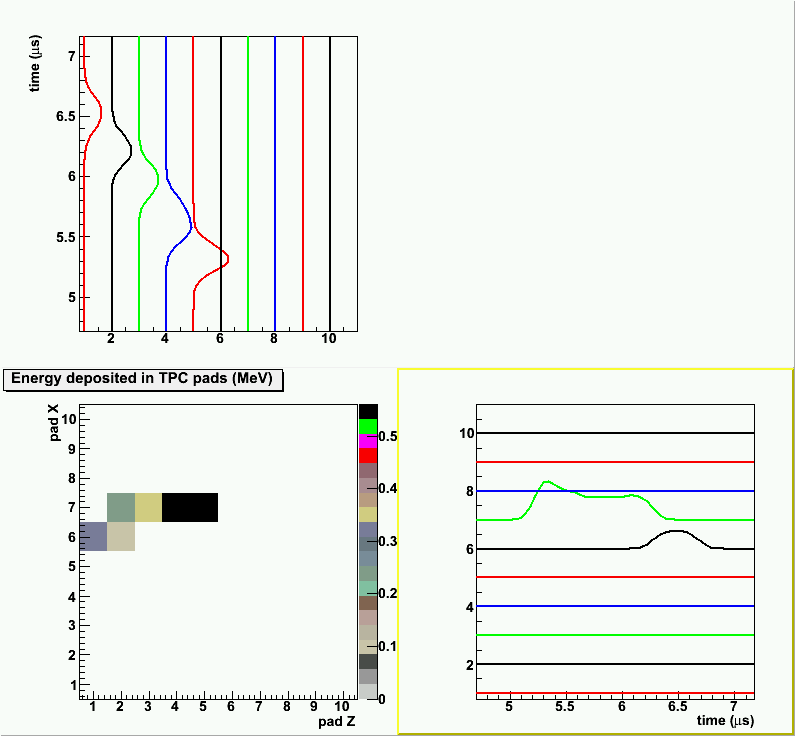}}
  \hspace{2mm}
  \fbox{\includegraphics[scale=0.24]{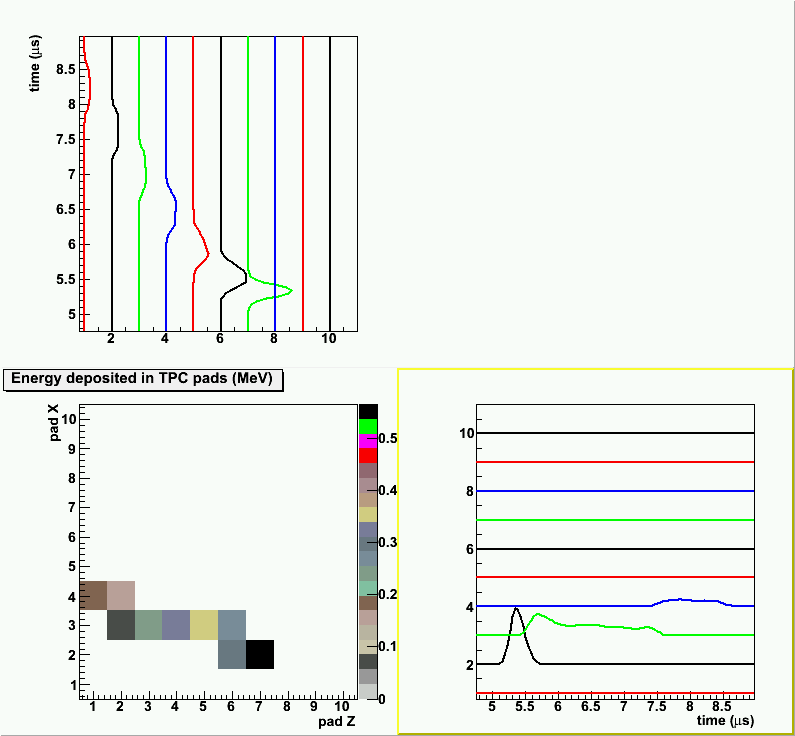}}
  \vspace{0mm}
  \caption{Four Monte Carlo generated muon stop signatures. The original fully three-dimensional information
on charge deposition in the TPC is displayed in the 2-dimensional projections.}
  \label{mustop.fig}
  \end{center}
\end{figure}

Fig.~\ref{mustop.fig} displays four Monte Carlo generated muon stop signatures in the \CIC. A pad TPC 
delivers a full three-dimensional sequence of the charge deposition in space, the figure shows the three two-dimensional
projections. We are working on the algorithm to derive the basic muon parameters, i.e. the vector leading to the
stopping point. The resolution in beam direction $z$ is expected to be a fraction of a pad, based on the \dedx\ information available. In $y$, a resolution of 1 mm, which corresponds to 250\,ns, can be achieved. The 
$x$ resolution
needs to be optimized by simulating the pixel arrangement. We note that a tracking resolution of 5 mm is
more than adequate for the experiment. 

\setlength{\fboxrule}{0.2mm}
 
On the other hand we need to be careful regarding misreconstruction and tracking losses. The typical event
signature of an incoming muon is very clean and simple, but some loss terms have to be controlled at the 100 ppm level. 
E.g.\ in the MuCap experiment, rare $\mu+p$ Coulomb scattering was carefully studied, as it potentially leads to a fake muon stop signature, where the large energy deposition expected at the end
of the muon Bragg curve is faked by the recoil proton signal and the scattered muon escapes undetected.
The full analog \dedx\ information available in MuSun should highly suppress these events. 
Nevertheless, we will study them by Monte Carlo simulations and experimental data. Similar considerations are relevant for 
the overlap between the inital muon track and a fusion signal, discussed below.

\subsubsection{Chemical Gas purity}

The purity requirements are  estimated based on the experiments~\cite{Andreev:2007wg,Bal07}. The literature values for the 
transfer rates are given in table~\ref{tab:transfer}.

\begin{table}[h]
\begin{center}
\begin{tabular}{cccc} 
\hline
Nucleus & E (eV) 	&$\lambda_{pZ}$ (10$^{10}$\ins)  	& $\lambda_{dZ}$ (10$^{10}$\ins)	\\
\hline
N 	&0.04 		&3.4 $\pm$ 0.7 		    	 	&14.5 $\pm$ 0.2 		      	\\
O 	&0.04 		&8.5 $\pm$ 0.2 		 		&6.3 $\pm$ 0.5 				\\
\hline
\end{tabular} 
\end{center}
\caption{Experimental muon transfer rates from $\mu p$ and $\mu d$ atoms to N and O, respectively. 
Transfer rates given for thermal energies, as thermalization is much faster than transfer 
at MuSun experimental conditions. The experimental references collected in the theoretical work~\cite{PhysRevLett.93.043401}.}
\label{tab:transfer}
\end{table}

Let us first estimate the requirements based on the MuCap experience.
The published MuCap data had an observed yield of $Y_Z=10.67$ ppm from the
production data observed with a sensitivity of about 0.1 ppm. Calibration runs were taken with nitrogen-doped protium to determine how the observed disappearance rate deviates with the observed capture yield. This parameter, $\beta=\frac{\Delta \lambda}{Y_{EVH}}$, was found to be $1.30 \pm 0.08 \; (s^{-1}$/ppm).  The efficieny for detecting a capture event was $\epsilon_N = 0.64$. The overall effect on the observed lifetime is $\alpha =\frac{\Delta \lambda}{c_N}$ = 96 Hz/ppm.  Accordingly the yield $Y_{EVH}$ has to be measured or constrained to better than 1.5 ppm to limit 
$\Delta \lambda \le$ 2 Hz. This level of precision was achieved in all production runs.
The main final error came from the  uncertainty in the H$_2$O contribution and the relative 
contribution of humidity and nitrogen, which does not apply to MuSun as no H$_2$O is expected
at cryogenic temperatures. 
It should be mentioned that the detection limit for detecting nitrogen by means of the chromatographic method was around 5 ppb. 

In terms of the nitrogen concentration $c_N$ the requirements are much harder in MuSun than in MuCap. 
According to $\alpha$ given above a precision $\Delta c_N \le$ 20 ppb is required for MuCap. 
In order to have the same correction $\Delta \lambda \propto Y_N$ for MuCap
and MuSun, the following condition must be fulfilled.
\beq
\phi^{MuSun} \Delta c_N^{MuSun} \lambda_{dN} = \phi^{MuCap} \Delta c_N^{MuCap} \lambda_{pN},
\eeq
i.e. $\Delta c_N^{MuSun} = \Delta c_N^{MuCap}/21 \approx$ 1 ppb. 

In summary, it is very likely that we will achieve the required 1 ppb purity at cryo temperatures. However, the
explicit verification of this fact will be hard. It requires to determine $\Delta Y_N$ to 1 ppm or 
alternatively $\Delta c_N$ to 1 ppb. The former condition was easily met in MuCap, the latter condition was
not reached and an upgraded getter for the chromatography system is envisioned.

\subsubsection{Gas Chromatography}

To control the cleaning conditions both chromatography and online detection method are planned. 
Online humidity detector will control the moisture level without design modifications. To increase 
the sensitivity and precision of gas chromatographic method an additional subsystem for accumulation 
of impurities under cryogenic conditions is proposed.  A N$_2$ getter will be placed directly in the
CHUPS flow. In this way several 1000 l of D$_2$ would be passed through this getter per day, 
dramatically increasing the sensitivity compared to our 20 l typical gas samples. The collected
impurity enriched gas will then be analyzed by the gas  chromatograph and a sensitivity
at sub-ppb level is expected.         

\subsubsection{Monitoring by Particle Detection}

\begin{figure}[ht]
  \begin{center}
  \includegraphics[scale=0.45]{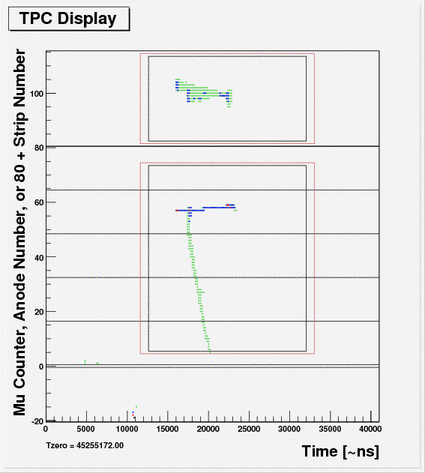}\includegraphics[scale=0.45]{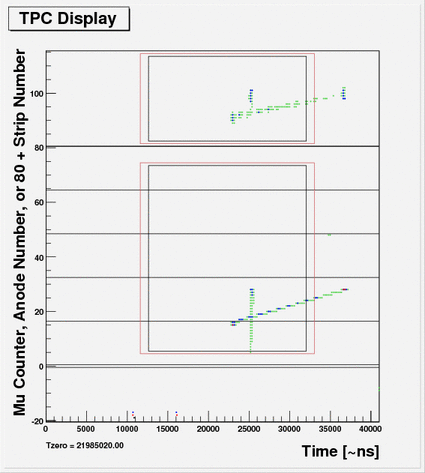}   \\
  \vspace{0mm}
  \caption{Two candidate events for a muon stopping in hydrogen with a subsequent muon capture on a nitrogen
nucleus, leading to a charged particle track emmitted from the point of capture. The information is taken from the MuCap data and shows the $z-y$ (lower box) and $x-y$ projections of tracks in the TPC obtained from the drift time, anode and cathode information.}
  \label{ltracks.fig}
  \end{center}
  \vspace{-0mm}
\end{figure}

MuCap~\cite{Ganzha:2007uk} developed a powerful method to detect capture recoils inside the 
TPC following muon transfer and capture on trace impurities in the protium gas. 
This allowed for continuous in-situ monitoring of the target purity over periods of several months.
Applying the same method for MuSun is a technical challenge. MuCap had the ideal situation that the main reactions
had pure neutral final states only and the maximum muon energy deposition on the TPC anode was below 250 keV,
leaving the 300-500 keV capture recoils cleanly separated from background. As shown in Fig.~\ref{recoil.fig}
a variety of charged fusion recoils are produced in pure D$_2$ and the muons can deposit energies up to 
1 MeV (Fig.~\ref{MC-stop-energy}). Thus excellent resolution and full analog readout are essential in trying to
identifying rare nitrogen capture events. Under the MuSun conditions, the $^3$He background (Fig.~\ref{recoil.fig}) 
can be reduced 14 fold by the combination of a delayed time window after muon stop (where the fusion intensity
has dropped) and by rejecting  capture candiates if a decay electron is observed.
It remains to be experimentally tested whether
1 ppb sensitivity to nitrogen can be achieved with such analysis cuts.

But more likely an additional tag (X-ray, capture topology in \CIC, capture neutron) is required.
If the tagging efficiency $\kappa \approx 0.01$, then we would expect some 1000 tagged
 capture events over the whole run. Probably a $^3$He suppression by the tag by an order of 
magnitude is sufficient. We are studying configuration a), where the additional tagging detector is part of 
the main setup, which has the price of reducing its solid angle, or configuration b), where it is 
positioned close to the TPC vessel wall and a dedicated run without the electron tracker is performed. We will modify 
the beam pipe such that the electron tracker can be rolled upstream, and the new detectors placed around
the TPC, which is rolled a bit downstream relative to its nominal detection position.
Additional detectors can be placed at R$_o$=390 mm and R$_i$=185 mm for configuration a) and b), respectively. 
Naively the ratio of solid angles is  $(\frac{R_o}{R_i})^2 \approx$4 and the signal/noise is favorable for 
configuration b). The tagging processes considered are the following:

\begin{itemize}

\item 
Capture recoil topology in the TPC. We have been analyzing charged particle emission after muon capture in
nitrogen doped data from MuCap. According to our knowledge, there is not reliable literature data on this process.
We find that about 14\% of capture events with a initial nuclear recoil above 300 keV exhibit long range
charged particle tracks (Fig.~\ref{ltracks.fig}), probably dominantly from protons. Currently these
selected events are being parametrized. They will be used as input for Monte Carlo simulations in order to understand the possibility of efficiently detecting and identifying such events with different \CIC\ pad geometries.

\item
X-ray and neutron emission during transfer and capture. During $\mu p$ to $\mu Z$ transfer muonic Lyman X-rays in the
range 102-131 keV are emitted with nearly 100\% probablity. These could be detected with gamma detectors, e.g.
a 10 mm thick NaI slab placed in the isolation vacuum would be an efficient detector. Additionally nuclear X-rays
and neutrons are emitted during the capture process, which also could serve as tags to discriminate against
fusion events.  

\end{itemize}    

While the solid angle for different detector options can be easily calculated, the signal to background ratio
is best explored experimentally. We plan to prepare several test detectors for our stage 1 measurement to optimize
the tagging method.

We also plan protium measurements during the final stage 2 measurements, where the required capture yield sensitivity
is easily obtained. That will allow for an improved understanding of the surface dependent cryo-pumping of our cold target
and will calibrate the various purity monitoring methods.

\subsubsection{Isotopic Gas Purity}

The up-to-now most precise $\mu d$ experiment~\cite{Bardin:1986} measured the effect of $^1$H impurities in
their liquid deuterium target and reported a shift in their measured decay rate of 12 \ins\ for a hydrogen
contamination of c$_p$ $\approx$ 1.6$\times$10$^{-3}$. As the MuSun target density $\phi$ is more than 10 times
smaller, the effect is expected to be reduced by an order of magnitude at the same c$_p$. Moreover, with
the Deuterium Separation Unit we will be able to produce deuterium with c$_p < 10^{-6}$, essentially eliminating this
correction. The use of hydrogen free deuterium will also suppress the $pd\mu$ low energy peak in Fig.~\ref{recoil.fig}
and thus improve our detection capablities for nitrogen capture.

\subsubsection{Uncertainties Introduced by the Muon-induced Kinetics}

The systematic impact of these kinetic effects has been carefully considered in the design of the experiment.
As discussed in section~\ref{targetoptimization.sub} in the optimized experiment at T = 30 K and $\phi$=0.05 the uncertainties
from the kinetics are reduced to less than $\delta\RD$ = 1 \ins, which is significantly below the total precision for this proposal and therefore of no concern.

\subsubsection{Fusion Processes}

As discussed above the muon-catalyzed-fusion processes serve as important monitors of the underlying muon-induced
kinetics. However, at the nominal $\phi$=0.05, the probability for emission of charged fusion 
products is several percent. Does the interference of these tracks with the muon track distort
the $\mu-e$ time distribution?

Let us assume that this interference leads to muon losses with the probablity $\eta(t_f)$, depending on the
time after muon stop $t_f$, when the fusion occurs. The probablity to lose an electron decay at 
time $t_e$ is then
\beq
P_{loss}(t_e)= \int_0^{t_e}  \eta(t_f) (\phi \qr N_q(t_f) + \phi \dr N_d(t_f)) e^{\rz t} dt_f.
\eeq
The time distribution under the integral is the solution of the kinetic equations Eq.~\ref{eq:sol}, setting \rz = 0 
(as the muon decays only at $t_e$). Conservative estimates of the impact of this time dependent effect
indicate that the loss probablity $\eta(t_f)$ should be kept below about 1\%. Muon stop reconstruction at this
level can be achieved with appropriately defined muon stop cuts. We plan to study
this question with Monte Carlo and during our test run. It is worth noting that the fusion neutron distribution
would directly reflect potential muon losses due to the $^3$He$+n$ branch of the fusion distribution. If muons 
are lost because of the above mentioned effect, the observed distribution $fus(t)$ would be multiplied by a factor
$(1-\eta(t_f))$, providing a direct experimental handle on  $\eta(t_f)$.

\subsubsection{Polarization of $\mu d$ atoms}

As described in detail in Appendix \ref{msr.tex}, 
a potential complication in the time spectrum 
of the decay electrons from the $\mu^-$d atoms 
is the presence of a $\mu$SR signal.
Using a Monte Carlo simulation of the $\mu$SR modulation 
-- with estimates of the muon beam polarization,
atomic cascade depolarization, D$_2$ collisional depolarization, 
and the anisotropy of the detection efficiency --
we studied the effects of the $\mu$SR signal on the
electron time spectrum. Based on our experience with
$\mu$SR effects in the MuLan experiments, 
we expect the effects on \Rd\ to be considerably smaller than
the proposed precision $\pm$6~s$^{-1}$.

\section{Measuring Program}

\subsection{Stage 1 - Room Temperature TPC}
\label{stage1.sub}

The experiment will proceed in two stages. First we will prepare a prototype of
the new pad TPC. The pad plane layout and Frisch grid will be identical to the
final TPC, but the chamber will operate at room temperature at density $\phi$=1\%
(MuCap conditions). This central detector will be a reconfiguration of the second TPC, which the 
collaboration prepared for the MuCap experiment~\footnote{The first MuCap TPC, 
used in our ultrapure protium measurements, will be left untouched in order to keep the possibility of running with the protium conditions again, if ever necessary.}. For this first stage, we request 10 
weeks of beam time for commissioning and physics running with this new chamber. Essential 
technical goals of this stage include:
\begin{itemize}
\item
Demonstrate excellent resolution and muon identification with new TPC operated as ionization chamber.
\item
Identification and separation of fusion recoils.
\item
Full analog readout of whole TPC in untriggered mode.
\end{itemize}

The physics goals are as follows
\begin{itemize}

\item
Measurement of the tranfer rate from deuterium to nitrogen.

\item
Attempt to monitor impurities by detection of capture events in the presence of fusion background
with dedicated set-up.

\item
Observation of residual polarization of muons in the $\mu d$ quartet state.

\item
In addition, depending on the advanced status of the setup, also an attempt for a first 
capture rate measurement can be envisioned although the items above are of more relevance in this period.

\end{itemize}
The general goal of this phase is to collect data required to optimize the final detector
in terms of performance and of systematic issues generated by physics background. Moreover,
new components should be tested and optimized before building the final detector which is integrated
in a complex cryo system. During the fall run, if the new pad TPC is ready, we would primarely work on 
commissioning this new detector and on dedicated experiments with auxilary detectors to 
develop the best method to tag impurities. 
The electron tracker then could be added during the shut-down period, to prepare for the 
systematically essential transfer rate experiment after the shutdown.
The full analog readout of the TPC in real time will be implemented already in 2008. 

The collaboration is working on an optimized schedule for 2008 and will discuss a more detailed plan at the PAC meeting.

\subsection{Stage 2 - Cryo-TPC and \Rd\ Determination}

By fall of 2009 the high density \CIC\ should be ready and a first commissioning run
is planned. The further requests will depend on the experience gained and whether a permanent experimental
cage can be established for the MuSun experiment in the $\pi$E3 area, which would very significantly
increase the scheduling flexibility. If the new detector works as expected and the systematic issues 
outlined in section~\ref{systematics.sec} are under control, we would focus on the determination of 
the capture rate \Rd\ to a precision of 1.2\%. The required statics can be achieved in two- to three 10-12 weeks
runs including set-up, distributed over a two-years period. As has been successful in MuCap we usually
split each run into main $\mu^-$ data taking, $\mu^+$ reference data and systematic studies, as significant
systematic cancellation occur when comparing data taking within a run period.

\section{Organization}

\subsection{Responsibilities and Budget}

The division of responsibilities between the participating
institutions will follow the lines depicted in
table~\ref{responsibilities.tab}, though there will be, as in the
past for MuCap, significant collaborative overlap. We do not
elaborate on the expertise and resources of the individual
institutions, as the present plan broadly follows the concept which
has successfully worked for MuCap over many years.

\begin{table}[ht]
\begin{center}
\begin{tabular}{cccccccc} \hline
{\bf System}    &\multicolumn{7}{c}{\bf responsible institutions}\\
       &\bf PNPI    &\bf UIUC    &\bf PSI    &\bf UKY    &\bf BU        &\bf UCL    &\bf RU     \\   \hline
Detectors    &         &$\bigodot$        &        &$\bigodot$        &        &$\bigodot$        &        \\
TPC        &$\bigodot$        &$\bigodot$        &        &        &       &        &        \\
Cryogenic system&$\bigodot$        &        &$\bigodot$        &        &        &        &        \\
Gas and purification
system        &$\bigodot$         &        &$\bigodot$        &        &        &        &        \\
Front end electronics &$\bigodot$   &$\bigodot$        &$\bigodot$        &        &  $\bigodot$  &$\bigodot$        &        \\
DAQ + computers           &        &        &        &$\bigodot$        & & &$\bigodot$        \\

\hline   \end{tabular}
\end{center}
\caption{Main hardware responsibilities of participating institutions.}
\label{responsibilities.tab}
\end{table}

For computing requirements, we plan to maintain and upgrade the
existing on-line analysis cluster developed by the MuCap and MuLan
Collaborations to cope with the very significant data volume. We
request about 100~TB of data storage space on the PSI archive over
the next three years. We expect that the bulk of the off-line
analysis will be performed at the US National Center for
Supercomputing Application (NCSA), where we have developed analysis
software structures and experience over the last years and have
obtained significant allocations of several 100k CPU hours on the
required multi-node system.

The equipment costs can be kept low, because MuSun benefits from
much higher investments by MuCap and MuLan, providing important
experimental infrastructure. This includes the MuCap detectors,
electronics, high vacuum and purification system and the MuLan
kicker, FADC electronics, as well as the DAQ infrastructure of both
experiments, to name a few major examples. We estimate the overall
new equipment expenses for this experiment to be of the order of
350k~CHF. The main cost driver will be the cryo-TPC, which also has
a significant labor-intensive component. Based on past MuCap/MuLan
experience, additional operating expenses for running the experiment
(travel, shipping, storage media, incidentals like chamber gas,
etc.) amount to about 100k~CHF per year. The collaborating institutions
have discussed the new project with their respective funding agencies
(mainly the National Science Foundation in the US and the Russian Academy
of Sciences in Russia) and received encouraging responses. Naturally,
acceptance of this proposal by the PAC is critical to go forward with
full funding proposals.

We welcome additional groups to join the MuSun experiment.

\subsection{Request to PSI\label{requesttopsi.sec}}

Our requests to PSI are based on the PSI expertise, which was
critical to the success of the MuCap experiment. Different from
MuCap, the PNPI group will assume the main responsibility for the
new cryo-TPC development, with direct support from UIUC and
infrastructure from PSI as is available.

The financial request to PSI is based on the MuCap experience and 
amounts to 20 kCHF/year for equipment costs and 30 kCHF/year for running costs.
The equiment budget includes contributions to the new
cryogenic system, the gas handling and purification system, and the
100~kV supply (15k~CHF), which is required for the main drift field.
The running costs involve magazine items, small orders and
partial support of the PNPI visits (guest house,
travel support).

As described earlier, MuSun will require several weeks of set-up
time before the data taking.  A permanent location of the experiment
in the annex of $\pi$E3 beamline area would greatly reduce technical
risks and enhance efficiency, allowing for a flexible work schedule
for the teaching faculty members within our collaboration.

\subsection{Project Schedule}

\begin{landscape}
\begin{figure}[ht]
\vspace{-10mm}
 \begin{center}
 \includegraphics[scale=0.8]{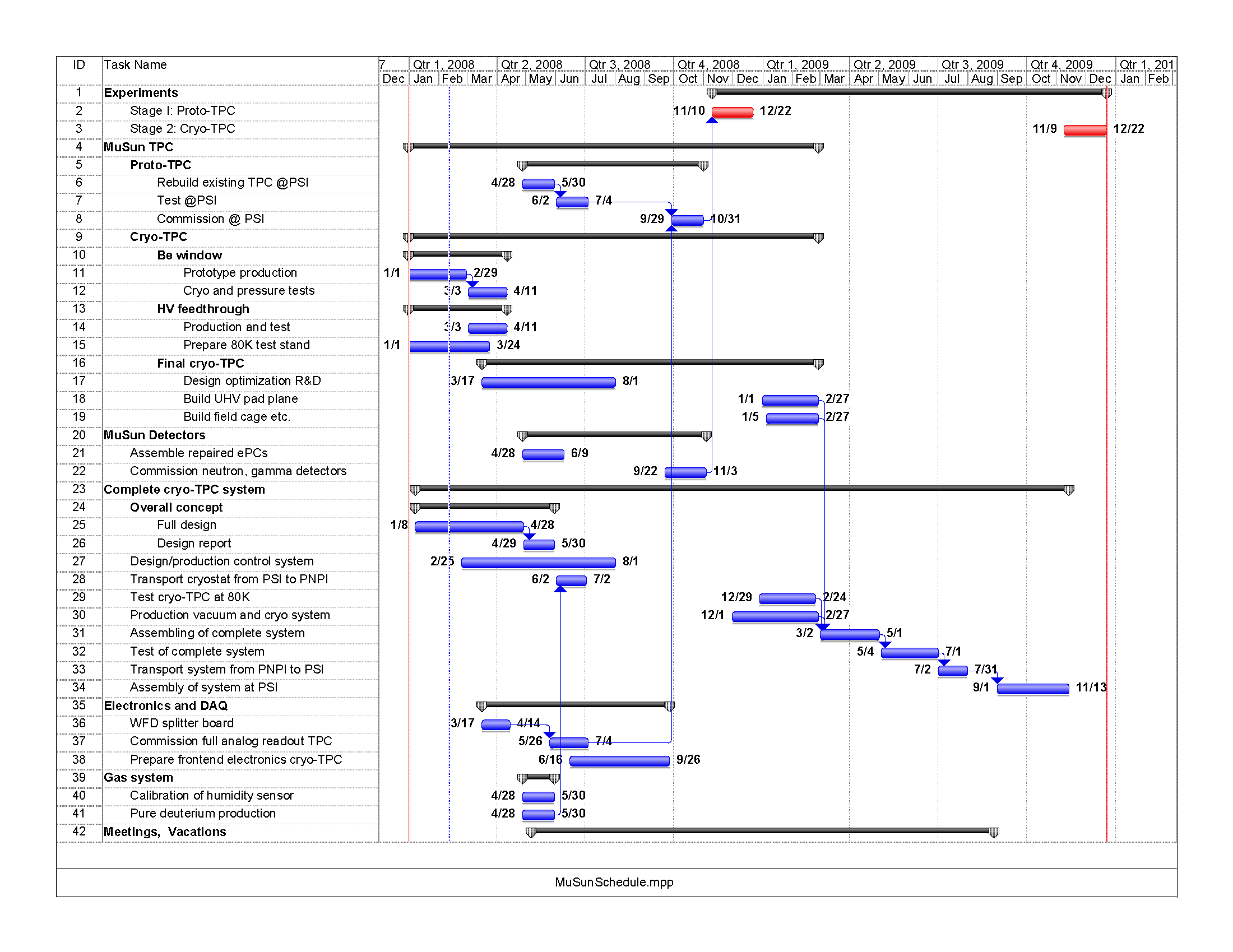} \\
 \vspace{0mm}
 \caption{Preliminary project schedule gantt chart}
 \label{gantt.fig}
 \end{center}
 \vspace{-0mm}
\end{figure}
\end{landscape}

Fig.~\ref{gantt.fig}  provides a preliminary overview of the project
schedule. The present version is a work-in-progress and will be
updated as the planning evolves. The project has two main
directions. In 2008, we plan to build the room temperature prototype
TPC (proto-TPC) and perform first measurements at the end of the
running year.  These measurements are essential for the optimization
of the final chamber design and for the overall experiment. We are
optimistic that we can achieve these goals within one run. In that
case we would not need the additional 5 weeks early 2009, which are
mentioned in the beam request. In parallel, we will perform basic
R\&D towards the cryo-TPC, including the detector itself, the
cryogenic vessel, window and feedthrough and the overall concept.
Based on the learning experience of the fall 2008 run, we plan to
proceed with the construction of the final apparatus early in 2009
and have a first commissioning run in fall 2009.

\newpage
\input{Reference.tex}

\newpage
\section{Appendix}

\subsection{Polarization and Muon Spin Rotation}
\label{msr.tex}

A potential complication in the time spectrum 
of the decay electrons from the $\mu^-$d atoms 
is the presence of a $\mu$SR signal.
If the $\mu^-$d atoms have a non-zero polarization,
their spins will precess and relax in the environmental magnetic field
of the muon stopping volume
(the precession frequencies are 
$\omega_{3/2}/B = 0.026$~$\mu$s/G for the quartet state
and $\omega_{1/2}/B =  0.034$~$\mu$s/G for the doublet state).
Due to the directional correlation between 
the muon spin vector and the electron momentum vector
this imparts a $\mu$SR modulation onto the
time spectrum of the decay electrons.
Note such effects were absent 
for the singlet $\mu^-$p atoms 
in the muCap experiment.


For a perfectly isotropic detector
this $\mu$SR signal would vanish 
in the summed time spectrum 
of electrons emitted in all directions.
However, any anisotropies in the detection efficiency
about the $\mu$-spin axis 
will lead to a $\mu$SR modulation
of the electron time spectrum.
Below we describe our simulations 
of the influence of the $\mu$SR signal, both 
the spin precession and the spin relaxation,
on the determination
of the $\mu^-$d effective lifetime
from the electron time spectrum. 
Our simulations are based on
estimates of: (1) the $\mu^-$ polarization of the 
cloud muon beam, (2) the initial $\mu^-$ depolarization 
in the atomic cascade process,
(3) the subsequent $\mu^-$ depolarization 
via collisions with D$_2$ molecules, and (4) the
anisotropy in the electron detection efficiency.

\begin{enumerate}

\item
for the $\pi$E3 channel 
the available data for negative cloud muons 
with momenta 35-45~MeV/c imply a polarization of about 25\% 
at 32.6 MeV/c. This result is consistent with our own determination 
of the beam polarization for positive cloud muons at 32.6 MeV/c, 
which yielded a value of 22\%. Herein, we assume a value
of 25\% for the negative cloud muon beam polarization. 

\item
the particular case of muon depolarization in $\mu^-$d cascade 
was considered by Uberall  \cite{Uberall:1959} and Dzhelepov 
and Fil'chenkov \cite{Dzhelepov:1983}. Using a spin-orbit 
depolarization factor of 1/6
from Ref.\ \cite{Dzhelepov:1983} and spin-spin depolarization factors  
of $10/27$ ($F = 3/2$ state) and $1/27$ ($F = 3/2$ state) 
from Ref.\ \cite{Uberall:1959}, we obtain overall depolarization factors 
of 0.17 and 0.03 for the $F = 3/2, 1/2$ hyperfine states,
respectively.
\footnote{Unfortunately, the experimental data on $\mu$SR in deuterium 
is limited and confusing. Bin'ko {\it et al.}\ \cite{Binko:1989wg}
studied $\mu^-$ depolarization in 300K, 10 Atm D$_2$
gas and reported an initial
polarization of quartet atoms of (7.2$\pm$2.1)\%.
Bystritskii {it et al.}\ \cite{Bystritskii:1981}
studied $\mu^-$ depolarization in 300K, 40 atm D$_2$
gas and reported an initial
polarization of quartet atoms of (1.0$\pm$0.9)\%.}
Herein, we assume a quartet state depolarization factor
of 0.17 and ignore the much smaller effects 
of the doublet state polarization. 

\item
after formation of ground state $\mu^-$d atoms
their initial polarization will further relax
due to exchange collisions with the surrounding D$_2$ molecules.
Note, both collisions which (a) change the $\mu^-$d 
atom's spin and (b) change the $\mu^-$ spin
projection will contribute to relaxation,
and therefore the rate of spin relaxation 
can exceed the measured hyperfine 
transition rate. Herein, we assume 
value of 2.3$\times$10$^6$~s$^{-1}$ at $\phi = 0.05$
from Ref.\ \cite{Dzhelepov:1983} for the relaxation rate of the quartet state.

\item
the detector anisotropy has contributions including
detector solid angle variations and detector intrinsic efficiency
variations and was estimated from our experience with 
the muCap setup. Herein, we shall assume an anisotropy 
of $\epsilon = 0.02$ in the detection efficiency.

\end{enumerate}

In order to estimate the influence of a
$\mu$SR signal on the determination of the 
$\mu^-$d lifetime we performed a Monte Carlo simulation. 
Electron time spectra were generated according
to a exponential decay law with a $\mu$SR 
modulation according by

\begin{equation}
N \exp( -t / \tau ) ( 1 + A \exp( -t/ \tau_R ) \cos( \omega t + \phi ) )
\end{equation}

where $\tau$ is the muon's effective lifetime 
and $A$, $\omega$, $\phi$ and $\tau_R$ are the 
amplitude, frequency, phase and relaxation constant
of the $\mu$SR signal. The amplitude $A = 0.013$ was determined
by the product of the initial beam polarization (0.25), 
atomic cascade depolarization (0.17) and the asymmetry coefficient
of the muon spin/electron momentum directional correlation (0.3).

To incorporate the detector anisotropy $\epsilon$
we  generated (i) a ``forward hemisphere'' time spectrum
with phase $\phi = 0$ and total counts $N_o / 2$ and (ii)
a ``backward hemisphere'' time spectrum
with phase $\phi = \pi$ and total counts ($N_o / 2$)$\times$($1 + \epsilon$),
and then summed the two time spectra.
For $\epsilon = 0$ the $\mu$SR signal vanishes
in the sum spectrum, whereas for $\epsilon > 0$ 
a diluted version of the individual $\mu$SR signals appears in the sum spectrum.
Next the sum spectra
were fit to extract the lifetime. In the first fitting procedure, 
the ``worst case'' scenario, we 
fit a single exponential, {\it i.e.}\ completely ignoring the time structure
of the $\mu$SR signal.   In the second fitting procedure, the 
``best case'' scenario, we fit a single exponential with
the $\mu$SR function, {\it i.e.}\
correctly incorporating the time structure of the $\mu$SR signal.

In the ``worst case'' scenario we found 
a lifetime shift of up to $\pm$10~ppm (equivalent to
a shift in the doublet rate of about $\pm$5~s$^{-1}$). 
Note the shift in the fitted lifetime was correlated with the fit start time,
the sinusoidal time dependence of the $\mu$SR signal inducing 
a sinusoidal time variation of the extracted lifetime with the fit start time. 
Also note that the omission of the $\mu$SR signal in the fits
was obvious in the poor $\chi^2$'s and the large residuals for these fits.

In the ``best case'' scenario we found 
no shift in the fitted lifetime at the level of about $\pm$1~ppm.
A small (sub ppm) increase in the statistical uncertainty on the 
fitted lifetime, that is presumably from the correlations with the
$\mu$SR parameters, was, however, observed. 
In this fitting procedure the fit $\chi^2$ was always acceptable.

In reality, we expect the true situation to fall between
the ``best case'' and ``best case'' scenarios. 
While its impossible to exactly know the time structure 
of the $\mu$SR signal, the difference spectra between 
forward/backward hemispheres  will obviously help 
in characterizing its time structure. In short, the
$\mu$SR effect is not expected to be significant problem 
in the lifetime determination.

\end{document}